%% file: main_arxiv.tex
\documentclass[sigconf, nonacm]{acmart}
\usepackage{pvldb}

\renewcommand\vldbdoi{XX.XX/XXX.XX}
\renewcommand\vldbpages{XXX-XXX}
\renewcommand\vldbavailabilityurl{https://github.com/imoumoulidou/perception-aware-sampling} 

\usepackage{optidef}
\usepackage{xspace}
\usepackage{xcolor}
\usepackage{amsfonts}
\usepackage{bm}

\usepackage{graphicx}
\usepackage{subcaption}
\usepackage{multicol}
\usepackage{multirow}
\usepackage{rotating}
\usepackage{enumitem}

\usepackage{booktabs}
\usepackage{float}

\usepackage{placeins}

\definecolor{randcolor}{HTML}{60A216} 
\definecolor{mmcolor}{HTML}{028A82} 
\definecolor{vascolor}{HTML}{F3C30F} 
\definecolor{bncolor}{HTML}{7570b3} 
\definecolor{dbscolor}{HTML}{E7298A} 
\definecolor{pacolor}{HTML}{04247C} 
\definecolor{approxcolor}{HTML}{FF9307} 

\definecolor{highcompcolor}{HTML}{FEE6CE}
\definecolor{medcompcolor}{HTML}{FDAE6B} 
\definecolor{lowcompcolor}{HTML}{04247C} 

\usepackage{algorithm}
\usepackage[noend]{algpseudocode}
\usepackage{setspace}

\usepackage[many]{tcolorbox}

\newcommand\mycommfont[1]{\ttfamily\textcolor{gray}{#1}}
\algnewcommand{\LineComment}[1]{\State{\scriptsize\mycommfont{$\triangleright${#1}}}}
\algnewcommand{\LineCommentx}[1]{\Statex{\scriptsize\mycommfont{$\triangleright${#1}}}}

\newcommand{\newcontent}[1]{{\color{black}#1}} 

\newcommand{\map}{\mathcal{M}\xspace} 

\newcommand{\universe}{\mathcal{D}\xspace}
\newcommand{\sample}{\mathcal{D'}\xspace}
\newcommand{\vis}{V\xspace}
\newcommand{\samplesize}{k}

\newcommand{\weights}{\mathcal{W}\xspace}

\newcommand{\nocolorpa}{\textsc{PAwS}\xspace}
\newcommand{\nocolorvas}{\textsc{Vas}\xspace}
\newcommand{\nocolorran}{\textsc{Random}\xspace}
\newcommand{\nocolordbs}{\textsc{DBS}\xspace}
\newcommand{\nocolorbn}{\textsc{Blue-noise}\xspace}
\newcommand{\nocolorMM}{\textsc{Max-Min}\xspace}
\newcommand{\nocolorapproxpa}{\textsc{ApproPAwS}\xspace}

\newcommand{\partitions}{\mathcal{P}\xspace}
\newcommand{\representatives}{\mathcal{R}\xspace}

\newtheorem{example}{Example}

\newtheorem{definition}{Definition}

\usepackage{tikz}
\newcommand\emphbox[2][fill=black!10]{%
    \tikz[baseline]\node[%
        inner ysep=0pt, 
        inner xsep=2pt, 
        anchor=text, 
        rectangle, 
        rounded corners=1mm,
        #1] {\strut#2};%
}

\newcommand\frbox[2][draw=black!30]{%
    \tikz[baseline]\node[%
        inner ysep=0pt, 
        inner xsep=2pt, 
        anchor=text, 
        rectangle, 
        rounded corners=1mm,
        #1] {\strut#2};%
}

\newcommand\paFr[2][draw=pacolor]{%
    \tikz[baseline]\node[%
        inner ysep=0pt, 
        inner xsep=2pt, 
        anchor=text, 
        rectangle, 
        rounded corners=1mm,
        #1] {\strut#2};%
}
\newcommand{\pa}{\paFr{\textsc{PAwS}}\xspace}

\newcommand\vasFr[2][draw=vascolor]{%
    \tikz[baseline]\node[%
        inner ysep=0pt, 
        inner xsep=2pt, 
        anchor=text, 
        rectangle, 
        rounded corners=1mm,
        #1] {\strut#2};%
}
\newcommand{\vas}{\vasFr{\textsc{Vas}}\xspace}

\newcommand\ranFr[2][draw=randcolor]{%
    \tikz[baseline]\node[%
        inner ysep=0pt, 
        inner xsep=2pt, 
        anchor=text, 
        rectangle, 
        rounded corners=1mm,
        #1] {\strut#2};%
}
\newcommand{\ran}{\ranFr{\textsc{Random}}\xspace}

\newcommand\dbsFr[2][draw=dbscolor]{%
    \tikz[baseline]\node[%
        inner ysep=0pt, 
        inner xsep=2pt, 
        anchor=text, 
        rectangle, 
        rounded corners=1mm,
        #1] {\strut#2};%
}
\newcommand{\dbs}{\dbsFr{\textsc{DBS}}\xspace}

\newcommand\bnFr[2][draw=bncolor]{%
    \tikz[baseline]\node[%
        inner ysep=0pt, 
        inner xsep=2pt, 
        anchor=text, 
        rectangle, 
        rounded corners=1mm,
        #1] {\strut#2};%
}
\newcommand{\bn}{\bnFr{\textsc{Blue-noise}}\xspace}

\newcommand\approxFr[2][draw=approxcolor]{%
    \tikz[baseline]\node[%
        inner ysep=0pt, 
        inner xsep=2pt, 
        anchor=text, 
        rectangle, 
        rounded corners=1mm,
        #1] {\strut#2};%
}
\newcommand{\approxpa}{\approxFr{\textsc{ApproPAwS}}\xspace}

\newcommand\mmFr[2][draw=mmcolor]{%
    \tikz[baseline]\node[%
        inner ysep=0pt, 
        inner xsep=2pt, 
        anchor=text, 
        rectangle, 
        rounded corners=1mm,
        #1] {\strut#2};%
}
\newcommand{\MM}{\mmFr{\textsc{Max-Min}}\xspace}

\usepackage{colortbl}

\usepackage{soul}

\newif\ifextended
\extendedtrue

\newcommand\appref[2]{%
  \ifextended
    Appendix~\ref{#1}%
  \else
    our technical report~\cite{#2}%
  \fi
}

\begin{document}

\title{Perception-Augmented Databases for Effective Visual Analytics}

\author{Zafeiria Moumoulidou}
\affiliation{%
  \institution{Meta Inc.}
   \country{}
}
\email{ir.moumoulidou@gmail.com}

\author{Hamza Elhamdadi}
\affiliation{%
  \institution{Stonehill College}
   \country{}
}
\email{helhamdadi@stonehill.edu}

\author{Redwan Ahmed Rizvee}
\affiliation{%
  \institution{UMass Amherst}
   \country{}
}
\email{rrizvee@umass.edu}

\author{Ke Yang}
\affiliation{%
  \institution{UT San Antonio}
   \country{}
}
\email{ke.yang@utsa.edu}

\author{Subrata Mitra}
\affiliation{%
  \institution{Adobe Research}
   \country{}
}
\email{subrata.mitra@adobe.com}

\author{Cindy Xiong Bearfield}
\affiliation{%
  \institution{Georgia Tech}
   \country{}
}
\email{cxiong@gatech.edu}

\author{Alexandra Meliou}
\affiliation{%
  \institution{UMass Amherst}
   \country{}
}
\email{ameli@cs.umass.edu}

\renewcommand{\shortauthors}{Moumoulidou et al.}

\begin{abstract}
    Analysts often use sampling in visual exploration to overcome the computational and perceptual bottlenecks of working with massive datasets.
    Yet, existing sampling methods optimize for statistical or geometric data properties, without explicit consideration for how humans \emph{perceive} a visualization. We propose \emph{perception-augmented databases}: a database framework that stores perceptual models of data as precomputed metadata and exposes a new query operator---\emph{perception-aware sampling}---that selects subsets optimized for how humans visually perceive the result.
    We make the following contributions: (1)~We introduce \emph{perception-augmented databases} and design \nocolorpa, a \emph{perception-aware} sampling operator for scatterplot visualizations that leverages saliency maps---a computer vision tool for predicting areas of visual attention---combined with density and coverage objectives. (2)~We design \nocolorapproxpa: a fast approximate visualization method that works directly over a perception-aware compressed synopsis of the data, achieving up to $100\times$ speed-ups with minimal loss in visual fidelity, and without accessing the underlying data at query time. (3)~We contribute a \emph{perceptual similarity} metric that evaluates the quality of a sample by comparing its saliency map to that of the original dataset, a novel evaluation methodology, applicable to any sampling method for scatterplots. (4)~Our extensive experimental evaluation over datasets up to 3.5M points demonstrates that our methods consistently outperform five state-of-the-art sampling algorithms. A user study confirms that \nocolorpa samples are preferred by humans, validating our quantitative findings.
\end{abstract}

\ccsdesc[500]{Information systems~Data management systems}
\ccsdesc[500]{Human-centered computing~Visual analytics}

\keywords{perception-augmented databases, sampling, visualization}

\maketitle

\vldbtopmatter

\input{1_Introduction_arxiv27}
\input{2a_Background_arxiv27}
\input{3_Algorithms_arxiv27}
\input{3a_Algorithms_arxiv27}
\input{4_Metrics_arxiv27}
\input{5_Experiments_arxiv27}
\input{2_RelatedWork_arxiv27}
\input{7_Conclusions_arxiv27}

\bibliographystyle{ACM-Reference-Format}
\bibliography{literature} 

\clearpage
\appendix
\input{6_Appendix_arxiv27}

\end{document}

%% file: 1_Introduction_arxiv27.tex
\section{Introduction}\label{sec:intro}
Data visualizations are a powerful tool commonly used by practitioners to understand, analyze, and find interesting patterns in their data. Scatterplots, in particular, are one of the most popular and frequently used visualization types~\cite{micallef2017towards, yuan2020evaluation, sarikaya2017scatterplots}. Their efficiency and usability have been evaluated in various visualization tasks, such as those identified by Amar et al.~\cite{amar2005low}: from identifying outliers and clusters to correlation trends~\cite{quadri2021survey, quadri2022automatic, rensink2010perception, harrison2014ranking, quadri2020modeling}.

However, data analysis and visualization challenges grow with the size of the data~\cite{chen2015may, chen2010information}. Interactive visual analytics systems must contend with two hard constraints simultaneously: computational scalability and \emph{perceptual} scalability.
The computational challenge is well-studied: rendering times increase with data size~\cite{liu2014effects, shneiderman1984response, park2016visualization}, and interactive latency critically degrades exploratory workflows~\cite{liu2014effects}. The perceptual challenge is equally fundamental: visual clutter affects the comprehensibility of dense visualizations~\cite{mayorga2013splatterplots, ellis2007taxonomy} and human attention has limited bandwidth. As a result, data analysts commonly work with samples of the data, to simultaneously address perceptual and interactivity challenges~\cite{battle2013dynamic, kwon2017sampling, masiane2020towards, ellis2007taxonomy}.

\looseness-1
Unfortunately, deciding which points to keep in a sample is non-trivial. An abundance of sampling methods exist---e.g., random sampling (identified as the method most accessible to data analysts in practice~\cite{rojas2017sampling, yuan2020evaluation}), density-biased sampling~\cite{palmer2000density}, visualization-aware sampling (VAS)~\cite{park2016visualization}, blue-noise sampling~\cite{yan2015survey}---but none solve this problem well. Each method optimizes a different data-driven proxy (relative density, spatial coverage, or minimum pairwise distance), but none directly models what a human user will perceive when viewing the result.

\begin{figure}[t]
  \centering
  \begin{subfigure}[t]{0.33\columnwidth}
    \centering
    \includegraphics[width=0.95\textwidth]{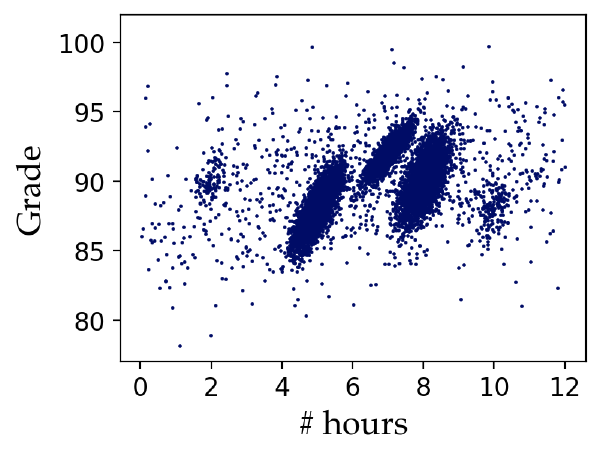}
    \vspace{-2mm}
    \caption{Original dataset}
    \label{fig:original}
  \end{subfigure}%
 \begin{subfigure}[t]{0.33\columnwidth}
      \centering
      \includegraphics[width=0.95\textwidth]{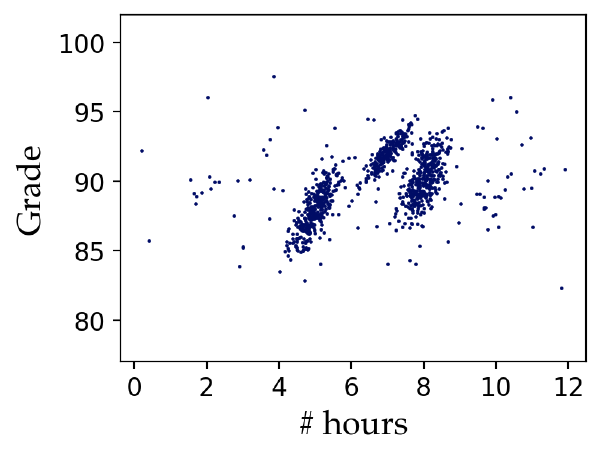}
      \vspace{-2mm}
      \caption{Random}
     \label{fig:random}
  \end{subfigure}%
   \begin{subfigure}[t]{0.33\columnwidth}
      \centering
      \includegraphics[width=0.97\textwidth]{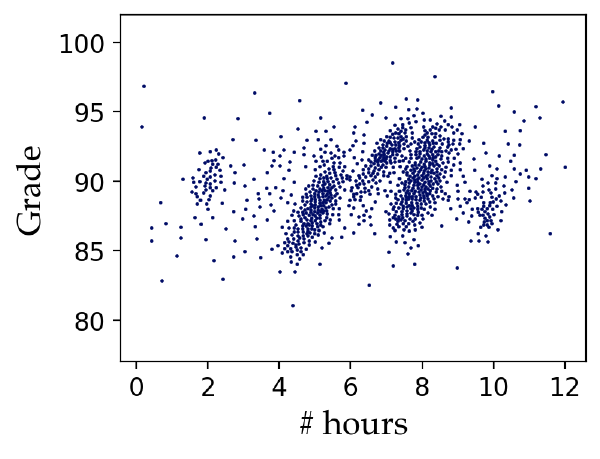}
      \vspace{-2mm}
      \caption{Perception-aware}
     \label{fig:pa}
  \end{subfigure}
  \vspace{-3mm}
  \caption{A random sample~(b) of the original dataset~(a) obscures the two smaller clusters. A sample of the same size, drawn with our perception-aware operator~(c), better preserves key features of the original dataset, including trends, clusters, and outlier spread.}
  \vspace{-3mm}
  \label{fig:random_ex1}
\end{figure}

\begin{example} \label{ex:clusters}
\looseness-1
Aysha, a data analyst, is exploring a dataset of historical student performance in three courses offered at a CS program.
She uses a scatterplot to analyze the relationship between the time students spend studying per week and their course performance. Aysha draws a sample using uniform random sampling, and gets the visualization of Figure~\ref{fig:random}. The visualization supports her expectation that more study time correlates with better grades.  However, she fails to notice two smaller student cohorts, clearly visible in the original data (Figure~\ref{fig:original}): one of students who achieve good grades with little effort, and another corresponding to a higher-workload, 4-credit course section. These cohorts are obscured in the sample, causing Aysha to miss important information. With a distorted perception of the data, she may fail at downstream tasks such as feature selection or model design.

\end{example}

This example illustrates a gap that is specifically a data management problem: the database returns a result set (the sample) that is technically valid---it satisfies a given size constraint and is drawn from the correct relation---but fails the user because it was selected without any model of how she will perceive it. \textbf{We argue that databases serving visual analytics workloads should maintain perceptual metadata about their data}, and use that metadata to answer sampling queries in a perception-aware way.

\smallskip

\looseness-1
We propose \textbf{perception-augmented databases}, a framework in which a database precomputes and stores perceptual models as first-class metadata alongside the underlying data. Designing an effective and scalable perceptual model is in itself a substantial research direction.  In our prototype implementation we design a \emph{perceptual model based on saliency maps}---heatmaps predicting where human attention will focus in a visualization---which have been established in the computer vision and visualization literature as a computational proxy of human attention. Saliency metadata does not grow with data size; e.g., a saliency map for a 1084×924 image requires 438KB and 42 seconds to compute, regardless of whether the underlying dataset has 10K or 3.5M records.
Critically, this cost is incurred once per visualization schema at preprocessing time, not at query time; the perception-augmented database serves all subsequent sampling queries from the stored synopsis with no additional saliency computation.
At query time, the perception-augmented database uses this metadata to evaluate \emph{perception-aware sampling queries}---a new query operator that selects a subset of $k$ tuples from a relation, optimized for perceptual fidelity under a target visualization.

This abstraction draws parallels to established database infrastructure. Just as approximate query processing (AQP) systems precompute synopses---histograms, wavelets, sketches---to answer aggregate queries efficiently~\cite{cormode2011synopses}, our perception-augmented database precomputes perceptual synopses to answer visualization queries faithfully. The key difference is the objective: AQP optimizes for statistical error bounds on aggregates, while perception-aware sampling optimizes for the fidelity of what a human analyst will see.

\smallskip
\noindent
\textbf{Why existing methods fall short.}
A recent study evaluating seven sampling methods~\cite{yuan2020evaluation} found that existing algorithms excel at preserving only one or two dimensions of data patterns, and none holistically accounts for human perception. Figure~\ref{fig:competing_techniques} illustrates this for three popular sampling methods on the data of Example~\ref{ex:clusters}. Density-biased sampling~\cite{palmer2000density} probabilistically over-samples sparse areas but, due to randomization, may still miss outliers and small clusters. VAS~\cite{park2016visualization} optimizes a coverage loss---the cost of not representing a point---resulting in good overall shape but extreme sensitivity to noise. Blue-noise sampling~\cite{yan2015survey} achieves a better balance between dense and sparse areas, but at prohibitive computational cost ($\sim 3.6$ hours to sample 250 points from 3.5M records in our analysis); this makes it impractical for interactive workflows.
\emph{Crucially, all three methods optimize for data-driven properties. None asks: does the sample faithfully convey the same visual information to a human as the original data?}

\smallskip
\noindent
\textbf{Our work} addresses this gap with a novel sampling operator, Perception-Aware Sampling (\nocolorpa), that uses saliency maps to directly model human visual attention as a sampling objective. Our approach relies on four key insights:

\begin{figure}[t]
  \centering
  \begin{subfigure}[t]{0.33\columnwidth}
    \centering
    \includegraphics[width=0.95\textwidth]{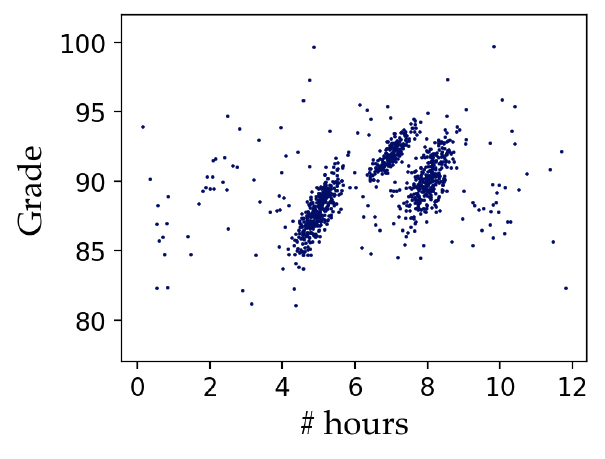}
    \vspace{-2mm}
    \caption{Density-biased}
    \label{fig:dbs}
  \end{subfigure}%
 \begin{subfigure}[t]{0.33\columnwidth}
      \centering
      \includegraphics[width=0.95\textwidth]{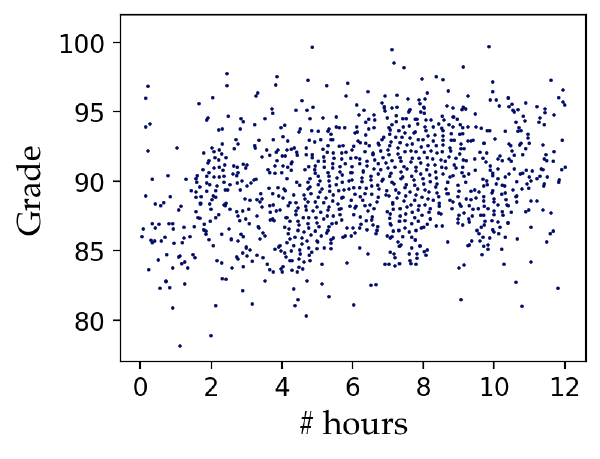}
      \vspace{-2mm}
      \caption{VAS}
     \label{fig:vas}
  \end{subfigure}%
  \begin{subfigure}[t]{0.33\columnwidth}
      \centering
      \includegraphics[width=0.95\textwidth]{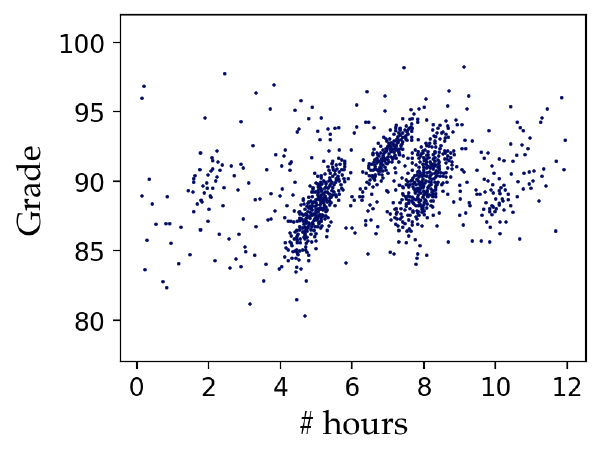}
      \vspace{-2mm}
      \caption{Blue-noise}
     \label{fig:blue}
  \end{subfigure}%
  \vspace{-4mm}
  \caption{State-of-the-art sampling methods fail to preserve key aspects of the original dataset (Figure~\ref{fig:original}). Density-biased and blue-noise miss the smaller clusters, while VAS fails to capture any trends.
  }
  \vspace{-5mm}
  \label{fig:competing_techniques}

\end{figure}

\begin{itemize}[leftmargin=7pt, topsep=1pt]
    \item \emph{Saliency maps are an actionable proxy for human perception.}
    Saliency models, well-established in computer vision, predict where viewers' eyes focus in an image~\cite{bylinskii2016should, bylinskii2018different, matzen2017data, shin2022scanner, itti2001computational}. Applied to a scatterplot, they produce a heatmap over the visualization canvas; projected onto the underlying data assigns each tuple a perception weight---its estimated importance to a human viewer.  Our perceptual model augments saliency with density information---a crucial element in how humans perceive a visualization~\cite{yuan2020evaluation}.

    \item \emph{A good sample minimizes perceptual distortion.}
    Intuitively, highly-salient areas are important in a visualization, because human attention is drawn to these areas.  If the saliency map of a sample differs from that of the original data, this indicates that a human \emph{perceives} this sample differently. We define sample quality as the similarity between these two saliency maps.

    \item \emph{Coverage is necessary to preserve saliency.}
    The saliency of an area is contextual; e.g., outliers draw attention precisely because surrounding data exists. A sample of only outlier points would not preserve this perception. Thus, perceptual information must be combined with coverage to preserve the global data shape.

    \item \emph{Human perceptual limits enable efficient approximate visualization.}
    Humans cannot perceive small spatial perturbations in a complex scatterplot~\cite{healey2011attention}. We exploit this to build \nocolorapproxpa, a method that generates \emph{approximate visualizations} from a compact, perception-aware synopsis of the data---a quad-tree partition of the visualization canvas---without accessing the original records at query time. \nocolorapproxpa achieves up to 100× speed-up over PAwS on large datasets, with minimal perceptual degradation.

\end{itemize}

\noindent
We provide an overview of our approach, including more discussion on these intuitions and design choices, in Section~\ref{sec:background}.

\smallskip
\noindent
\textbf{Contributions and outline.}
We make the following contributions:

\begin{itemize}[wide,labelwidth=!,labelindent=0pt,leftmargin=\parindent, itemsep=1pt, topsep=1pt]
    \item We introduce \textbf{perception-augmented databases} and define the \textbf{perception-aware sampling query}, a new database operator that selects $k$ tuples from a relation to maximize perceptual fidelity under a target visualization. We present an overview and discuss important background for our framework.~[Section ~\ref{sec:background}]

    \item We present the \textbf{perception-augmented database architecture}: the precomputation and storage of aggregate saliency metadata, the derivation of per-tuple perception weights, and the design of \nocolorpa, a hybrid sampling algorithm based on Max-Min diversification that employs perception weights through a distance rescaling mechanism.~[Section ~\ref{sec:algo}]

    \item We introduce \textbf{\nocolorapproxpa}, a fast approximate visualization mechanism built on a novel perception-aware compressed synopsis (a quad-tree partition with data uniformity and perception homogeneity constraints). \nocolorapproxpa serves approximate visualization queries directly from the synopsis without accessing the underlying data, achieving $100\times$ runtime gains on our largest dataset ($\sim3.5$M points).~[Section~\ref{sec:approx}]

    \item We contribute a \textbf{perceptual similarity evaluation methodology} with a detailed discussion of five metrics for evaluating the quality of samples, including an established image similarity metric that measures the difference between two visualizations in the saliency dimension.  This methodology is applicable to any scatterplot sampling method.~[Section ~\ref{sec:metrics}]

    \item We present an \textbf{extensive experimental evaluation} of our methods against five state-of-the-art baselines, over six datasets with diverse trends, ranging from 11.5K to 3.5M points.  \nocolorpa and \nocolorapproxpa outperform the alternatives and demonstrate robust performance over the five metrics.  A user study further validates that \nocolorpa samples are preferred by humans.~[Section ~\ref{sec:experiments}]

\end{itemize}

%% file: 2a_Background_arxiv27.tex
\section{Overview and background} \label{sec:background}

\begin{figure}[t!]
     \centering
     \includegraphics[width=\columnwidth]{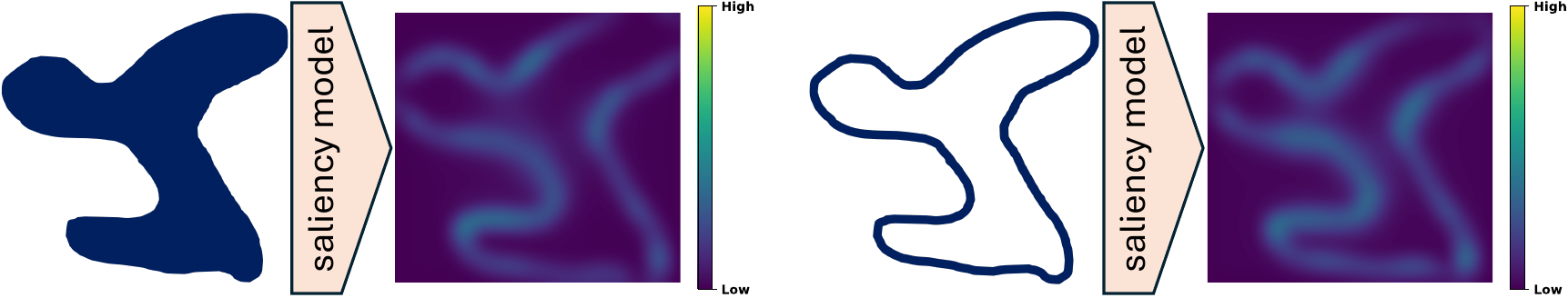}
    \vspace{-6mm}
   \caption{\newcontent{Human attention focuses on the outline of highly-dense areas.  As a result, the saliency maps of the two shapes above---one solid and one hollow---are remarkably similar; yet, the shapes are perceptually different.  Our perceptual model avoids this pitfall by augmenting saliency with density information.}}
   \vspace{-3mm}
   \label{fig:saliencyDensity}
\end{figure}

In this section, we present an overview of perception-augmented databases, including the perceptual model, the formal problem definition, and the algorithmic intuitions driving our design.
We start by describing our insights on modeling perception, and how to measure the perceptual quality of a sample.  We proceed to formally define the perception-aware sampling query. We then discuss the intuitions driving the design of our perception-aware sampling algorithm.  We pursue further speed-ups through approximate visualizations based on the design of a perception-aware synopsis.

\subsection{Modeling perception through saliency}
\newcontent{Efforts in human perception research have leveraged eye-tracking technologies}
to model how people perceive images in fields such as computer vision~\cite{shanmuga2015eye, shi2017gaze} and natural language processing~\cite{bonhage2015combined, conklin2016using}. Such technologies, however, are complex, not as readily available, and hard to incorporate into a sampling workflow.

We propose \newcontent{a new perceptual model based on} \emph{saliency models}, commonly used in computer vision, that \emph{predict} areas of a visualization that attract viewers' attention~\cite{matzen2017data}. Saliency models take as input a visualization, and produce \emph{saliency maps}---essentially heatmaps---predicting eye-gaze locations in human interactions with a visualization~\cite{bylinskii2016should, bylinskii2018different, matzen2017data, shin2022scanner, itti2001computational}.
Various computer vision tasks have successfully leveraged saliency models: from automatically learning how to zoom in image classification tasks to designing data augmentation techniques for deep learning models~\cite{recasens2018learning, sanghyeok2022}.
In this paper, we use saliency \newcontent{(augmented with density information)} to direct sampling, so that the resulting sample preserves the attention points that saliency maps identify on the original data.
\textbf{Saliency maps can be viewed as perceptual synopses of the data}---compact, visualization-space representations of what matters to a human viewer, analogous to the role histograms play in AQP as statistical synopses.

\begin{figure}[t!]
     \centering
     \includegraphics[width=\columnwidth]{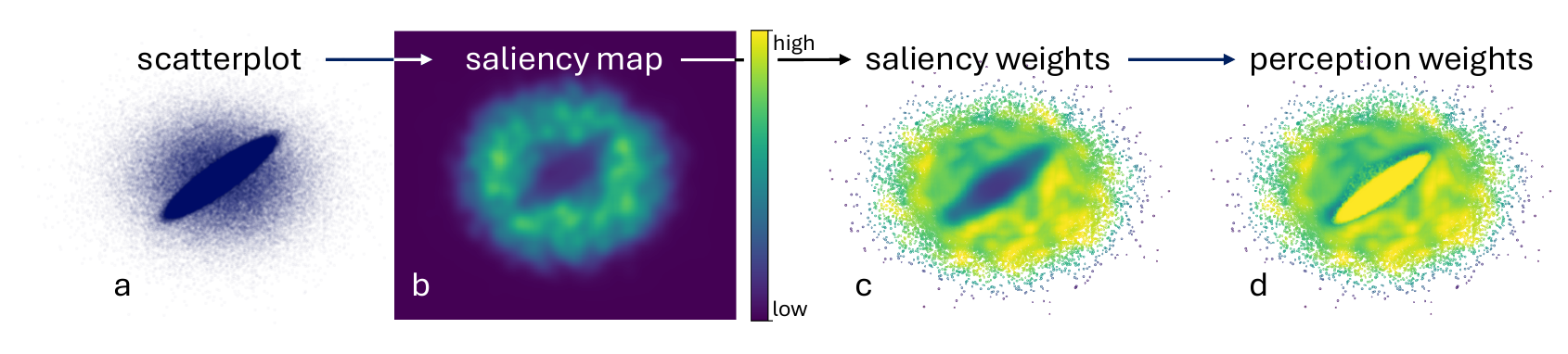}
    \vspace{-5mm}
    \caption{\newcontent{The DVS model~\cite{matzen2017data} derives a saliency map~(b) of a scatterplot visualization~(a).   Human attention typically does not focus on the interior of dense areas, so the center of the map has low saliency.  This heatmap can be projected on the data to assign \emph{saliency weights}~(c). Perception weights~(d) assign high perceptual significance for areas that are highly salient \emph{or} highly dense.}}
    \vspace{-3mm}
   \label{fig:densityAugmentation}
\end{figure}

\smallskip
\noindent
\textbf{Saliency.}
Given a saliency model $\map$ and a target visualization $\vis$ of $m \times n$ pixels, a saliency map $\map(V): \vis^{m \times n}$ $\rightarrow \ [0, 1]^{m \times n}$ is defined as a mapping function from the pixel space to continuous values within the $[0, 1]$ range, i.e., higher values indicate more \emph{salient} pixels. In our implementation, we use an established, state-of-the-art saliency model for visualizations: Data Visualization Saliency~(DVS)~\cite{matzen2017data}. Other saliency or perceptual models can be considered in alternative implementations, as this is a blackbox component in our framework. Figure~\ref{fig:densityAugmentation}b shows the saliency map produced by the DVS model over the data in Figure~\ref{fig:densityAugmentation}a.

\smallskip
\noindent
\newcontent{
\textbf{Accounting for high-density areas.}
Saliency models are designed to represent human attention, which is an intuitive proxy for perception, with one caveat: human focus is drawn on the perimeter of highly-dense elements, ignoring their interior.  As a result, the saliency of a solid object is hardly indistinguishable from that of a hollow object (Figure~\ref{fig:saliencyDensity}); yet, most would agree that these shapes are perceived differently. Thus, we augment saliency models with density information, implementing the following intuition: \emph{an area of high saliency or high density should have a high perceptual value.} Figure~\ref{fig:densityAugmentation}d shows the heatmap of \emph{perception weights} computed over the data of Figure~\ref{fig:densityAugmentation}a; these combine the saliency weights (Figure~\ref{fig:densityAugmentation}c) with density information. We discuss the details in Section~\ref{sec:algo}.}

\smallskip
\noindent
\textbf{Measuring quality.}  \looseness-1
There are no established measures for evaluating the quality of a sample, and in our case quality is based on a novel objective: perception-awareness. \newcontent{Intuitively, a good sample should draw human attention to the same salient areas as if the person was viewing the original dataset.  Thus, the saliency map of the sample, should closely resemble the saliency map of the original data.} We use \emph{Structural Similarity Index Measure (SSIM)}~\cite{wang2004image}, a well-established metric for comparing images, as our primary measure: given the saliency map of the visualization of a sample and the saliency map over the original data, the SSIM score should be high.

\subsection{The perception-aware sampling operator}

We now define the perception-aware sampling query formally as a database operator.
\begin{definition}[Perception-Aware Sampling Query]\label{def:pa}
    Assume a dataset $\universe$ augmented with perceptual meta-data in the form of a saliency model $\map$ corresponding to a target visualization $\vis$, and an image-based perceptual similarity function $f$. Given an integer $k< |\universe|$, the output of a perception-aware sampling query $Q_p(\universe,k,\vis)$ is a dataset $\sample$, such that $\sample \subset \universe$, $|\sample|=k$, and $f (\map(\vis_\universe), \map(\vis_{\sample}))$ is maximized, where $\vis_\universe$ and $\vis_{\sample}$ are the visualizations of $\universe$ and $\sample$, respectively, over the target visualization.
\end{definition}

This definition positions perception-aware sampling as a new class of query, distinct from aggregate queries (which optimize for statistical accuracy) and from coverage queries (which optimize for representational completeness).

\smallskip
\noindent
\textbf{Solving perception-aware sampling queries.}
Definition~\ref{def:pa} seeks the sample of size $k$ that results in the most similar saliency map as that of the original data. However, since the saliency model and the SSIM measure (which is our perceptual similarity function) are blackbox components, the objective cannot be optimized directly.  Iterative approaches may be possible, but the computational cost of the saliency model makes repeated calls to it impractical.  Thus, we explore heuristic methods.

\begin{figure}[t!]
     \begin{subfigure}[t]{0.33\columnwidth}
     \centering
     \includegraphics[width=0.93\textwidth]{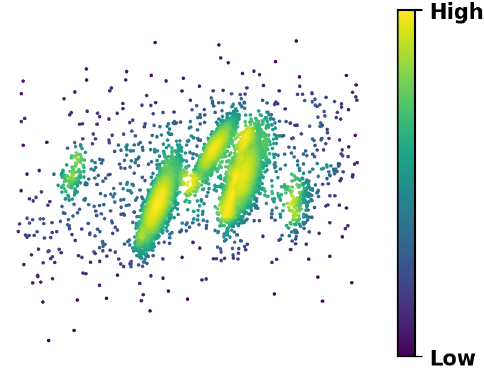}
     \vspace{-2mm}
     \caption{Perception weights}
     \label{fig:saliencyMap}
   \end{subfigure}%
   \begin{subfigure}[t]{0.33\columnwidth}
     \centering
     \includegraphics[width=0.93\textwidth]{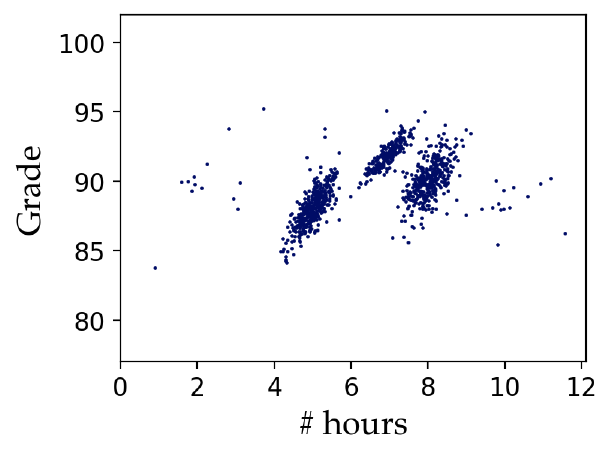}
      \vspace{-2mm}
     \caption{\newcontent{Weight-based}}
     \label{fig:saliencyBias}
   \end{subfigure}%
   \begin{subfigure}[t]{0.33\columnwidth}
     \centering
     \includegraphics[width=0.93\textwidth]{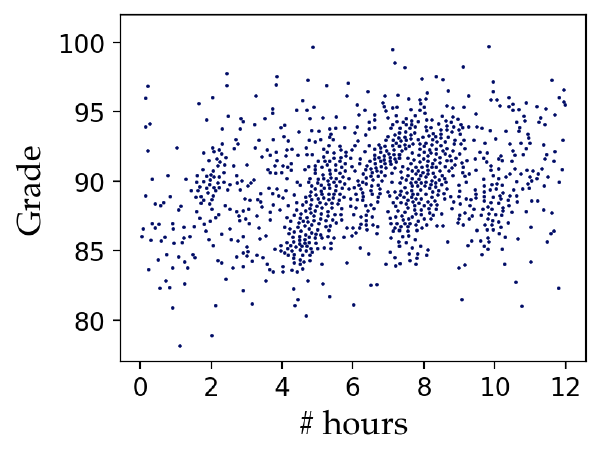}
      \vspace{-2mm}
     \caption{Max-Min}
     \label{fig:MMex}
   \end{subfigure}
    \vspace{-2mm}
   \caption{\newcontent{Perception weights for the dataset in Figure~\ref{fig:original}.} (b)~A \newcontent{weight-based} sample, which selects points with probabilities proportional to their \newcontent{perception} weights, obscures the outlier trend and the two smaller clusters. (c)~Max-Min provides coverage but fails to reveal clusters and trends in the data that are apparent in the saliency map.}
   \vspace{-2mm}
   \label{fig:saliency_example}
\end{figure}

\looseness-1
The naive approach of using \newcontent{perception} weights as sampling probabilities performs poorly (Figure~\ref{fig:saliencyBias}). The reason is that saliency is highly contextual and is impacted by data in surrounding areas. For example, outliers may draw human attention, but if the sample only contained outlier points, they would no longer be perceived as outliers and the attention patterns could change. Based on this intuition, perception weights should be used in unison with coverage objectives to reduce perceptual distortion in the sample.

\paragraph{Max-Min diversification.}
\newcontent{Diversity and coverage are related objectives in data selection, well-studied by prior work~\cite{drosou2017diversity, moumoulidou21, AddankiMMM2022, borodin2017max, Abbassi:2013:DMU:2487575.2487636}.} We focus on Max-Min, which is one of the most well-established and frequently-used models~\cite{drosou2017diversity}.

Max-Min diversification, also known as \emph{farthest point} sampling in the visualization and computer vision communities~\cite{quadri2022automatic, eldar1997farthest, lin2022task}, selects points that are \emph{uniformly} dispersed across the data space, and thus provides coverage and preserves the original data shape~\cite{drosou2013diverse, wang2018rc, moumoulidou21}. The objective of Max-Min is to maximize the diversity of a sample, defined as the minimum pairwise distance of sample points.  As with other coverage-based objectives, Max-Min is susceptible to noise (Figure~\ref{fig:MMex}), but it is computationally more efficient than VAS.

\begin{figure}[t!]
  \centering
  \begin{subfigure}[t]{0.33\columnwidth}
    \centering
    \includegraphics[width=0.95\textwidth]{images/example_for_introduction.pdf}
    \vspace{-2mm}
    \caption{Original data}
    \label{fig:original_2}
  \end{subfigure}%
 \begin{subfigure}[t]{0.33\columnwidth}
      \centering
      \includegraphics[width=0.97\textwidth]{images/example_for_introduction_paws_density.pdf}
      \vspace{-2mm}
      \caption{Perception-aware}
     \label{fig:pa_2}
  \end{subfigure}%
   \begin{subfigure}[t]{0.33\columnwidth}
      \centering
      \includegraphics[width=0.95\textwidth]{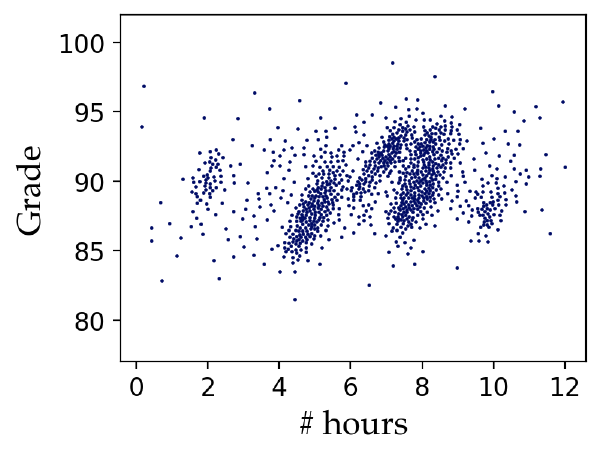}
      \vspace{-2mm}
      \caption{Approximate}
     \label{fig:approxvis}
  \end{subfigure}
  \vspace{-3mm}
  \caption{A human is unlikely to detect that the approximate visualization~(c), powered by our perception-aware compression, is automatically constructed and does not sample the underlying data~(a).  Our perception-aware sampling algorithm generates an actual sample~(b), with similar properties as the approximate visualization.}
  \label{fig:approxpa_intro}
\end{figure}

\paragraph{Perception-awareness and coverage.}
\looseness-1
GMM, a simple approximation algorithm for Max-Min diversification, relies on the \emph{farthest-first} traversal heuristic~\cite{1994:HSC:2753204.2753214}. GMM starts with a random point and keeps selecting the point whose minimum distance to previously selected points is maximized. \newcontent{We design a hybrid method based on GMM that employs perception weights through a distance re-scaling mechanism.} Our perception-aware sampling algorithm, \nocolorpa (Section~\ref{sec:algo}), inherits the simplicity and intuition of Max-Min sampling, but re-scales the pairwise similarity scores of data points using the \newcontent{perception weights, thus balancing coverage and prioritizing points with high perception value}.
Our distance-rescaling mechanism is analogous to weighted synopsis construction: the perception weights serve a role similar to importance weights in weighted random sampling for AQP~\cite{minos2011approximate, chaudhuri2007optimized}, but applied to a coverage-based rather than a probabilistic selection algorithm.

\smallskip
\noindent
\emph{Although saliency, density estimation, and Max-Min are individually established, they play distinct roles under the perceptual objective:} saliency identifies attention, density corrects saliency's blind spots in dense structures, and Max-Min preserves the spatial context that gives salient regions their perceptual meaning. Figures~\ref{fig:saliencyDensity},~\ref{fig:saliencyBias}, and~\ref{fig:MMex} illustrate what is lost when any component is used in isolation.

\subsection{Approximate visualizations}

In visual data exploration, analysts typically look for patterns, trends, and other insights, but they do not typically look at visualizations to \emph{read} individual data points.  In scatterplots in particular, humans are unlikely to visually perceive small value perturbations in the data, which means that we can allow inaccuracies in the presented data, as long as the visualization preserves the overall trends~\cite{healey2011attention}. Our key insight is that we can generate an \emph{approximate visualization}, that does not access or sample the real data at all. We achieve this through a \textbf{novel perception-aware synopsis}, which works by fragmenting the visual canvas into \newcontent{rectangular regions with similar perception weights}; considerations of area size and data density allow for adjusting the level of compression.

Figure~\ref{fig:approxvis} demonstrates the approximate visualization generated by our \nocolorapproxpa strategy. It displays the same number of points as our \nocolorpa sample (Figure~\ref{fig:pa_2}), but these points were not drawn from the original data; rather, \nocolorapproxpa modifies \nocolorpa to select boxes from the compressed representation to sample from, and then draws points uniformly at random within a box. Yet, a human would be extremely unlikely to recognize that this is not an actual sample from the original data (Figure~\ref{fig:original_2}).  Our \nocolorapproxpa strategy closely emulates the behavior of \nocolorpa, but because it works on the compressed representation and does not access the data directly, it can result in significant runtime gains in large datasets.  We discuss the details of this method in Section~\ref{sec:approx}.

\subsection{Scope and practical considerations}\label{sec:scope}

Our goal in this paper is to establish that perceptual models stored as first-class meta-data can help achieve more effective data samples compared to state-of-the-art sampling methods.  We provide proof-of-concept implementation and demonstrate the effectiveness of perception-aware sampling through extensive quantitative experiments and a user study (Section~\ref{sec:experiments}).  It is \emph{not} our goal to evaluate how well state-of-the-art saliency models capture visual attention.  This has already been established in prior work~\cite{bylinskii2016should, bylinskii2018different, matzen2017data, shin2022scanner, itti2001computational}, and we use these state-of-the-art tools as blackboxes.

Conceptually, saliency maps are abstract ``sketches'' of the underlying data, its shape, and trends. Similar to sketching techniques for deriving synopses of data~\cite{cormode2011synopses}, saliency maps form synopses of \emph{the looks} of a dataset. While it may be tempting to consider storing visualizations of the original data as images,  this does not address the limitations on the amount of data a user can process nor overplotting, to which saliency models are more robust. Moreover, such images would be static, but data analysts rely on interactive visual exploration that allows for filtering, brushing, linking, or zooming operations, and sampling is a fundamental approach in the literature for designing such interactive visualization systems~\cite{kwon2017sampling}.

\smallskip
\noindent
\textbf{Preprocessing cost and amortization.}
The perception-augmented database requires a one-time precomputation of saliency metadata per target visualization schema. For a 1084×924 image, this costs 42 seconds and 438KB of storage per configuration, independent of data size. Across the 16 visual configurations used in our evaluation, this totals approximately 11 minutes of precomputation; storage requirement remains at 438KB as the database only needs to maintain an aggregate saliency heatmap of all the configurations (Section~\ref{sec:aggregatesaliency}).
This one-time cost is amortized across all subsequent sampling queries, and compares favorably to index structures routinely built in database systems (e.g., B-trees, R-trees, VIS-trees~\cite{el2016vistrees}) and to the precomputation costs of AQP synopses. In settings with frequent sampling queries or interactive dashboards, the amortized per-query cost is negligible.

We note that saliency is not the only viable perceptual model within our framework. The black-box design of the perceptual metadata store (Section~\ref{sec:aggregatesaliency}) means that any model predicting visual attention can serve as a drop-in replacement. We have developed such an alternative  perceptual model that does not require visualization rendering, runs up to 200$\times$ faster than DVS, and produces samples within ~3\% SSIM of the saliency-based samples. But we use saliency on this first prototype of our perception-aware database framework because it is the literature-established computational proxy for human attention~\cite{bylinskii2018different, bylinskii2016should, itti2001computational, matzen2017data, shin2022scanner}, thus providing the strongest empirical grounding for our results; however, the framework's effectiveness is not contingent on this specific choice.

\smallskip
\noindent
\textbf{Perceptual synopses as database objects.} The aggregate saliency map is a compact per-attribute-pair synopsis of perceptual importance, reused across all perception-aware sampling queries, directly analogous to how a histogram synopsis is built once and reused for many aggregate queries in AQP. The quad-tree synopsis of \nocolorapproxpa extends this idea by partitioning the visualization canvas on data-distribution and perceptual-homogeneity criteria. Having the perceptual synopsis as a structured database object opens the door for DB-specific capabilities: query optimization can exploit the \nocolorpa/\nocolorapproxpa fidelity-latency tradeoff to select the right operator given latency constraints, and incremental synopsis updates are analogous to statistics maintenance in query optimizers. Crucially, our sampling operator returns a relation: $k$ tuples that can be requeried at a different sample size, filtered, joined, and explored interactively across sessions.

\smallskip
\noindent
\textbf{Scope.} The perception-augmented database framework is general, but its concrete realization in this paper is scatterplot-specific. Different visualization types rely on different perceptual cues (scatterplots on spatial coverage, density, and shape; bar charts on relative heights; line charts on continuity) and extending the framework would require visualization-specific perceptual models and operators, much as prior AQP-based sampling techniques are tailored per chart type~\cite{alabi2016pfunk}. Our focus is on scatterplots; other visualization types are an important direction for future work.

%% file: 3_Algorithms_arxiv27.tex
\section{The Perception-Augmented Database}\label{sec:algo}

\begin{figure}[t!]
    \centering
    \includegraphics[width=1.02\linewidth]{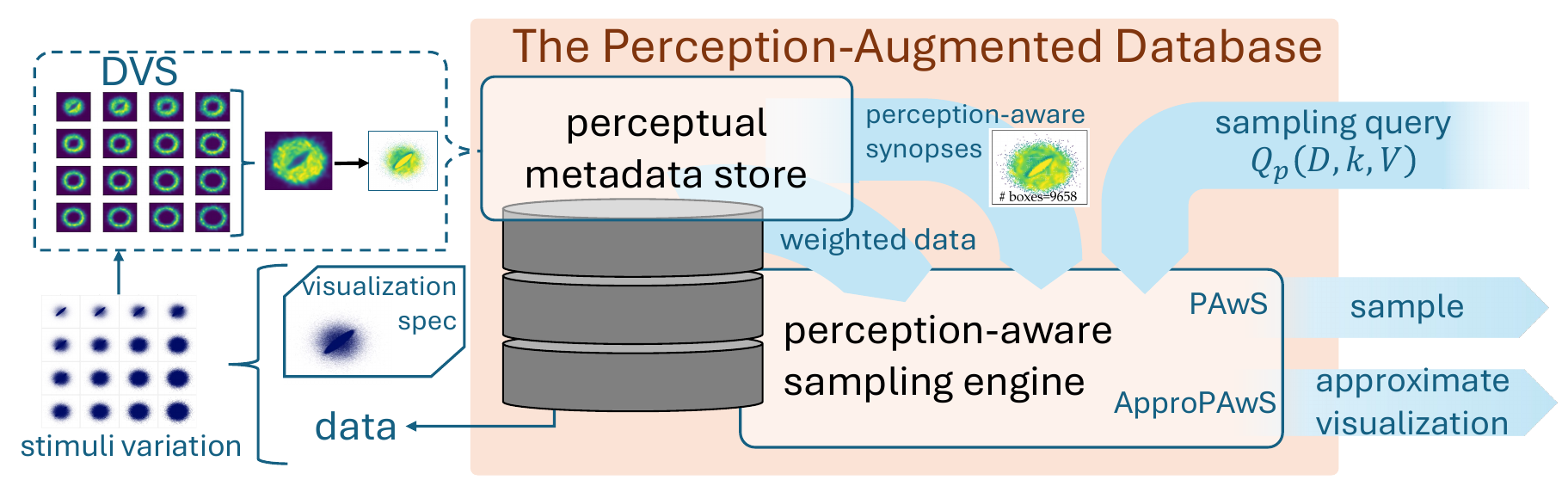}
    \vspace{-7mm}
    \caption{High-level architecture of the perception-augmented database. The perceptual metadata store keeps a perceptual synopsis---an aggregate saliency heatmap---generated from varied target visualization stimuli, and augmented with density information. Given a sampling query for a target visualization, the perception-aware query engine generates a weighting scheme for the underlying data and executes the \nocolorpa operator (Algorithm~\ref{algo:PA}) to produce a perception-aware sample of the specified size.  The \nocolorapproxpa operator (Section~\ref{sec:approx}, Algorithm~\ref{algo:approxpa}) uses perception-aware synopses---a quad-tree partitioning of the visualization canvas---to return approximate visualizations without accessing the actual records at query time.
    }
    \vspace{-3mm}
    \label{fig:pipeline}
\end{figure}

The perception-augmented database (Figure~\ref{fig:pipeline}) extends a conventional relational database with two components: (1)~a perceptual metadata store that holds precomputed saliency information for target visualizations, and (2)~a perception-aware sampling engine that uses this metadata to answer sampling queries.

\subsection{Perceptual metadata store}\label{sec:aggregatesaliency}

The perceptual metadata store precomputes and indexes a perceptual model of the data for each target visualization schema.

\begin{figure}[t!]
  \centering
  \begin{subfigure}[t]{0.33\columnwidth}
    \centering
    \includegraphics[width=\textwidth]{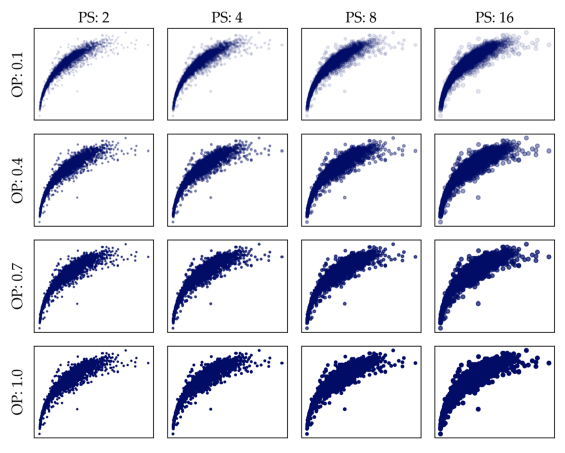}
    \caption{Stimuli variations}
    \label{fig:vis_grid}
  \end{subfigure}%
 \begin{subfigure}[t]{0.33\columnwidth}
      \centering
      \includegraphics[width=\textwidth]{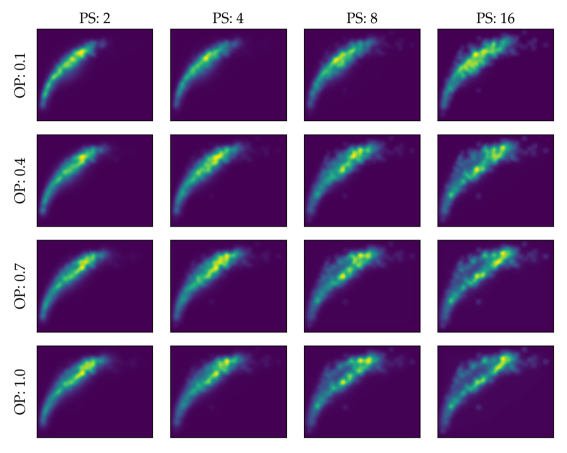}
      \caption{Saliency maps}
     \label{fig:saliency_grid}
  \end{subfigure}%
  \begin{subfigure}[t]{0.33\columnwidth}
      \centering
      \includegraphics[width=0.85\textwidth]{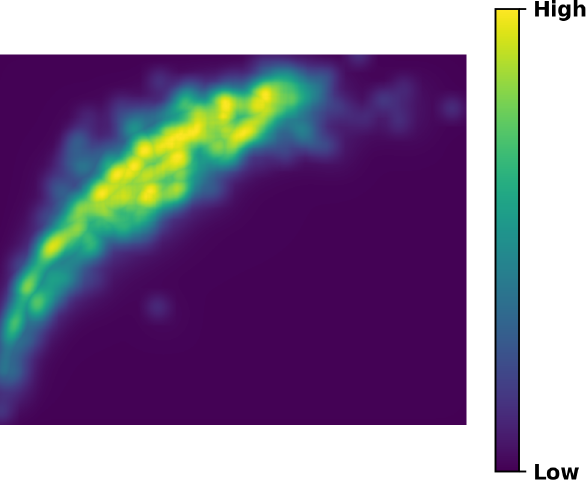}
      \caption{Aggregate saliency}
     \label{fig:aggr_map}
  \end{subfigure}
  \vspace{-3mm}
  \caption{\newcontent{The same dataset will look different with different point size and opacity configurations (a), which will translate to different saliency maps (b).  Our model aggregates the saliency across varied stimuli to capture all these perspectives (c).}}
  \vspace{-3mm}
  \label{fig:saliency_behavior}
\end{figure}

\smallskip
\noindent
\textbf{Saliency aggregation.}
Different configurations of visual features (e.g., point size, opacity, canvas aspect ratio) can affect the appearance of a visualization and thus its saliency~\cite{micallef2017towards, kwon2017sampling}.  Selecting a \emph{good} configuration for a scatterplot is non-trivial and it is an active problem in visualization research~\cite{micallef2017towards, quadri2022automatic, kwon2017sampling, quadri2020modeling}.  In Figure~\ref{fig:vis_grid}, we render the same dataset with varying configurations for the point size (2--16) and opacity (0.1--1), with an aspect ratio of 1.25:1.  Figure~\ref{fig:saliency_grid} shows the saliency map that corresponds to each rendering.  We observe that for small point sizes and opacity, the visualization highlights areas of higher population density, and the corresponding saliency maps identify these areas as more salient. For large point sizes and opacity, more areas of the visualization appear dense, and the attention focus tends to shift towards the outskirts and outliers.
Focusing on a particular configuration and saliency would therefore potentially miss important perspectives in the data that become salient under different configurations. For a more robust perception model, we \emph{aggregate} saliency information across a broad range of configurations (Figure~\ref{fig:aggr_map}). In our implementation, we use the maximum saliency value of each pixel across all configurations, but other aggregation options are possible.

\smallskip
\noindent
\textbf{Deriving perception weights.} \looseness-1
The perception-augmented database stores two pieces of information as meta data: the aggregate saliency map (a 2D matrix assigning saliency values to pixels) for the target visualization, and a density score $q_d$ for each data point.  The system derives a per-tuple saliency score $q_s$ on the fly (with negligible overhead), by assigning to each datapoint the corresponding pixel value.  The density score is precomputed using non-parametric kernel density estimators~\cite{scott2015multivariate, silverman2018density, parzen1962estimation}, assigning higher density weights to points in highly dense areas.  The perception-augmented DB then derives a perception weight for each point as $w_p= \max(q_s, \gamma \cdot q_d)$, where $\gamma \in [0, 1]$ controls the influence of density. The system adaptively sets $\gamma$  based on the variance of density weights: when variance is low, density adds little information as all areas have similar density ($\gamma$ approaches zero); when variance is high, density plays a larger role. In our experiments, we map the density variance of a dataset into a sigmoid function, assigning $\gamma$ values closer to one for datasets with larger density variance.

\subsection{The \nocolorpa sampling operator}

Users can query the perception-augmented DB to retrieve a sample of the desired size for their target visualization.  The perception-aware DB invokes \nocolorpa  as a query operator: given a relation $\universe$, a target visualization schema $V$, and a desired sample size $k$, the perception-augmented database retrieves the appropriate perception weights from the metadata store and executes Algorithm~\ref{algo:PA}. We assume a metric distance function $d: \universe \times \universe \rightarrow \mathbb{R}_{0}^{+}$ representing the pairwise similarities of points over the two visualization attributes and a set of perception weights $\weights = \{w_{p_1}, w_{p_2}, \ldots, w_{p_n}\}$ that the DB computes on the fly.

Our goal is to select a sample $\sample \subseteq \universe$ of size $k$ that maximizes the \emph{perceptual} similarity of $\sample$ to the original dataset $\universe$, i.e., the similarity of the saliency maps of $\sample$ and $\universe$ should be maximized.  We use coverage and perception scores as proxies towards this objective. \nocolorpa follows a greedy strategy, adapted from Max-Min sampling:
It initially selects a random point, and proceeds to augment the sample $\sample$ greedily, by adding the point $p^*$ with the largest multiplicative score of its perception weight ($w_{p^*}$) and distance from the current sample ($\min_{x \in \sample}d(p^*, x)$).
Thus, the objective penalizes points that have low \newcontent{perception weights}, or points that are too close to the existing sample. This strategy is analogous to data augmentation techniques in computer vision that balance saliency and coverage~\cite{sanghyeok2022}.

\smallskip
\noindent
\textbf{Running time.} A naive implementation requires $O(k^2n)$ time, but with an optimized approach\cite{wang2018rc, kleindessner2019fair}, it reduces to $O(kn)$.

\begin{algorithm}[t]
    \caption{\nocolorpa (Perception-Aware Sampling operator)}\label{algo:PA}
	{\footnotesize
		\begin{algorithmic}[1]
			\Statex
			\begin{description}
				\item[\rlap{Input:}\phantom{Output:}] $\universe$: Dataset of $n$ points
                    \item[\phantom{Output:}] $\mathcal{W}=\{w_{p}: p\in\universe\}$: Perception weights of points in $\universe$
				\item[\phantom{Output:}] $\samplesize \in \mathbb{Z}^{+}$
				\item[Output:] $\sample \subseteq \universe$ of size $k$
			\end{description}

			\Procedure{\nocolorpa}{$\universe,\mathcal{W},k$}

			     \State $\sample \gets$ a randomly chosen point in $\universe$

			\While{$|\sample|< k$}
                    \State $p^{*} \gets \underset{y \in \mathcal{D}}{\text{argmax}} \ w_{y} \cdot  \underset{x \in \sample}{\text{min}} \ d(y, x)$  \label{ln:dist}
    			\State $\sample \gets \sample \cup \{p^{*}\}$
			\EndWhile
                \Statex
			\Return $\sample$
			\EndProcedure
		\end{algorithmic}
	}
\end{algorithm}

%% file: 3a_Algorithms_arxiv27.tex
\section{Approximate Visualization Queries}\label{sec:approx}

Database systems routinely trade accuracy for performance through approximation: AQP systems answer aggregate queries with bounded error in a fraction of the time~\cite{agarwal2013blinkdb, hellerstein1999interactive, rahman2017ve}. We introduce an analogous mechanism for visualization queries. The \textbf{approximate visualization query} returns a set of $k$ points that, when rendered, is perceptually indistinguishable from a true sample---but does so without accessing the underlying data at query time, operating instead over a compact synopsis.

The perception-augmented DB uses a quad-tree partitioning scheme to produce a compressed representation (synopsis) of the data based on \newcontent{the perception model}. Roughly, the synopsis fragments the canvas area into bounding boxes of different sizes, so that the data that falls within each box is somewhat evenly distributed and has similar \newcontent{perception weights}. Our adapted sampling algorithm, \nocolorapproxpa (Algorithm~\ref{algo:approxpa}), generates a set of representative points for each box, drawn uniformly at random within the box area. It then greedily selects a point at a time, based on a combination of \newcontent{perception} and coverage objectives (like \nocolorpa), replacing the point with another random one in the same box.  \nocolorapproxpa achieves runtime gains simply on the premise of working over a small dataset of randomly-drawn points, so these gains are more pronounced the larger the original data is.

\subsection{Perception-aware synopsis construction}\label{sec:pa_compression}

We create a compressed representation (synopsis) of the dataset $\universe$ over the two dimensions relevant to the target visualization.
This synopsis plays the same architectural role as a histogram or sample-based synopsis in AQP~\cite{cormode2011synopses}, \textbf{but is constructed to minimize perceptual distortion rather than statistical error.}
Our compression mechanism splits the visualization canvas in a quad-tree hierarchical partitioning fashion: \newcontent{any cell that fails to meet certain data distribution or perception homogeneity requirements (discussed below) is split into four smaller cells.}
The partitioning terminates with a collection of non-overlapping bounding boxes of different sizes, such that (1)~data is close to uniformly distributed within each box, and (2)~data points within each box have roughly similar \newcontent{perception weights}. We explain these criteria below.

\smallskip
\noindent
\textbf{Data approximation.}
Our approximation scheme draws uniform random samples from a box in the compressed representation, instead of accessing the actual data.  The compression scheme ensures that, for each box, a sample drawn uniformly at random is close enough to true data points.  We use the \emph{Chamfer} (pseudo)-distance (CD), a commonly-used metric in the computer vision literature for designing generative neural network models for reconstructing 3D point clouds~\cite{achlioptas2018learning, fan2017point, wu2021density}. It compares the similarity of two unordered sets of points $\mathcal{S}_1$, $\mathcal{S}_2$ as follows:
{\small
\[
d_{CD}(\mathcal{S}_1 , \mathcal{S}_2) = \dfrac{1}{2} \left( \dfrac{1}{|\mathcal{S}_1 |}\sum_{x \in \mathcal{S}_1} \underset{y \in\mathcal{S}_2}{\text{min}} ||x - y||_{2} \ + \ \dfrac{1}{|\mathcal{S}_2|}\sum_{y \in \mathcal{S}_2} \underset{x \in \mathcal{S}_1}{\text{min}} ||x - y||_{2} \right)
\] }

The Chamfer distance measures how close every point in the actual data $\mathcal{S}_1$ is to their nearest neighbor in the random sample $\mathcal{S}_2$, and vice versa, and computes the average. A box is split if $d_{CD}$ is above a preset threshold $\lambda$.

\smallskip
\noindent
\newcontent{\textbf{Perception homogeneity.}}
Similarly, because we want the generated data points to appropriately represent the true data, the compression also ensures that data points within each box have similar perception weights. We split a box if the variance of perception weights within it is higher than a preset threshold $\sigma$.

\smallskip

\begin{figure}[t]
    \centering
    \begin{subfigure}[t]{0.48\columnwidth}
        \centering
        \includegraphics[width=\textwidth]{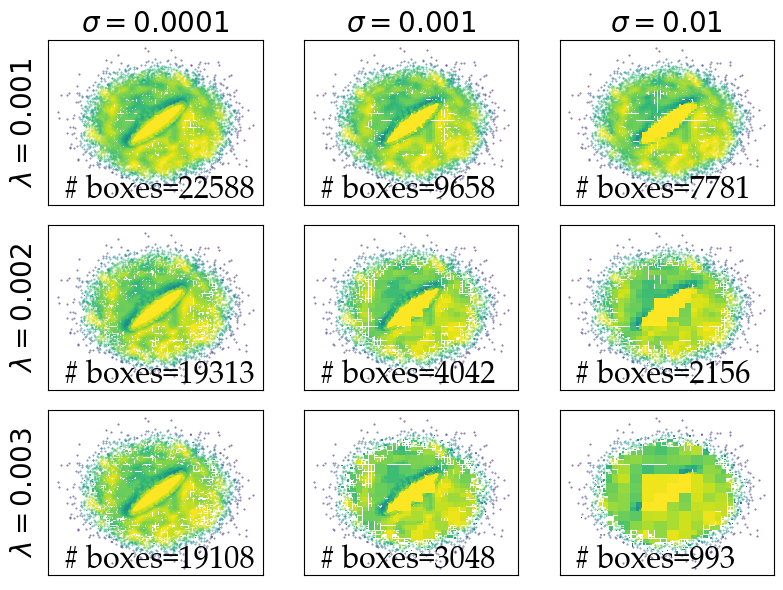}
        \vspace{-6mm}
        \caption{}
        \label{fig:hidden_schemes}
    \end{subfigure}%
   \hspace{1mm}
    \begin{subfigure}[t]{0.48\columnwidth}
        \centering
        \includegraphics[width=\textwidth]{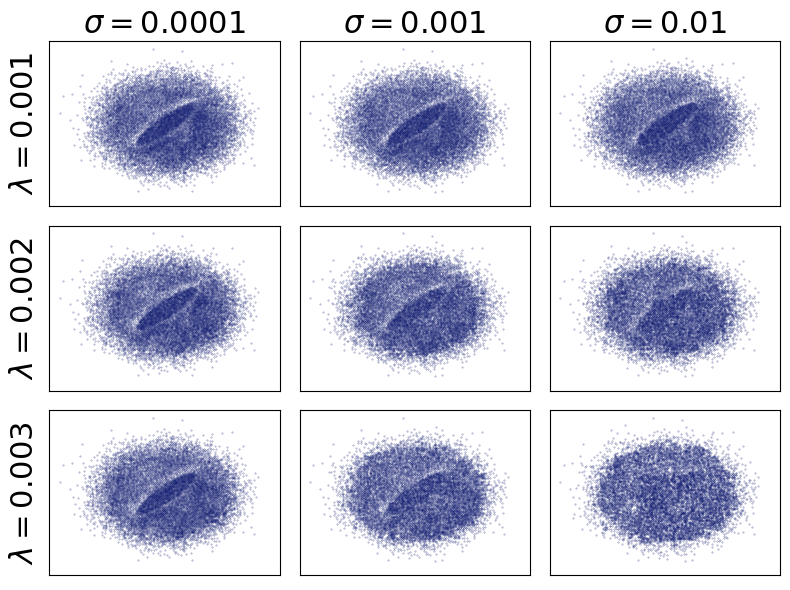}
        \vspace{-6mm}
        \caption{}
        \label{fig:hidden_qual}
    \end{subfigure}
    \vspace{-4mm}
    \caption{Varying the thresholds \bm{$\lambda$} and \bm{$\sigma$} results in different compression rates (a).  Lower compression rates (lower values for \bm{$\lambda$} and \bm{$\sigma$}) lead to lower distortion in the approximate visualizations by \nocolorapproxpa (b), shown for a sample size of $\bm{k=9,611}$ points.
    }
    \vspace{-3mm}
    \label{fig:hidden_corr_approx_scheme}
\end{figure}

Figure~\ref{fig:hidden_schemes} shows a visual representation of the compression of a dataset for varied values of $\lambda$ and $\sigma$.  These thresholds affect the number of boxes in the final compression.  Intuitively, more boxes result in lower compression and lower distortion of the final visualization.  In this case, we see the lowest compression in the upper left corner and the highest compression in the lower right corner of Figure~\ref{fig:hidden_schemes}. Figure~\ref{fig:hidden_qual} shows the corresponding approximate visualizations for each compression (produced by \nocolorapproxpa, discussed next).  We note that even at the lowest compression (most boxes), the size of the synopsis is only \newcontent{1.2~MB}; the resulting visualization achieves perceptual similarity on par with \nocolorpa, with a runtime that is about $100\times$ faster (Section~\ref{sec:exp2}).

\begin{algorithm}[t]
    \caption{\nocolorapproxpa (Approximate PAwS)}\label{algo:approxpa}
	{\footnotesize
		\begin{algorithmic}[1]
			\Statex
			\begin{description}
				\item[\rlap{Input:} \phantom{Output:}] $\partitions$: partitions in the compressed representation of $\universe$
                    \item[\phantom{Output:}] $\mathcal{W}_{\partitions}=\{w_b: b\in\partitions\}$: Perception weights of boxes in $\partitions$
				\item[\phantom{Output:}] $\samplesize \in \mathbb{Z}^{+}$
				\item[Output:] $\sample $ of size $k$
			\end{description}

			\Procedure{\nocolorapproxpa}{$\partitions, \mathcal{W}_{\partitions}, k$}
                    \State $R\gets\emptyset$
                    \LineCommentx{Construct a set of representatives with $C$ random points from each box}
                    \For{$\forall b\in\partitions$} \label{ln:represStart}
                        \State $\representatives_b\gets$ $C$ random points in $b$, $\forall r\in \representatives_b$, $r.w=w_b$
                        \State $\representatives \gets \representatives \cup \representatives_b$ \label{ln:represEnd}
                    \EndFor
                    \State $\sample \gets$ random point in $\partitions$ \Comment{Initialize the sample with a random point}

			\While{$|\sample|< k$}
                    \State $v \gets \underset{r \in \representatives}{\text{argmax}} \ r.w \cdot  \underset{x \in \sample}{\text{min}} \ d(r, x)$~\label{ln:selection}
    			\State $\sample \gets \sample \cup \{v\}$

                    \State $r'\gets$ random point from $b\in\partitions$, where $v\in b$ \Comment{Replace $v$ in $\representatives$}
                    \State $\representatives \gets$ $(\representatives \setminus v) \ \cup \ r' $ \label{ln:replace}
			\EndWhile
                \Statex
			\Return $\sample$
			\EndProcedure

            \end{algorithmic}
        }
\end{algorithm}

\subsection{The \nocolorapproxpa operator}

Given a perception-aware synopsis, \nocolorapproxpa (Algorithm ~\ref{algo:approxpa}) generates a sample with points drawn from the compressed representation, rather than from the actual data. Intuitively, \nocolorapproxpa operates exactly like \nocolorpa, but instead of performing data selection over the actual data, it generates and maintains a small representative dataset $\representatives$ on the fly. The algorithm initializes $\representatives$ by generating uniformly at random a small, constant number $C$ of points from every box in the compressed representation (lines~\ref{ln:represStart}--\ref{ln:represEnd}).  Using $C>1$ makes the algorithm robust to poor random draws (e.g., a box never getting picked due to a poor representative). When a point is added to the sample, a new representative from the same box replaces that point in $\representatives$ (line~\ref{ln:replace}).  Even though \nocolorapproxpa does not access the actual data, the construction of the compressed representation ensures that the random points drawn from the boxes conform to the data distribution and \newcontent{perception levels} of the original data.

\smallskip
\noindent
\textbf{Running Time.} \looseness-1
Algorithm ~\ref{algo:approxpa} has runtime complexity $O(k^2 + k \ C \ |\mathcal{P}|)$.  The first factor is due to computing the distance of a new representative to the current sample at each step.  The second factor corresponds to \nocolorpa's $O(kn)$ complexity, but replaces $n$ with $C \ |\mathcal{P}|$, where $C$ is the number of points we draw from each box (a small constant), and $|\mathcal{P}|$ is the number of boxes in the compression.  \nocolorapproxpa has pronounced gains for large datasets ($|\mathcal{P}| << n$); if $n$ is small its performance may degrade due to the additional factors.

%% file: 4_Metrics_arxiv27.tex
\section{Perceptual Similarity Metrics}\label{sec:metrics}
Evaluating the quality of a sample for visualization differs fundamentally from evaluating a sample for AQP: there is no ground-truth aggregate to compare against, and no statistical error bound to compute. We instead measure perceptual similarity---the degree to which a human would perceive a sample as equivalent to the original data. We operationalize this via saliency map comparison:
a key insight in our work is that a sample should distort saliency as little as possible.  We use well-established image- and distribution-based metrics to measure how similar the saliency of a sample is to the saliency of the original data.

\smallskip
\noindent
\textbf{Measuring perceptual similarity.}
Saliency maps are stored and visualized as images; thus, we use a well-established image-based metric as our primary metric: Structural Similarity Index Measure (SSIM)~\cite{wang2004image}. We further evaluate four distribution-based metrics from the saliency literature~\cite{bylinskii2018different}. These metrics treat a saliency map as a distribution over pixels and measure how accurately a saliency model predicts the attention focus areas of a dataset.

\smallskip
\noindent
\emphbox{Structural Similarity Index Measure (SSIM)} [image-based] is a well-known image similarity metric, frequently used as a loss function in computer vision problems~\cite{zhao2016loss, ledig2017photo}. With two images split into fixed-size windows, the SSIM score is the mean of similarity scores across windows measured over luminance~$(\ell)$, contrast~$(c)$, and structure~$(s)$, i.e., $\textup{SSIM}(\theta_1, \theta_2) = \ell(\theta_1, \theta_2)^{\alpha} \cdot \c(\theta_1, \theta_2)^{\beta} \cdot s(\theta_1, \theta_2)^{\gamma}$. If $\alpha=\beta=\gamma=1$, all three terms affect the score equally, while each term in the product depends on statistical information like the mean pixel value and variance of pixel values of an image window.

\smallskip
\noindent
\emphbox{Pearson’s Correlation Coefficient (CC)} [distribution-based] measures linear correlation across two random variables.
We measure the correlation of the saliency maps $\map(\vis_{\sample})$ and $\map(\vis_{\universe})$ as:
\[\textup{CC}(\map(\vis_\sample), \map(\vis_\universe)) = \dfrac{\textup{COV}(\map(\vis_\sample), \map(\vis_\universe))}{\sigma(\map(\vis_\sample)) \cdot \sigma(\map(\vis_\universe))}\]
CC takes values in the $[-1, 1]$ range where higher values indicate higher correlation (higher similarity). A value of 0 indicates no correlation (low similarity).

\smallskip
\noindent
\emphbox{Similarity (SIM)} [distribution-based] measures the similarity of two distributions by computing a histogram intersection. We normalize the saliency maps to derive the distributions
$p_{\sample}$ and $q_{\universe}$ over $I$ pixels, and compute SIM as follows:
$$\textup{SIM}(p_{\sample}, q_{\universe}) = \sum_{i \in [I]} \textup{min}\big(p_{\sample}(i), q_{\universe}(i)\big)$$
For every pixel of the saliency maps, we take the minimum probability value between the sample and the dataset distribution. SIM takes values in the [0, 1] range: a value equal to $1$ indicates a perfect distribution match, and $0$ no match at all.

\begin{figure}[t!]
    \centering\resizebox{0.95\columnwidth}{!}{
    \small
        \begin{tabular}{l c c c}
            \toprule
            \textbf{Metric} & \textbf{Metric Type} & \textbf{Range} & \textbf{Normalization} \\
            \midrule
            \rowcolor{black!10} SSIM~($\uparrow$)& image-based & [-1, 1] &  -- \\
            \rowcolor{black!10} CC~($\uparrow$) & distribution-based & [-1, 1] & by variance \\
            \rowcolor{black!10} SIM~($\uparrow$)& distribution-based & [0, 1] & by sum \\
            \midrule

             JSD~($\downarrow$)& distribution-based & [0, 1] & by sum \\

             EMD~($\downarrow$)& distribution-based & unbounded & by sum \\

            \bottomrule
        \end{tabular}
    }
    \vspace{-3mm}
    \caption{We use several metrics to evaluate the \emph{perceptual similarity} of samples to the original data. The top three metrics measure \emphbox{similarity}, and the last two dissimilarity. We report $1-JSD$ in our experiments, which is a \emphbox{similarity} metric.
    }
    \label{tab:mytable}
\end{figure}

\smallskip
\noindent
\emphbox{Jensen-Shannon Divergence (JSD)} [distribution-based] measures \emph{dissimilarity} between two probability distributions. We normalize the saliency maps to derive distributions
$p_{\sample}$ and $q_{\universe}$ and compute: $$JSD(p_{\sample}, q_{\universe})= \tfrac{1}{2}D_{KL}(p_{\sample}||\mu) + \tfrac{1}{2}D_{KL}(p_{\sample}||\mu)$$
where $\mu$ is the point-wise mean of $p_{\sample}$ and $q_{\universe}$. $D_{KL}$ is the KL-divergence, defined as:
$$D_{KL}(p_{\sample}||\mu) =\sum_{i \in [I]} p_{\sample}(i) \log\left(\tfrac{p_{\sample}(i)}{\mu(i)}\right)$$
JSD is a symmetric version of KL-divergence and takes values in the $[0, 1]$ range. Lower values indicate higher similarity: e.g., the JSD of two identical distributions is equal to zero.

\smallskip
\noindent
\emphbox{Earth Mover's Distance (EMD)} [distribution-based] %
measures spatial dissimilarity between two distributions as the minimal cost of moving probability mass between histogram bins. Computing EMD involves solving an optimal transport problem via linear programming~\cite{bylinskii2018different, pele2008linear, pele2009}---a computationally expensive step for high-resolution images. Following Bylinskii et al.~\cite{bylinskii2018different}, we downsample images to ${\sim}1/32$ of their original size and use an open-source implementation~\cite{EMDMetric}; even so, EMD remains computationally expensive.

\smallskip
\noindent \textbf{Discussion.}
All the above metrics are well-established in the literature.  We note that interpreting their behavior and the mechanism based on which they penalize mismatches across saliency maps is a separate research area and is not straightforward~\cite{nilsson2020understanding, bylinskii2018different}.  For example, CC is affected symmetrically by false positives and negatives (indicating an area is salient when it is not and vice versa), while SIM is mostly affected by false negatives~\cite{bylinskii2018different}. We refer the reader to the in-depth experimental evaluation by Bylinskii et al.~\cite{bylinskii2018different} for a better understanding of the metrics' behavior, and guidelines on how to use them in practice.

A separate research question is how well these metrics align with human perception~\cite{li2015data}. For instance, despite popular belief that SSIM is a perceptual-based similarity metric for images, there have been known cases in the literature where human judgment does not align well with the SSIM score~\cite{nilsson2020understanding}. Thus, designing better metrics for image similarity is still an active research problem~\cite{zhang2018unreasonable}. \textbf{It is not the focus of this work to contribute better and more robust similarity metrics; we lean on the state-of-the-art and use a variety of metrics to guide our evaluation.}

%% file: 5_Experiments_arxiv27.tex
\section{Experimental Evaluation}\label{sec:experiments}

We present an extensive evaluation of our perception-aware sampling and approximate visualization operators.
We assess perceptual quality, scalability, synopsis efficiency, and human preference, and describe our datasets, baselines, and experimental methodology below.
We used Python 3.8 for the code implementation and ran the experiments on one cluster node with a 2.3 GHz 8-Core Intel Core i9 and 16 GB RAM.

\medskip
\noindent
\textbf{Datasets.} We collected six datasets with a diversity of patterns, underlying trends, and sizes ranging from 11.5K to 3.5M points. We embed all datasets into a 2D space for the scatterplot visualizations, normalized in the $[0, 1]\times[0, 1]$ range (shown in \appref{sec:appendix}{onlineAppendix}).

\smallskip
\noindent
\frbox{MNIST}~\cite{lecun1998gradient} consists of $70,000$ images for different handwritten digits and is commonly used in the visualization literature for clustering analysis and perception research~\cite{yuan2020evaluation, quadri2022automatic}. We use a version of the dataset projected onto $2D$ space using the t-SNE dimensionality reduction method~\cite{mnistData}.

\smallskip
\noindent
\frbox{Estate Correlation} is a dataset of New York real estate pricing~\cite{estateData}, which consists of $17,614$ entries and 8 continuous attributes. We selected two moderately correlated attributes for the visualization.

\smallskip
\noindent
\frbox{Estate GIS} reuses the New York real estate pricing dataset ($17,614$ entries) with two different visualization attributes resulting in a shape
most suitable for outlier detection.

\smallskip
\noindent
\frbox{Epileptic Seizure}~\cite{andrzejak2001indications} describes EEG information~\cite{kaggle:seizure} ($11, 500$ entries, 178 attributes). We visualize two uncorrelated attributes.

\smallskip
\noindent
\frbox{ACSI}~\cite{ding2021retiring} is one of the ACS datasets for research in fair machine learning. It consists of $250,847$ entries and 6 continuous attributes. We chose two attributes that visualize into an abacus-like shape with no specific trend.

\smallskip
\noindent
\frbox{Hidden Correlation} is a synthetic two-dimensional dataset that consists of ${\sim}3.5$ million records. $97.5\%$ of the dataset consists of highly correlated points ($\rho=0.9$) while $2.5\%$ of non-correlated points ($\rho=0$). Due to high visual cluster, the correlation trend is often obscured in the visualization.

\medskip
\noindent
\textbf{Methods.}
Our evaluation compares our algorithms (\nocolorpa and \nocolorapproxpa) against five baselines, including probabilistic-based methods (\nocolorran and \nocolordbs), diversity-based methods (\nocolorMM and \nocolorvas), and other state-of-the-art (\nocolorbn):

\smallskip
\noindent
\pa is our primary algorithm for selecting perception-aware samples using the saliency models stored in a perception-augmented database (Section~\ref{sec:algo}).

\smallskip
\noindent
\approxpa is our approximate visualization method, which generates samples through perception-aware compression (Section~\ref{sec:approx}).

\smallskip
\noindent
\ran is standard uniform random sampling, an extremely common method, owing to its simplicity and accessibility~\cite{rojas2017sampling, yuan2020evaluation}.

\smallskip
\noindent
\MM, also known as Farthest-point sampling~\cite{1994:HSC:2753204.2753214, eldar1997farthest}, selects a sample $\sample \subseteq \universe$ of size $k$ that maximizes the minimum distance across the selected points. We use GMM~\cite{1994:HSC:2753204.2753214}, a greedy approximation algorithm ($O(kn)$) that starts with a randomly-selected point and adds the point farthest from the sample at every step.

\smallskip
\noindent
\vas is a visualization-aware sampling method for scatterplots~\cite{park2016visualization}. It selects a sample $\sample \subseteq \universe$ of size $k$ such that for any point $x \in \universe$ $\exists \ y \in \sample$ that is close enough to $x$.
 Closeness is controlled through a user-specified parameter $\epsilon$.\footnote{Since the data lies in a unit square after normalization, we set $\epsilon = \sqrt{2}/100$ because $\sqrt{2}$ is an upper bound on the diameter value of $\universe$.} \nocolorvas uses the Expand + Shrink (ES) local search algorithm, which starts with a candidate solution of $k$ points and keeps swapping a point at a time until no swap offers improvement. Typically, ES converges to a solution in less than half an hour~\cite{park2016visualization}. Park et al.~\cite{park2016visualization} discuss a runtime optimization using R-trees, but only recommend it for sample sizes of more than 10K points. For smaller sizes, the overhead of maintaining the R-tree dominates the running time. In this work, we implement the vanilla version of the ES algorithm and refer to the original paper for analyzing improvements in running time.

\smallskip
\noindent
\dbs probabilistically over-samples points in sparse areas and under-samples points in dense areas~\cite{palmer2000density}. We use an open-source implementation~\cite{yuan2020evaluation} that employs a KNN algorithm and samples a point with a probability proportional
to its K-farthest neighbor distance~\cite{samplingLibrary}.
A large distance implies that the point is likely in a sparse area and thus has higher probability of being selected.

\smallskip
\noindent
\bn sampling selects a set of points that are at least some radius $r$ far apart from each other, for some pre-specified value of $r$~\cite{yan2015survey}. A common implementation uses the dart-throwing technique~\cite{cook1986stochastic}: given a sampling radius $r$, the algorithm randomly selects a point; if the point's distance from the sample is greater than $r$, then it is added to the sample, and it is rejected otherwise. After a certain number of rejections, the algorithm reduces the value of $r$ to reach the desired sample size. We note that the setting of $r$ affects the algorithm's behavior: the smaller the $r$, the more the algorithm focuses on prioritizing dense areas. \nocolorbn aims for a balanced representation across sparse and dense areas, and thus optimizes for data shape and coverage. We use the open-source implementation of Yuan et al.~\cite{yuan2020evaluation}.

\begin{figure*}[t]
  \centering
  \begin{subfigure}[t]{0.5\textwidth}
    \centering
    \includegraphics[width=0.95\textwidth]{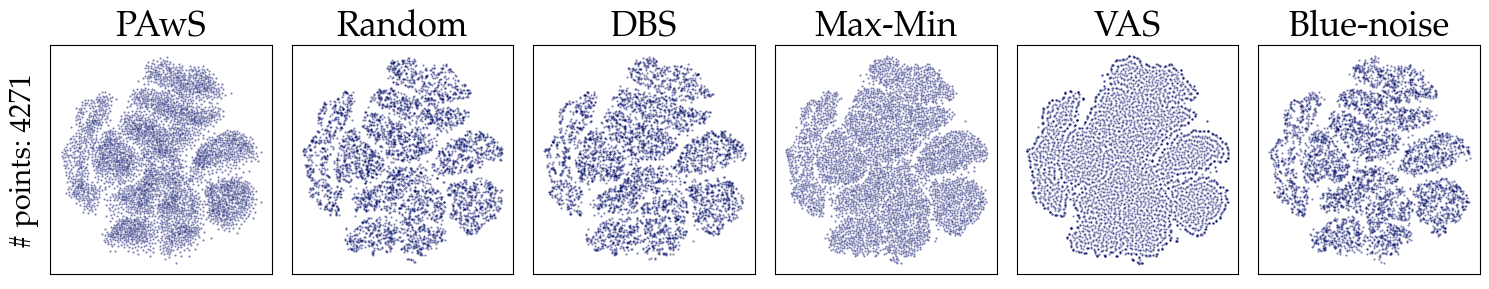}
    \vspace{-2mm}
    \caption{MNIST}
    \label{fig:qual_mnist}
  \end{subfigure}%
  \begin{subfigure}[t]{0.5\textwidth}
      \centering
      \includegraphics[width=0.95\textwidth]{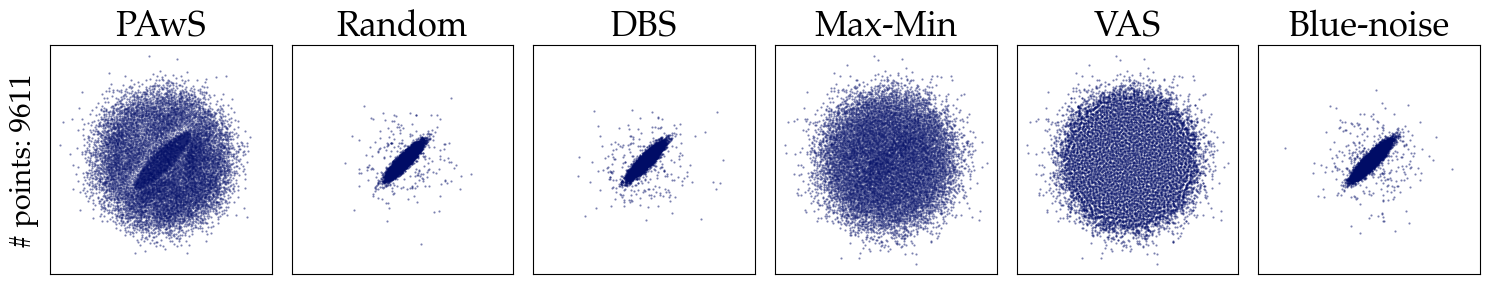}
      \vspace{-2mm}
      \caption{Hidden correlation}
      \label{fig:qual_hidden}
  \end{subfigure}
  \vspace{-4mm}
  \caption{\nocolorpa prioritizes the selection of data points in salient and high-density areas, while providing data coverage. \nocolorMM and \nocolorvas optimize for coverage, and may fail to preserve trends (e.g., in hidden correlation). \nocolorran and \nocolordbs preserve relative density, while \nocolorbn aims for a balanced representation among dense and sparse areas, but its behavior varies and sometimes behaves like random sampling.}
  \label{fig:qualitative_results}
  \vspace{-3mm}
\end{figure*}

\begin{figure*}[t!]
    \centering
    \includegraphics[width=0.9\textwidth]{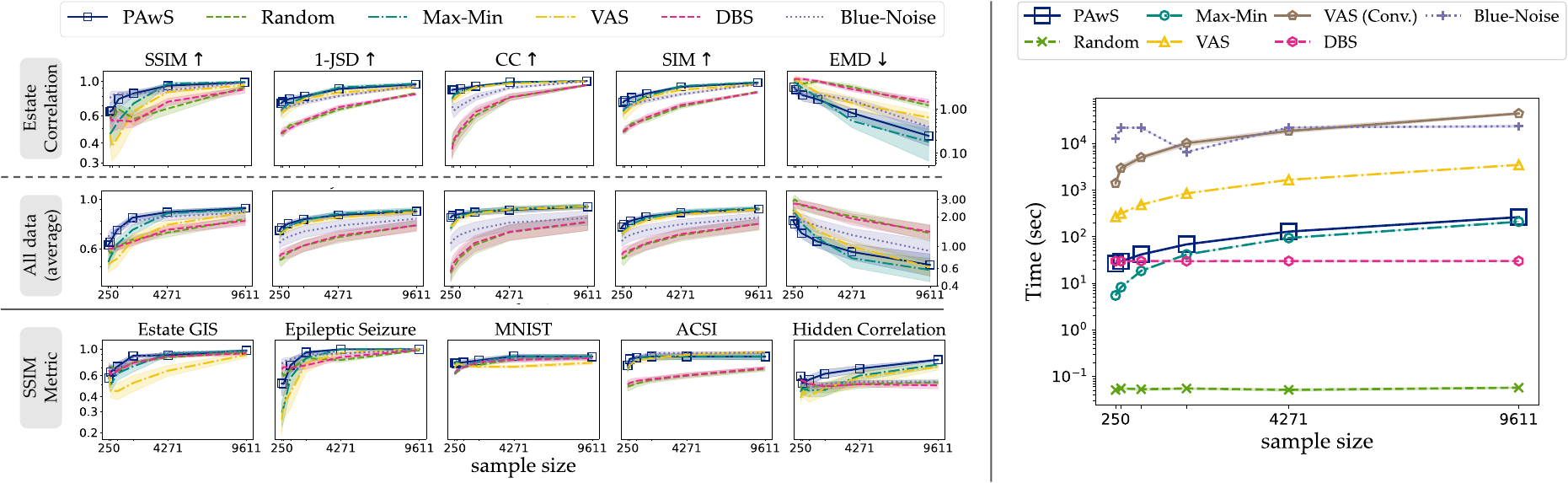}
    \vspace{-4mm}
    \caption{\nocolorpa shows strong performance across all metrics and is also more scalable than other state-of-the-art on hidden correlation, a \bm{$\sim 3.5$}M dataset. Only probabilistic-based methods are faster but perform poorly at producing perceptually good samples.}
    \label{fig:grouped_results}
    \vspace{-5pt}
\end{figure*}

\smallskip
\noindent
\textbf{Experimental design.} \newcontent{Here, we discuss our methodology, sample sizes, and other parameter and design choices.}

\smallskip
\noindent
\emph{Scatterplot rendering and \newcontent{perception weights}.}
As we discussed in Section~\ref{sec:aggregatesaliency}, visual features, such as point size and opacity, affect the \newcontent{saliency} of a visualization. The effect can be significant; e.g., in hidden correlation, the correlation trend is only visible for small marker size and low opacity.
Our perception-augmented DB stores \emph{aggregate} saliency maps over visual configurations varying point size (PS: \{2px, 4px, 8px, 16px\}) and opacity (OP: \{10\%, 40\%, 70\%, 100\%\}), on a fixed canvas size $14 \ \textup{in} \times 12 \ \textup{in}$ (i.e., aspect ratio of $1.16$ and resolution $1,085 \times 924$). \newcontent{The aggregate saliency map is projected onto the data to derive  weights, which are then augmented with density information.\footnote{Our implementation employs a non-parametric, state-of-the-art, tree-based, density estimator from the scikit-learn library~\cite{bentley1975multidimensional, pedregosa2011scikit}. Off-the-shelf density estimators were not scalable for hidden correlation ($\sim 3.5$M), and we approximated the behavior by rendering the original data using a small point (2px) \& opacity (10\%) configuration .}}

\smallskip
\noindent
\emph{Perceptual similarity.}
We express the perceptual similarity of two visualizations as the similarity between their saliencies.  Since visual features affect perception and saliency, we use 16 configurations with the point size and opacity settings discussed above to generate saliency maps for the original data and each sample.  We compare the saliency similarity of each pair (original and sample) with the same configuration, and compute the final perceptual similarity as the average across all 16 configurations. We report this average score along with a $95\%$ confidence interval across the 16 configurations, and do the same for all five metrics (Section~\ref{sec:metrics}).

\smallskip
\noindent
\emph{Sample sizes.} We select sample sizes using the Weber-Fechner Law, which states the relationship between change in stimulus and perception is logarithmic~\cite{portugal2011weber}---this methodology is favored in the visualization literature. We follow Yuan et al.~\cite{yuan2020evaluation} and experiment with sample sizes $\{250, 375, 844, 1898, 4271, 9611\}$, defined by the geometric series $\{250 \times 1.5^0, 250 \times 1.5^1, 250 \times  1.5^3, 250 \times 1.5^5, 250 \times 1.5^7, 250 \times 1.5^9\}$, where $1.5$ is a constant value in the methodology proposed by the authors~\cite{yuan2020evaluation}.

\subsection{Perceptual quality}\label{sec:exp1}
In this section, we evaluate all methods with respect to their ability to avoid perceptual distortion in the resulting visualizations. We first discuss the behavior of different algorithms, as this is showcased through some qualitative results, and then report on the perceptual similarity scores samples achieve across all metrics. \newcontent{We further evaluate humans' preferences over samples through a user study,} and put results in context with their runtime performance.

\newcontent{Figure~\ref{fig:qualitative_results} showcases samples produced by each method on two of our datasets.} Since \nocolorMM and \nocolorvas optimize for coverage, they select points that are evenly distributed across the data space, and can fail to capture trends at small sample sizes.  On the other hand, \nocolorran and \nocolordbs prioritize areas of density, thus failing to capture the overall shape. \nocolorbn achieves better balance by adaptively prioritizing the selection of points in sparse and dense areas, but behaves like random sampling on the hidden correlation dataset. \nocolorpa prioritizes points in areas of high \newcontent{perceptual significance} without sacrificing coverage, resulting in samples that represent well the overall shape and trends. \newcontent{In Figure~\ref{fig:qual_hidden}, \nocolorpa is the only method that preserves both the data shape and correlation trend.}

Figure~\ref{fig:grouped_results} reports quantitative results on the efficacy and efficiency of all algorithms. Due to space limitations, we omit or summarize some results; we refer the interested reader to \appref{sec:appendix}{onlineAppendix}. On the left side of Figure~\ref{fig:grouped_results}, we report on each sampling algorithm's effectiveness at producing perceptually good samples, using our five metrics.  The first row of plots (\emphbox{Estate Correlation}) shows the perceptual similarity achieved by all algorithms in samples generated over the \emph{Estate Correlation} dataset.   \nocolorpa achieves high scores across all metrics and sample sizes.  The performance of other methods varies a bit across metrics, but generally, \nocolorMM, \nocolorvas, and \nocolorbn lag in perceptual similarity scores until the sample size gets larger. \nocolorran and \nocolordbs achieve much worse scores than the other algorithms.

In the second row (\emphbox{All data}), we average the perceptual similarity scores for each method across all datasets, to provide a more holistic view of the algorithms' behavior.  The probabilistic methods consistently underperform, with respect to all metrics.  The relative performance of \nocolorMM, \nocolorvas, and \nocolorbn varies across the different metrics, but what remains constant is that \nocolorpa demonstrates superior performance with respect to all metrics.  These results also demonstrate that \nocolorpa needs smaller sample sizes to achieve a certain level of perception similarity.  E.g., \nocolorpa scores 0.75 SSIM similarity on average for a sample of 844 points, when \nocolorvas needs 5x bigger sample.

\begin{figure*}[t]
    \centering
    \begin{subfigure}[t]{0.23\textwidth}
        \centering
        \includegraphics[width=0.95\textwidth]{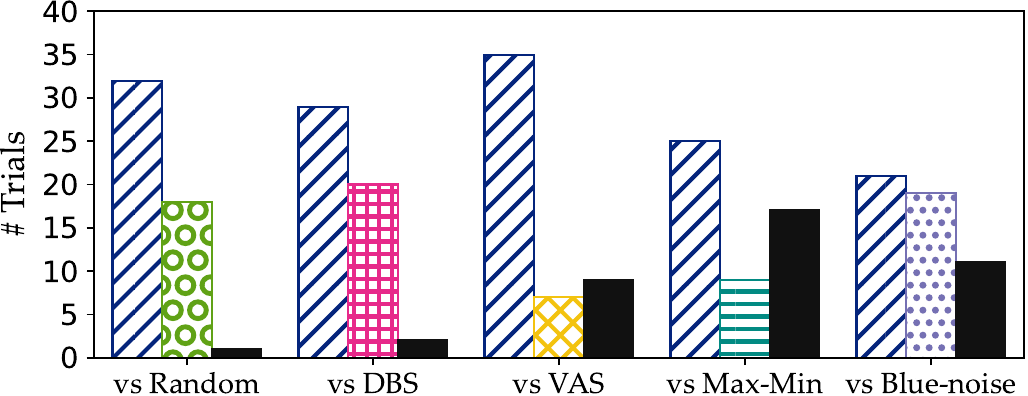}
        \vspace{-1mm}
        \caption{Trends and patterns}
    \end{subfigure}%
    \begin{subfigure}[t]{0.23\textwidth}
        \centering
        \includegraphics[width=0.95\textwidth]{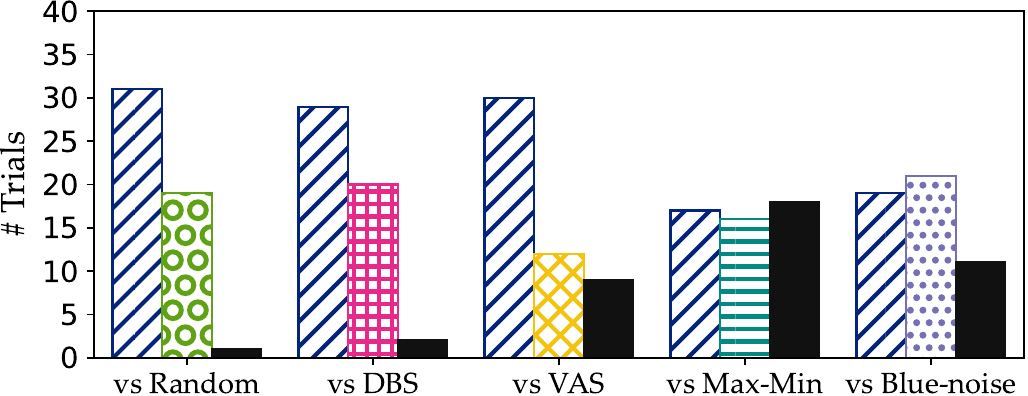}
        \vspace{-1mm}
        \caption{Outliers and data shape}
    \end{subfigure}%
    \begin{subfigure}[t]{0.23\textwidth}
        \centering
        \includegraphics[width=0.95\textwidth]{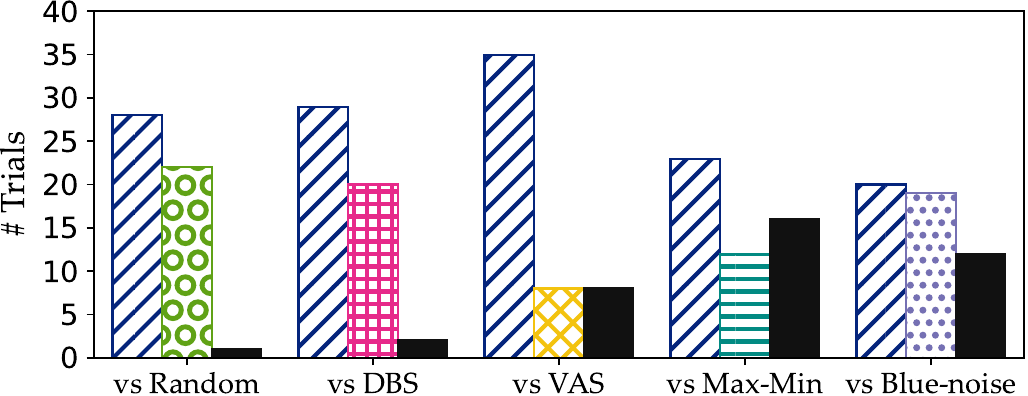}
        \vspace{-1mm}
        \caption{Density preservation}
    \end{subfigure}%
    \begin{subfigure}[t]{0.23\textwidth}
        \centering
        \includegraphics[width=0.95\textwidth]{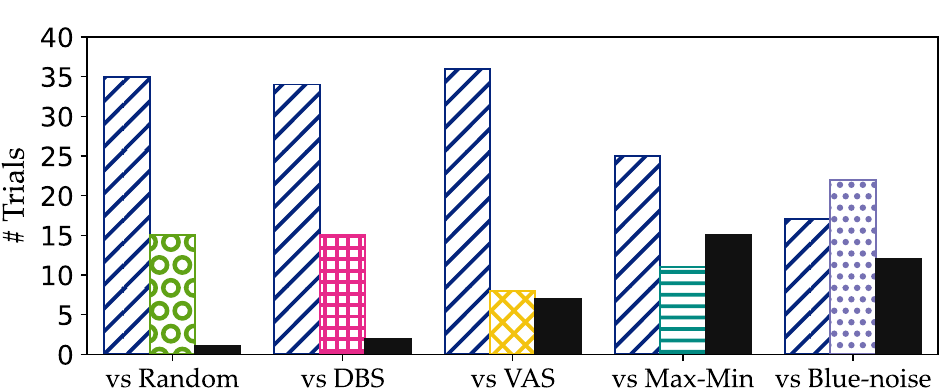}
        \vspace{-1mm}
        \caption{Best overall}\label{fig:user_overall}
    \end{subfigure}
    \begin{subfigure}[t]{0.07\textwidth}
        \centering
        \includegraphics[width=\textwidth]{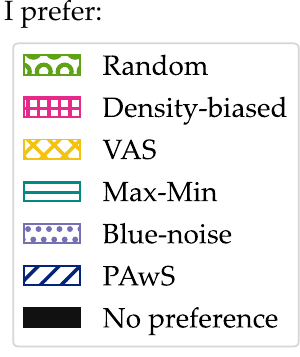}
        \vspace{-1mm}
    \end{subfigure}
    \vspace{-4mm}
    \caption{In one-to-one matchups against the state-of-the-art, users largely prefer \nocolorpa with respect to (a)~representation of correlation trends, clusters, and other patterns, (b)~representation of outliers and overall data shape, and (c)~representation of density variations. \nocolorbn is slightly preferred in the ``best overall'' category (d), but it is computationally prohibitive. \nocolorpa is preferred overall against all other methods.}
    \label{fig:user_study}
    \vspace{-3mm}
\end{figure*}

The third row (\emphbox{SSIM Metric}) reports the algorithms' performance across five datasets (the corresponding plot for the Estate Correlation dataset appears in the first row), focusing on the SSIM metric, which is our only image-based metric.  We note that it is normal for sampling behavior to vary across datasets, given the different shapes and trends, or lack thereof.  However, we observe that \nocolorpa robustly achieves high scores across all datasets, and outperforms other methods in most cases.  We note that, intuitively, perceptual similarity increases as sample sizes grow for all methods, so differences are often more pronounced at smaller sizes. On the MNIST data, all methods are close in similarity scores; this is due to the clustering structure and even density across data regions, which allows perception-agnostic methods to perform well despite their limitations. In ACSI, the probabilistic methods are distinctly worse, but the others achieve similar results.

\smallskip
\noindent
\textbf{Runtime performance.}
We discuss the runtime performance of the algorithms on the Hidden Correlation dataset ($\sim\! 3.5M$), which is the largest one in our evaluation (Figure~\ref{fig:grouped_results}, right). \nocolorpa and \nocolorMM are the fastest among algorithms with reasonable perceptual efficacy.  Their runtimes are similar, because they implement the same algorithm with modified objectives.
While \nocolorran and \nocolordbs are faster, they typically produce samples of poor perceptual quality.  The KNN algorithm, which is used by \nocolordbs
to identify points in sparse areas, dominates its running time, which stays constant across sample sizes.
The runtimes of \nocolorvas and \nocolorbn are significantly longer, by 2--3 orders of magnitude.  \nocolorvas takes a long time to converge (${\sim}23$ minutes for a sample of 250 points), though we have observed that it produces samples of similar quality after one pass over the data (${\sim}5$ minutes for a sample of 250 points).  Figure~~\ref{fig:grouped_results} reports both runtimes (till full convergence and for a single pass).
We note that \nocolorbn is extremely inefficient (${\sim}3.6$ \emph{hours} for a sample of 250 points), which aligns with prior analyses~\cite{quadri2022automatic, chen2019recursive}.

\smallskip
\noindent
\newcontent{\textbf{User study.}
We conducted a user study to assess how humans perceive \nocolorpa samples against each of the other baselines, across all datasets in our study. We recruited 17 Computer Science students ($\textup{Mean}_{\textup{age}}=27$, $\textup{SD}_{\textup{age}}=3.4$) familiar with statistical concepts, verified through a few simple questions. In each task, participants viewed the full dataset---where they could adjust marker transparency---and compared a \nocolorpa sample against one alternative baseline at a time (a total of five sample pairs).  For each sample pair, we asked participants to select the sample that is best at (a)~preserving correlation trends, clusters, and other patterns, (b)~retaining outliers and overall data shape, (c)~preserving density variations, and (d)~best overall (Figure~\ref{fig:user_overall}).
For each question, participants could select ``sample 1'', ``sample 2'', or ``no preference''. For each dataset, we chose sample sizes between 0.3 and 16\% of the data size (more details in \appref{sec:appendix}{onlineAppendix}).

Figure~\ref{fig:user_study} presents aggregated results across all datasets and participants: i.e., $\sim\!50$ trials per sample pair.
For each matchup, we report how often users preferred \nocolorpa, how often they preferred the alternative baseline, and how often they declared no preference.   The closest in user preference is \nocolorbn, which ends up slightly overtaking \nocolorpa in the ``overall best'' category, but, as we discussed, \nocolorbn is computationally prohibitive. Users show strong preference for \nocolorpa against all other alternative methods.  In particular, probabilistic-based methods generally rank below \nocolorpa, but we observed participants only decisively prefer \nocolorpa when those methods fail to capture shape and outliers (as is the case on Hidden Correlation and ACSI).  We show a per-dataset analysis of the results in \appref{sec:appendix}{onlineAppendix}.
}

\subsection{\nocolorapproxpa synopsis evaluation}\label{sec:exp2}

\looseness-1
In this section, we evaluate the effectiveness of approximate visualizations generated by \nocolorapproxpa (Algorithm~\ref{algo:approxpa}), using our perception-aware compression (Section~\ref{sec:pa_compression}). Compression splits the canvas area into non-overlapping bounding boxes based on thresholds $\lambda$ and $\sigma$, controlling data approximation and homogeneity \newcontent{of perception weights}, respectively.  These parameters affect the compression level implicitly: lower thresholds result in more and smaller boxes (lower compression), thus, sample points drawn randomly within a box match better the underlying data and lead to lower distortion (example in Figure~\ref{fig:hidden_corr_approx_scheme}).  We present results at three compression levels: (1)~low: $\langle \lambda, \sigma \rangle= \langle 0.001,  0.001 \rangle$, (2)~medium: $\langle \lambda, \sigma \rangle= \langle 0.002, 0.001 \rangle$, and (3)~high: $\langle \lambda, \sigma \rangle= \langle 0.003,  0.01 \rangle$. For the ACSI dataset, we alter the medium and high levels slightly, to $\langle \lambda, \sigma \rangle= \langle 0.005, 0.001 \rangle$ and $\langle \lambda, \sigma \rangle= \langle 0.01,  0.01 \rangle$, respectively.  We note that the derived compressions may differ in the number of boxes: e.g., hidden correlation has 9$,$658, 4$,$042, and 993 boxes across the three levels (Figure~\ref{fig:hidden_schemes}), while estate correlation has 3$,$491, 1$,$184, and 448 boxes.
We give more details and context in \appref{sec:appendix}{onlineAppendix}. In practice, the thresholds can be tuned by data analysts using a grid search approach to reach the desired level of compression.

\begin{figure*}[t]
    \centering
    \includegraphics[width=0.9\columnwidth]{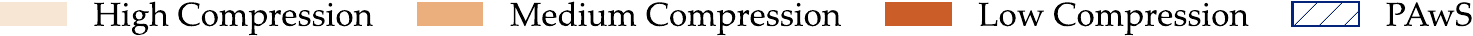}\\
    \begin{subfigure}[t]{0.19\textwidth}
        \centering
        \includegraphics[width=1.02\textwidth]{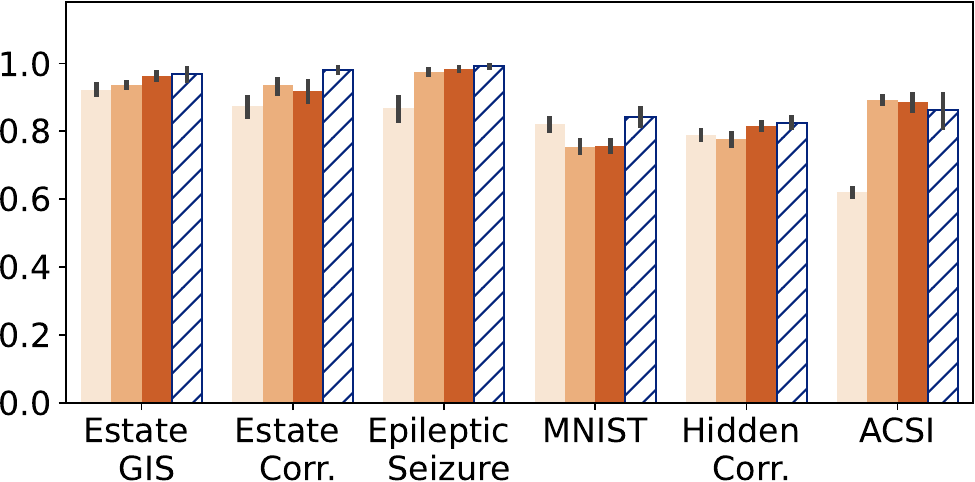}
        \vspace{-5mm}
        \caption{SSIM$\uparrow$}
    \end{subfigure}\hfill
    \begin{subfigure}[t]{0.19\textwidth}
        \centering
        \includegraphics[width=1.02\textwidth]{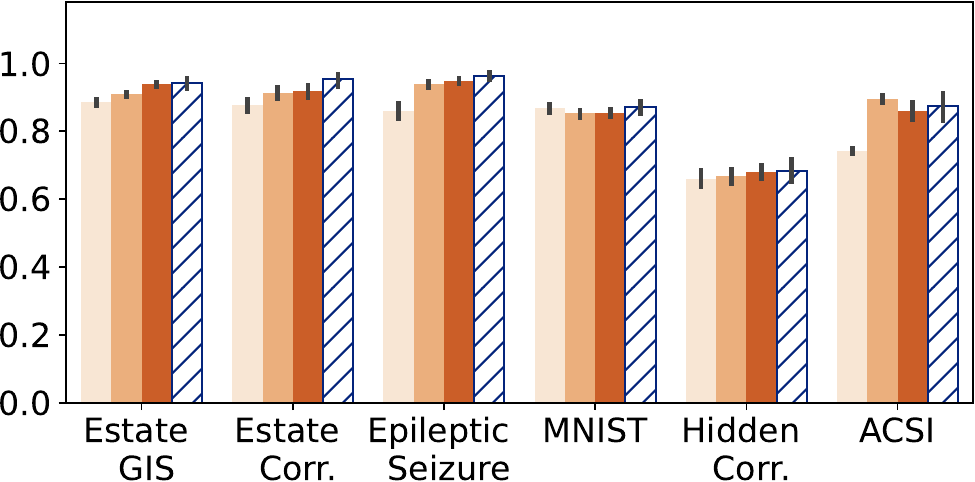}
        \vspace{-5mm}
        \caption{1-JSD$\uparrow$}
    \end{subfigure}\hfill
    \begin{subfigure}[t]{0.19\textwidth}
        \centering
        \includegraphics[width=1.02\textwidth]{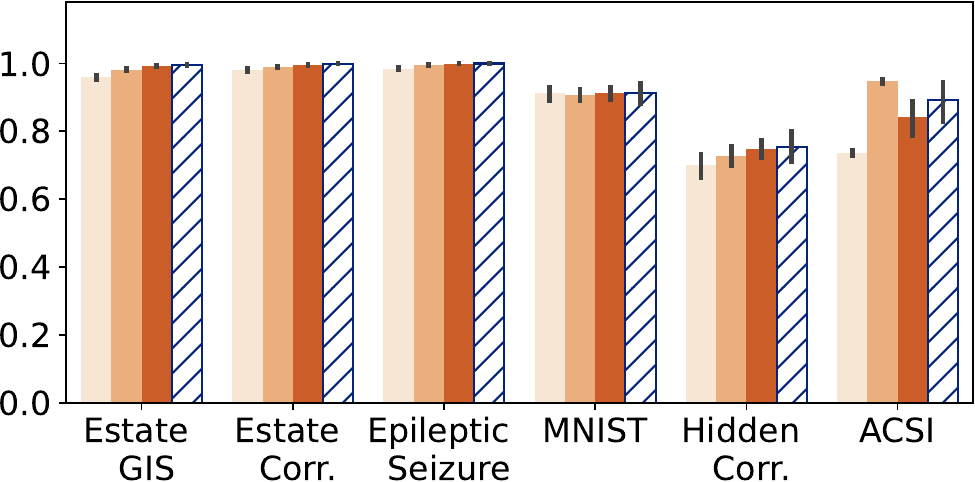}
        \vspace{-5mm}
        \caption{CC$\uparrow$}
    \end{subfigure}\hfill
    \begin{subfigure}[t]{0.19\textwidth}
        \centering
        \includegraphics[width=1.02\textwidth]{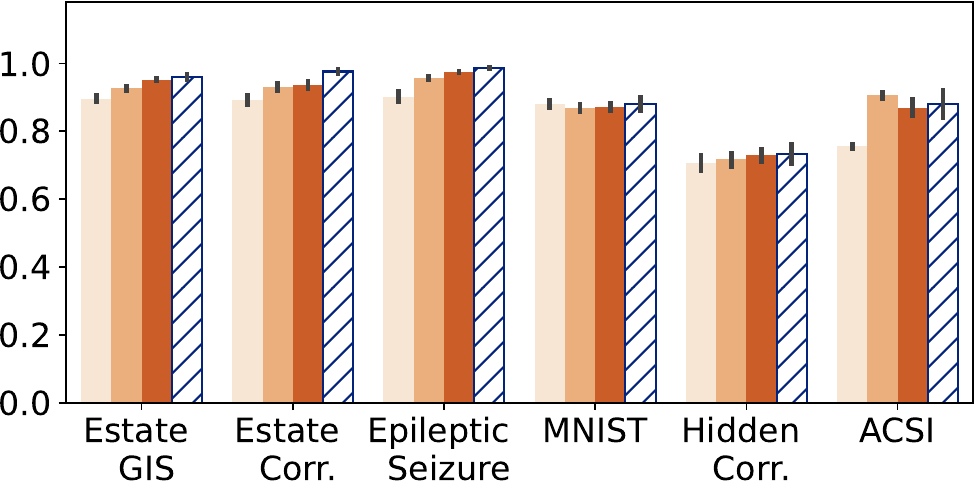}
        \vspace{-5mm}
        \caption{SIM$\uparrow$}
    \end{subfigure}\hfill
    \begin{subfigure}[t]{0.19\textwidth}
        \centering
        \includegraphics[width=1.02\textwidth]{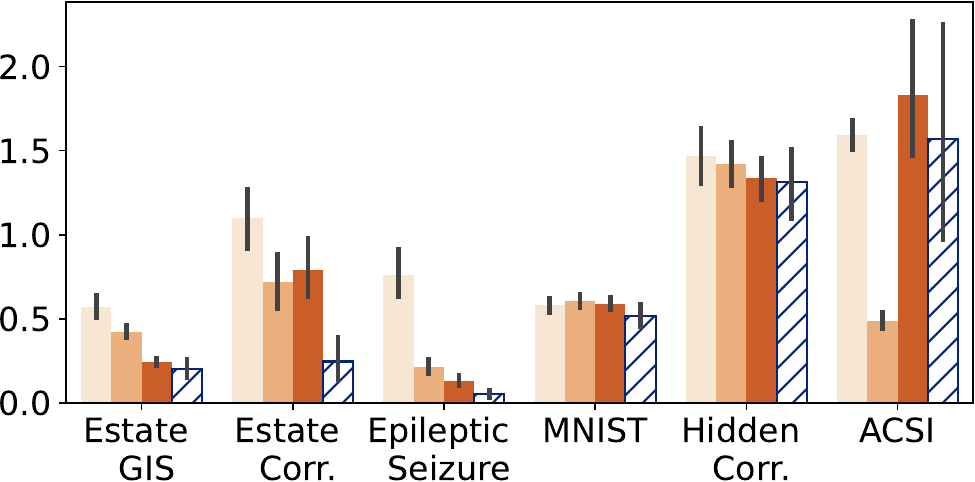}
        \vspace{-5mm}
        \caption{EMD$\downarrow$}
    \end{subfigure}
    \vspace{-4mm}
    \caption{\nocolorapproxpa shows similar performance to \nocolorpa across all metrics and various levels of compression.}
    \label{fig:approx_metrics}
    \vspace{-3mm}
\end{figure*}

Figure~\ref{fig:approx_metrics} contrasts the perceptual quality of \nocolorapproxpa samples at 3 compression levels with \nocolorpa.
We report the methods' similarity scores at sample size $k=9,611$---intuitively, distortion due to compression becomes more visible in larger sample sizes.

We note that \nocolorapproxpa achieves excellent scores on the first four metrics, closely matching the performance of \nocolorpa. We achieve high performance even at the highest compression level. Still, the trend that perceptual quality increases as compression decreases remains visible.  We note that values better than \nocolorpa are possible, though uncommon; we observe them on the ACSI dataset, which models a corner case in scatterplots, and the EMD metric, which is known in the literature to have less consistent behavior, and to be more difficult to optimize for, and ``\emph{prefers sparser predictions, even if they do not perfectly align with fixated regions}''~\cite{bylinskii2018different}.

\begin{figure}[t]
    \centering
    \includegraphics[width=0.4\textwidth]{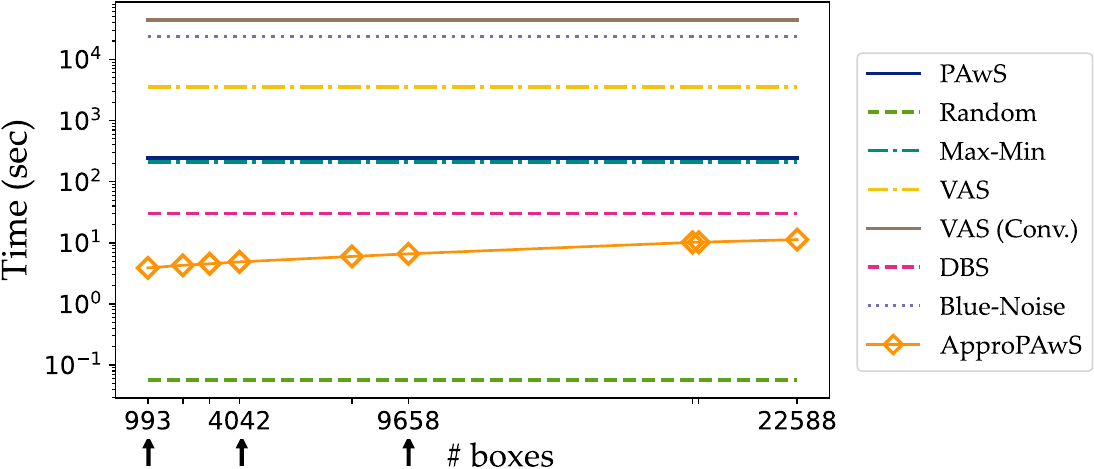}
    \vspace{-4mm}
    \caption{\nocolorapproxpa is significantly more efficient than \nocolorpa and state-of-the-art, even at low compression levels. $\uparrow$ marks the compression levels used in the results of Figure~\ref{fig:approx_metrics}.
    }
    \vspace{-3mm}
    \label{fig:approx_running_times}
\end{figure}

\smallskip
\noindent
\textbf{Runtime performance.} \looseness-1
Figure~\ref{fig:approx_running_times} demonstrates how the running time of \nocolorapproxpa scales as compression decreases.  We use the largest sample size ($k$=9$,$611) and largest dataset (hidden correlation) in our evaluation, to test 9 levels of compression, resulting in representations with 993 boxes, up to 22$,$588.

The running time of \nocolorapproxpa increases with the number of boxes (as compression decreases), but not drastically. It is faster than all methods (except \nocolorran, which has poor perceptual quality), and  100x faster than \nocolorpa. Note that at 9$,$658 boxes (lowest compression in Figure~\ref{fig:approx_metrics}), \nocolorapproxpa produces approximate visualizations of high quality---comparable to \nocolorpa but way faster. Intuitively, the gains in efficiency are higher in larger datasets and higher compression rates.

%% file: 2_RelatedWork_arxiv27.tex
\section{Related Work}

Our work sits at the intersection of database systems for interactive visualization, approximate query processing, and sampling for scatterplots. We position our contributions relative to each thread.

\noindent
\textbf{Database systems for interactive visualization.}
Supporting interactive visual analytics at scale is a well-studied problem in the database community. Data prefetching and precomputation\cite{battle2016dynamic} reduce latency by anticipating user interactions; specialized index structures---VIS-Trees~\cite{el2016vistrees}, AID*~\cite{ghosh2020aid}, and series indexes~\cite{zoumpatianos2014indexing}---accelerate data retrieval for visual exploration workloads.
Our perception-augmented database complements these approaches: it contributes a new metadata layer (perceptual synopses) and a new operator class (perception-aware sampling queries) that optimize for what a human analyst perceives, rather than for retrieval speed alone.
Kwon et al.~\cite{kwon2017sampling} survey the role of sampling broadly in scalable visual analytics, providing useful context for our work.

\noindent
\textbf{Approximate query processing and perceptual error.} \looseness-1
AQP systems answer aggregate queries with bounded statistical error using precomputed synopses—histograms, wavelets, sketches, and samples~\cite{cormode2011synopses, hellerstein1999interactive, rahman2017ve}.
Related work has also explored incremental and progressive approaches~\cite{fisher2012trust, rahman2017ve} and optimized stratified sampling~\cite{chaudhuri2007optimized, ding2016sample+}. The closest work to ours is Alabi and Wu~\cite{alabi2016pfunk}, who introduce perceptual functions to determine when statistical error becomes visually indiscernible in bar charts, improving sampling rates for approximate aggregation queries. Our work extends this perceptual intuition to scatterplots and to a richer objective: rather than bounding when error is imperceptible, we directly optimize sampling to minimize perceptual distortion. The \nocolorapproxpa synopsis---a perception-aware quad-tree partition of the visualization canvas---plays the same architectural role as a synopsis in AQP~\cite{cormode2011synopses}, but is constructed to minimize perceptual rather than statistical error.

\noindent
\textbf{Sampling for visualization.} %
Sampling is a primary technique for interactive visual analytics in both the visualization and database communities~\cite{qin2020making}.
For scatterplots, prior methods can be characterized by their end goal:
preserving (1)~relative densities, (2)~spatial separation and data shape, or (3)~outliers~\cite{quadri2022automatic, yuan2020evaluation}.
Density-oriented methods include non-uniform sampling in the data and pixel space~\cite{bertini2004chance, bertini2006give}, SVD-based approaches~\cite{joia2015uncovering}, Z-order space-filling curves for kernel density estimation~\cite{hu2019data, zheng2013quality}, and KD-tree structures for balanced area coverage~\cite{chen2010information}.
Visualization-aware sampling (VAS)~\cite{park2016visualization} and Max-Min (farthest-point) sampling~\cite{1994:HSC:2753204.2753214, eldar1997farthest} optimize for spatial separation and data shape, while blue-noise sampling~\cite{yan2015survey} seeks a balanced representation between dense and sparse areas at high computational cost.
A recent evaluation of seven methods~\cite{yuan2020evaluation} finds that existing algorithms each excel at preserving only one or two dimensions of data patterns; none holistically models human perception, which is the novel objective of our work.
Cluster-based methods (e.g., HDBSCAN~\cite{HDBScan}) first discover clusters and then allocate sampling budget accordingly. The clustering parameters can significantly impact the outcome, and they don't work well in non-cluster settings.
For bars, pies, heatmaps, and line charts, the database community has developed a rich body of AQP-based sampling techniques~\cite{moritz2017trust, ding2016sample+, fisher2012trust, rahman2017ve, kim2015rapid, alabi2016pfunk, agarwal2013blinkdb, chaudhuri2007optimized, minos2011approximate, gibbons1998new, hellerstein1999interactive, acharya1999aqua};
we refer the reader to Qin et al.~\cite{qin2020making} for a comprehensive survey. For multi-class scatterplots, set-cover optimization~\cite{hu2019data} and outlier-aware extensions of known methods~\cite{liu2017visual, wei2010multi, xiang2019interactive} have been proposed;
our work focuses on single-class scatterplots, consistent with the most effective methods identified by Yuan et al.~\cite{yuan2020evaluation}.
Orthogonal to sampling, other approaches that aim to improve the usability of visualization systems include improving user experience and optimizing visualization design~\cite{quadri2022automatic, kim2018assessing, micallef2017towards, szafir2017modeling, ellis2007taxonomy, demiralp2014learning},  alternative data representations~\cite{keim2010generalized}, and designing recommendation-based systems for data exploration and visualization~\cite{vartak2015seedb, dimitriadou2016aide, porwal2022efficient, garg2023reinforced, qian2021learning}.

\noindent
\textbf{Information theory and probabilistic methods.}
Entropy and mutual information have been studied for visualizations of simulation and spatiotemporal data, targeting preservation of statistical associations~\cite{chen2010information, dutta2019multivariate, biswas2013information}.
Probabilistic methods---stratified~\cite{aoyama1954study}, systematic~\cite{madow1944theory}, and cluster sampling~\cite{henderson1982cluster}---are computationally efficient but have known limitations for visualization tasks~\cite{park2016visualization}.

\noindent
\textbf{Saliency maps and perception}  have seen interest in computer vision~\cite{shanmuga2015eye, shi2017gaze} and natural language processing~\cite{bonhage2015combined, conklin2016using}. However, data visualization does not always follow the rules of perception in the natural world~\cite{franconeri2021science}, thus, improving on these models and adapting them to predict human focus in visualizations is an active problem in visualization research~\cite{matzen2017data, shin2022scanner, kim2017bubbleview}. Saliency models for visualizations are still in their infancy but already demonstrate good performance at predicting attention-focus areas~\cite{matzen2017data, shin2022scanner}.
We use the DVS model~\cite{matzen2017data}, which builds upon Itti et al.~\cite{itti1998model}, as a black box. Future advances in visual perception tools can directly improve our methods through more effective saliency data.

%% file: 7_Conclusions_arxiv27.tex
\section{Summary and Future Directions}

We introduced perception-augmented databases: a framework that extends conventional databases with perceptual metadata, enabling a new class of query operators that optimize for how humans perceive visualized data. We proposed two operators: \nocolorpa, a perception-aware sampling query that leverages saliency maps and Max-Min diversification, and \nocolorapproxpa, an approximate visualization query that operates directly over a perception-aware quad-tree synopsis without accessing the underlying data. Our evaluation demonstrated significant quality and runtime gains over five state-of-the-art methods across six datasets up to 3.5M points, and a user study confirmed that humans prefer \nocolorpa samples.

\looseness-1
There are interesting challenges in optimizing the overhead of storing saliency for multiple candidate visualizations, which are called scatterplot matrices (SPLOMs) in the visualization literature~\cite{hu2019data}. Future work could leverage insights from the computer vision literature---such as saliency mix-up techniques~\cite{kim2020}, which blend salient regions across multiple images---as a potential direction for synthesizing saliency information across multiple views. Synopsis maintenance (e.g., recomputing the saliency heatmap when the underlying data changes significantly) is analogous to incremental statistics maintenance in database systems, and is an interesting direction for future work.
On-the-fly sample augmentation can be useful to support interactivity in visualizations (e.g., zooming in), and the biggest challenge is to achieve interactive sampling times.  With eye-tracking technologies, one may also aim for on-the-fly sample adaptation, based on user interactions and eye-gaze.  Importantly, evaluating sampling methods in the context of visualizations is challenging, as there is no established benchmark.  Creating such a benchmark would facilitate future research and make the comparison among different techniques easier and more robust.
On the database systems side, an important direction is the integration of perception-aware operators into existing query engines and visual analytics platforms, and the design of query optimization strategies that select between \nocolorpa and \nocolorapproxpa based on data size, available preprocessing budget, and latency constraints.

%% file: 6_Appendix_arxiv27.tex
\section{Appendix}\label{sec:appendix}

\subsection{Perception-awareness vs cluster-preservation}

Several datasets in our evaluation, including the dataset from Example 1 (Figure~\ref{fig:random_ex1}), contain visually distinct groups of points. This raises the question of how perception-awareness relates to cluster preservation and, whether \nocolorpa should be viewed as another cluster-aware sampling technique. In this section, we show why this is not the case.

Cluster-aware techniques assume that the data contain clusters and aim to preserve that structure. \nocolorpa makes no such assumption: it aims to preserve whatever structures are perceptually salient in the visualization, and therefore adapts to different structures.

In this section, we first show how cluster-aware sampling breaks down once there are no clusters to preserve. We then turn to datasets with clear cluster structure and show that \nocolorpa performs well even though it is not explicitly designed to preserve clusters. %

\noindent
\textbf{Cluster-aware sampling.} Cluster-aware sampling aims to preserve the cluster structure of the underlying dataset in the sample. It first identifies clusters and then distributes the sampling budget across them, for example equally across clusters or in proportion to their sizes, before selecting representative points within each cluster uniformly at random. We demonstrate this approach using HDBSCAN~\cite{campelloDBSCAN}, a widely used density-based clustering algorithm that does not require the number of clusters as input and labels points in low-density regions as noise.

\begin{figure}[t]
  \centering
  \begin{subfigure}[t]{0.25\columnwidth}
    \centering
    \includegraphics[width=0.95\textwidth]{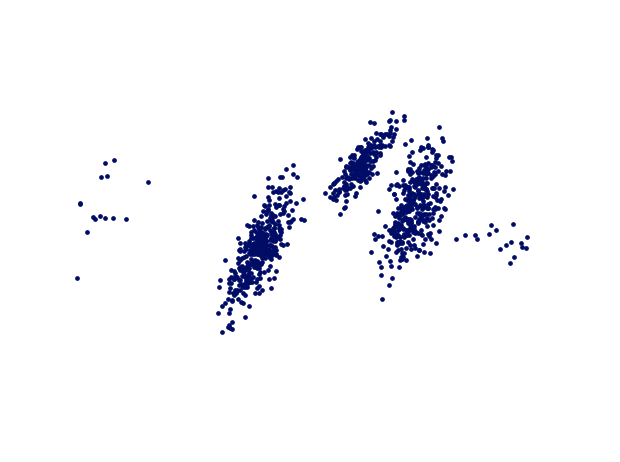}
    \vspace{-6mm}
    \caption{Proportional}
    \label{fig:intro_proportional}
  \end{subfigure}%
 \begin{subfigure}[t]{0.25\columnwidth}
      \centering
      \includegraphics[width=0.95\textwidth]{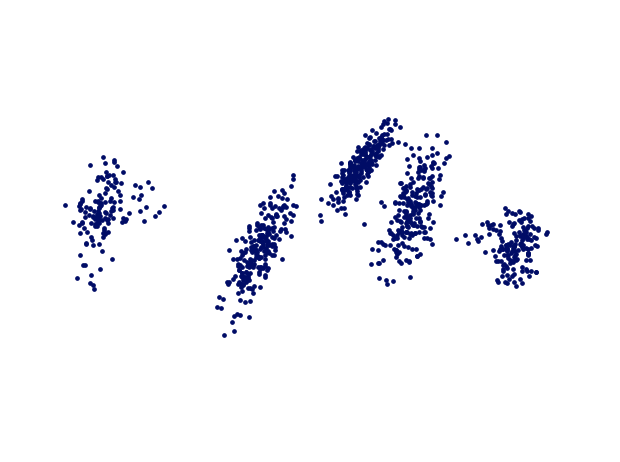}
      \vspace{-6mm}
      \caption{Equal}
     \label{fig:intro_proportional}
  \end{subfigure}%
   \begin{subfigure}[t]{0.25\columnwidth}
      \centering
      \includegraphics[width=0.95\textwidth]{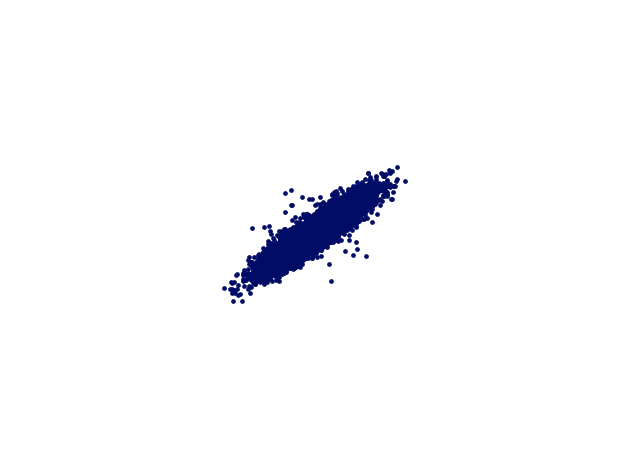}
      \vspace{-6mm}
      \caption{Proportional}
     \label{fig:pa}
  \end{subfigure}%
   \begin{subfigure}[t]{0.25\columnwidth}
      \centering
      \includegraphics[width=0.95\textwidth]{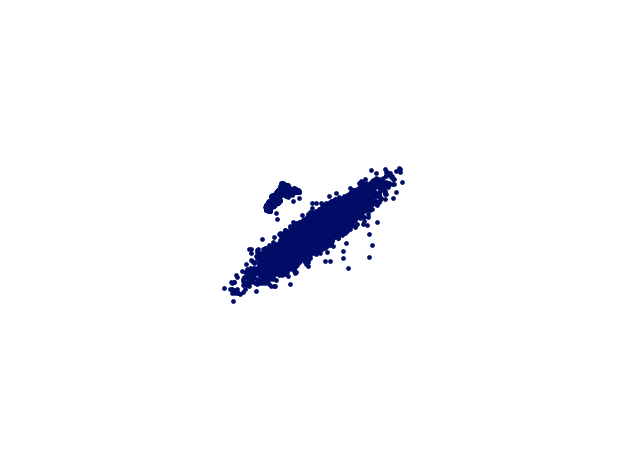}
      \vspace{-6mm}
      \caption{Equal}
     \label{fig:pa}
  \end{subfigure}
  \vspace{-8mm}
  \caption{Cluster-aware sampling (HDBSCAN, then proportional or equal allocation) on Example 1 and Hidden Correlation. }
  \label{fig:samples}
\end{figure}

HDBSCAN works on the data of Example~\ref{ex:clusters} (Figure~\ref{fig:samples}), which has cluster structure,  but the allocation parameter has no right value: equal allocation renders the two small cohorts as dense as the main trends, while proportional allocation preserves densities but under-represents the small cohorts, much like random sampling. In both cases the outlier spread is lost, since points HDBSCAN labels as noise receive no budget. On Hidden Correlation, which has no cluster structure, proportional allocation behaves like random sampling, and equal allocation inflates a small region HDBSCAN found in the noise into a cluster that the original data does not visually contain. Neither recovers the overall shape that \nocolorpa preserves. Cluster-aware sampling adapts to what the clustering finds, while perception-aware sampling adapts seamlessly to different data structures.

These results clarify that \nocolorpa is not a cluster-preservation method and targets no specific task. They further show that cluster-aware sampling is not competitive, especially in non-clustered data patterns.

\textbf{Cluster Separation: \nocolorpa vs \nocolorMM.} Although \nocolorpa is not designed for clustering, we now examine how well it preserves cluster structure when such structure exists. We complement our saliency-based evaluation with a data-based assessment of cluster separation.

We evaluate cluster separation on two datasets with clear cluster structure: the dataset from Example~1 (Figure~\ref{fig:original}), which contains five groups, and MNIST from Section~\ref{sec:experiments}, whose ten digit classes form visually distinct clusters. Because each sampled point comes directly from the original dataset, it retains its ground-truth label: the generating group for Example~1 and the digit class for MNIST.

For each sample, we measure cluster quality in the normalized coordinate space of the visualization using the Silhouette Coefficient (SC)~\cite{rousseeuw1987silhouettes} and Davies--Bouldin Index (DBI)~\cite{davies1979cluster}. Both metrics reward compact, well-separated groups. Let $X$ denote the evaluated points, partitioned into $m$ nonempty clusters $C_1,\ldots,C_m$. All distances are Euclidean.

\begin{figure}[t]
    \centering
    \begin{subfigure}[t]{0.5\linewidth}
        \centering
        \resizebox{0.8\linewidth}{!}{%
        \begin{tabular}{@{}rrrrr@{}}
            \toprule
            & \multicolumn{2}{c}{SC $\uparrow$}
            & \multicolumn{2}{c}{DBI $\downarrow$} \\
            \cmidrule(lr){2-3}\cmidrule(l){4-5}
            $k$ & \nocolorpa & \nocolorMM & \nocolorpa & \nocolorMM \\
            \midrule
            250   & \textbf{0.142} & $-0.223$ & \textbf{1.435} & 2.039 \\
            500   & \textbf{0.124} & $-0.150$ & \textbf{1.487} & 2.059 \\
            1,000 & \textbf{0.150} &  0.054   & \textbf{1.437} & 1.846 \\
            1,500 & \textbf{0.169} &  0.029   & \textbf{1.379} & 1.652 \\
            3,000 & \textbf{0.227} &  0.168   & \textbf{1.196} & 1.321 \\
            4,271 & \textbf{0.281} &  0.243   & \textbf{1.056} & 1.144 \\
            \bottomrule
        \end{tabular}}
        \caption{Dataset of Example~\ref{ex:clusters} (Figure~\ref{fig:random_ex1})}
        \label{fig:cluster-separation-synthetic}
    \end{subfigure}\hfill
    \begin{subfigure}[t]{0.5\linewidth}
        \centering
        \resizebox{0.8\linewidth}{!}{%
        \begin{tabular}{@{}rrrrr@{}}
            \toprule
            & \multicolumn{2}{c}{SC $\uparrow$}
            & \multicolumn{2}{c}{DBI $\downarrow$} \\
            \cmidrule(lr){2-3}\cmidrule(l){4-5}
            $k$ & \nocolorpa & \nocolorMM & \nocolorpa & \nocolorMM \\
            \midrule
            250   & \textbf{0.26} & 0.15 & \textbf{1.35} & 1.90 \\
            375   & \textbf{0.26} & 0.15 & \textbf{1.30} & 1.81 \\
            844   & \textbf{0.25} & 0.17 & \textbf{1.48} & 1.94 \\
            1,898 & \textbf{0.25} & 0.19 & \textbf{1.49} & 1.73 \\
            4,271 & \textbf{0.25} & 0.21 & \textbf{1.46} & 1.63 \\
            9,611 & \textbf{0.27} & 0.24 & \textbf{1.44} & 1.53 \\
            \bottomrule
        \end{tabular}}
        \caption{MNIST}
        \label{fig:cluster-separation-mnist}
    \end{subfigure}
    \vspace{-3mm}
    \caption{\nocolorpa preserves cluster structure better than \nocolorMM at every sample size $k$, especially at small $k$.}
    \label{fig:cluster-separation}
\end{figure}

\emphbox{Silhouette Coefficient (SC).} For a point $x$, let $a(x)$ denote its mean distance to points in its own group, and let $b(x)$ denote the smallest mean distance from $x$ to the points in any other cluster:

\[
\mathrm{SC} = \frac{1}{|X|}\sum_{x \in X} s(x), \quad
s(x) = \frac{b(x)-a(x)}{\max\{a(x),\,b(x)\}}
\]
For points in singleton clusters, $s(x)$ is defined as zero. SC lies in $[-1,1]$, with higher values indicating greater within-cluster cohesion and between-cluster separation.

\emphbox{Davies--Bouldin Index (DBI).} Let $s_i$ denote the mean distance of points in $C_i$ from its centroid, $\delta_{ij}$ the distance between the centroids of $C_i$ and $C_j$. Then
\[
\mathrm{DBI}= \frac{1}{m}\sum_{i=1}^{m} \max_{j \neq i} \frac{s_i+s_j}{\delta_{ij}}
\]
DBI is nonnegative, with lower values indicating more compact and better-separated groups.

Figure~\ref{fig:cluster-separation} shows that \nocolorpa achieves higher SC and lower DBI values than \nocolorMM at every reported sample size on both datasets. The differences are more pronounced at small
sample sizes. On the synthetic data, at $k=250$, \nocolorpa achieves SC$=0.142$, compared to $-0.223$ for \nocolorMM, while reducing DBI from $2.039$ to $1.435$. On MNIST, at the same sample size, SC increases from $0.15$ to $0.26$ and DBI decreases from $1.90$ to $1.35$. The gap narrows at larger sample sizes: at $k=4,271$ on the synthetic data, the SC difference is $0.038$; at $k=9,611$ on MNIST, it is $0.03$. These observations provide complementary evidence that \nocolorpa achieves a better balance of cluster cohesion and separation than \nocolorMM in the tested samples, particularly under small sampling budgets.

\subsection{Experimental Details}

\begin{figure*}[t!]
    \begin{subfigure}[t]{0.5\textwidth}
        \centering
        \includegraphics[width=0.95\textwidth]{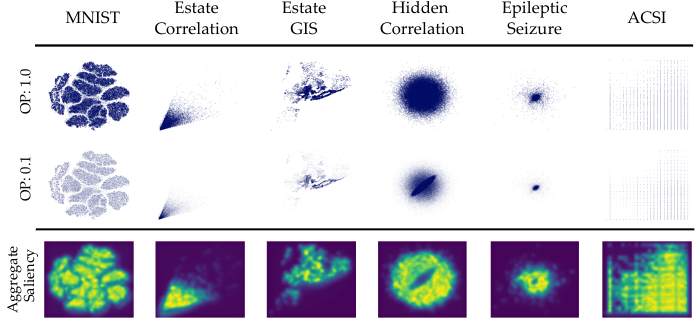}
        \caption{Dataset visualizations and saliency stored in the perception-augmented DB}
        \label{fig:datasets_summary}
   \end{subfigure}%
   \hspace{1mm}
       \begin{subfigure}[t]{0.45\textwidth}
            \centering
            \includegraphics[width=1\textwidth]{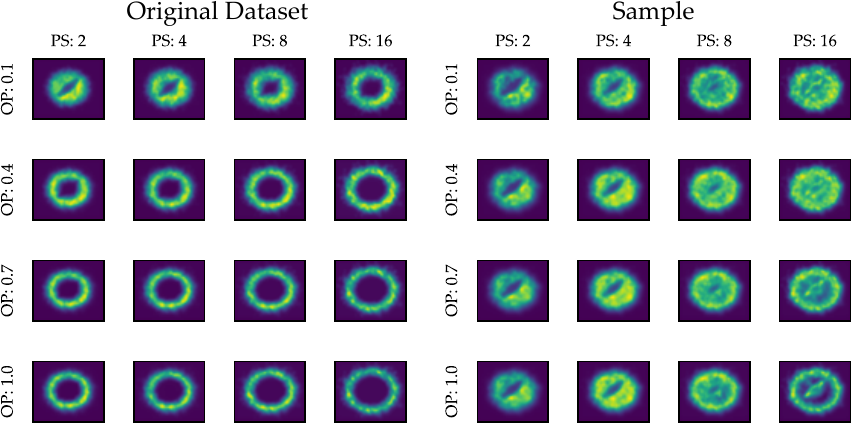}
            \caption{Perceptual similarity is averaged across all settings}
            \label{fig:maps}
       \end{subfigure}
    \vspace{-3mm}
    \caption{(a)~Visualizations of the datasets we use in the experimental evaluation and their aggregate saliency maps.
    (b)~We demonstrate the evaluation methodology on the hidden correlation dataset. We render the original dataset using 16 configurations by varying the opacity (OP) and point size (PS) and generate the saliency maps using DVS~\cite{matzen2017data} (left). Given a sample, we use the same configurations to generate 16 saliency maps for the sample (right). We compare the saliency similarity of each pair (original and sample) with the same configuration, and compute the final perceptual similarity as the average across all 16 configurations.
    }
    \label{fig:summary}
    \vspace{-3mm}
\end{figure*}

\begin{figure*}[t!]
     \begin{subfigure}[t]{0.33\columnwidth}
     \centering
     \includegraphics[width=0.95\textwidth]{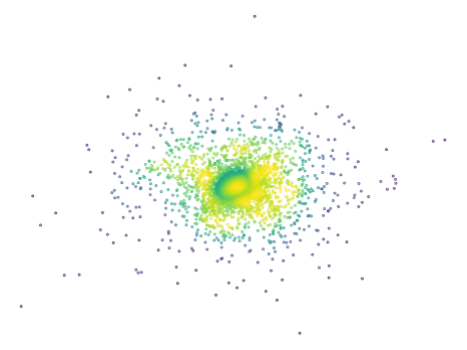}
     \vspace{-1mm}
     \caption{Epileptic Seizure}
     \label{fig:epileptic_percMap}
   \end{subfigure}%
   \begin{subfigure}[t]{0.33\columnwidth}
     \centering
     \includegraphics[width=0.95\textwidth]{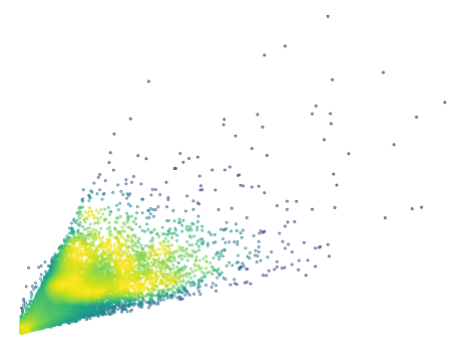}
      \vspace{-1mm}
     \caption{Estate Correlation}
     \label{fig:estateCorr_percMap}
   \end{subfigure}%
   \begin{subfigure}[t]{0.33\columnwidth}
     \centering
     \includegraphics[width=0.95\textwidth]{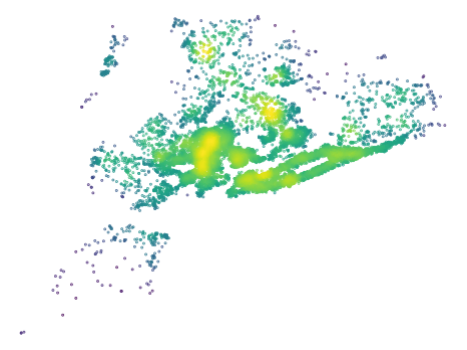}
      \vspace{-1mm}
     \caption{Estate GIS}
     \label{fig:anomalies_percMap}
   \end{subfigure}%
   \begin{subfigure}[t]{0.33\columnwidth}
     \centering
     \includegraphics[width=0.95\textwidth]{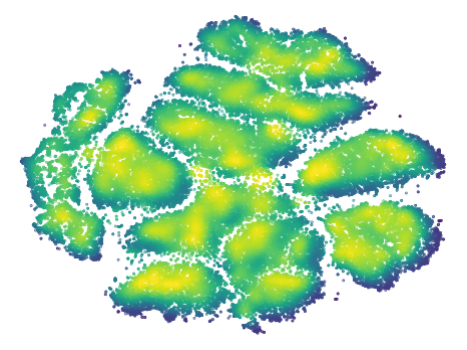}
      \vspace{-1mm}
     \caption{MNIST}
     \label{fig:mnist}
   \end{subfigure}%
   \begin{subfigure}[t]{0.33\columnwidth}
     \centering
     \includegraphics[width=0.95\textwidth]{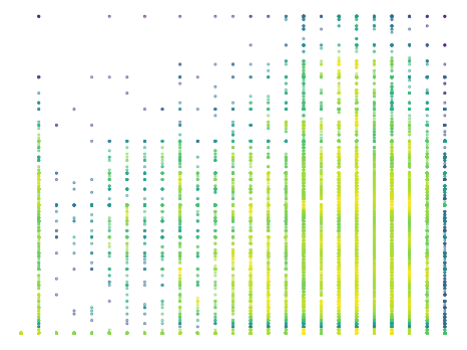}
      \vspace{-1mm}
     \caption{ACSI}
     \label{fig:acsi_percMap}
   \end{subfigure}%
   \begin{subfigure}[t]{0.33\columnwidth}
     \centering
     \includegraphics[width=0.95\textwidth]{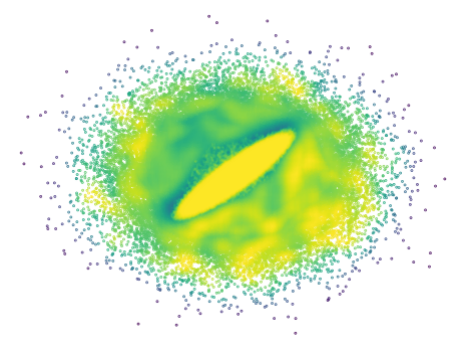}
      \vspace{-1mm}
     \caption{Hidden Correlation}
     \label{fig:hidden_percMap}
   \end{subfigure}
    \vspace{-3mm}
   \caption{Perception weights for the datasets in our experimental evaluation.}
   \label{fig:perceptionMaps}
\end{figure*}

\begin{figure}[t!]
     \begin{subfigure}[t]{0.33\columnwidth}
     \centering
     \includegraphics[width=0.95\textwidth]{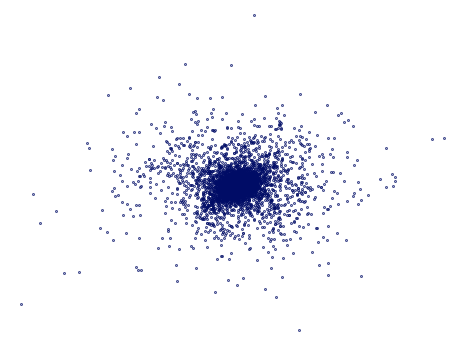}
     \vspace{-1mm}
     \caption{Epileptic Seizure}
     \label{fig:epileptic}
   \end{subfigure}%
   \begin{subfigure}[t]{0.33\columnwidth}
     \centering
     \includegraphics[width=0.95\textwidth]{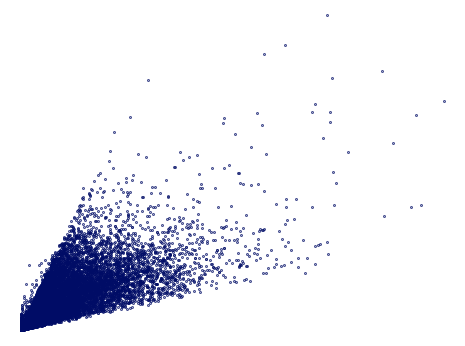}
      \vspace{-1mm}
     \caption{Estate Correlation}
     \label{fig:estateCorr}
   \end{subfigure}%
   \begin{subfigure}[t]{0.33\columnwidth}
     \centering
     \includegraphics[width=0.95\textwidth]{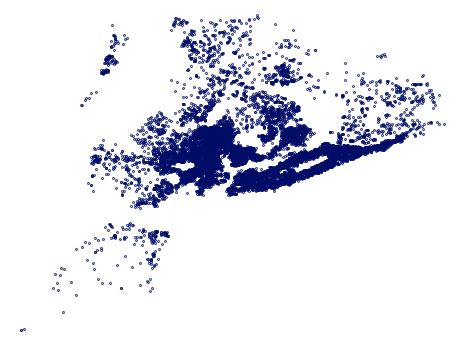}
      \vspace{-1mm}
     \caption{Estate GIS}
     \label{fig:anomalies}
   \end{subfigure}%

   \begin{subfigure}[t]{0.33\columnwidth}
     \centering
     \includegraphics[width=0.95\textwidth]{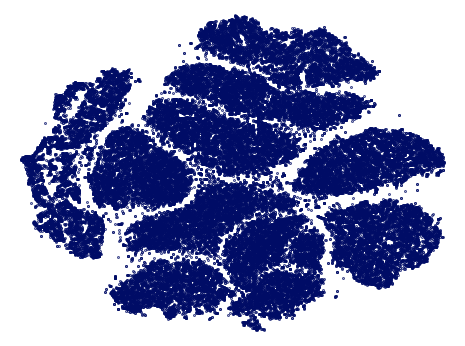}
      \vspace{-1mm}
     \caption{MNIST}
     \label{fig:mnist}
   \end{subfigure}%
   \begin{subfigure}[t]{0.33\columnwidth}
     \centering
     \includegraphics[width=0.95\textwidth]{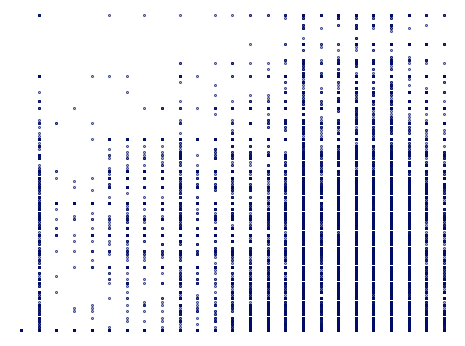}
      \vspace{-1mm}
     \caption{ACSI}
     \label{fig:acsi}
   \end{subfigure}%
   \begin{subfigure}[t]{0.33\columnwidth}
     \centering
     \includegraphics[width=0.95\textwidth]{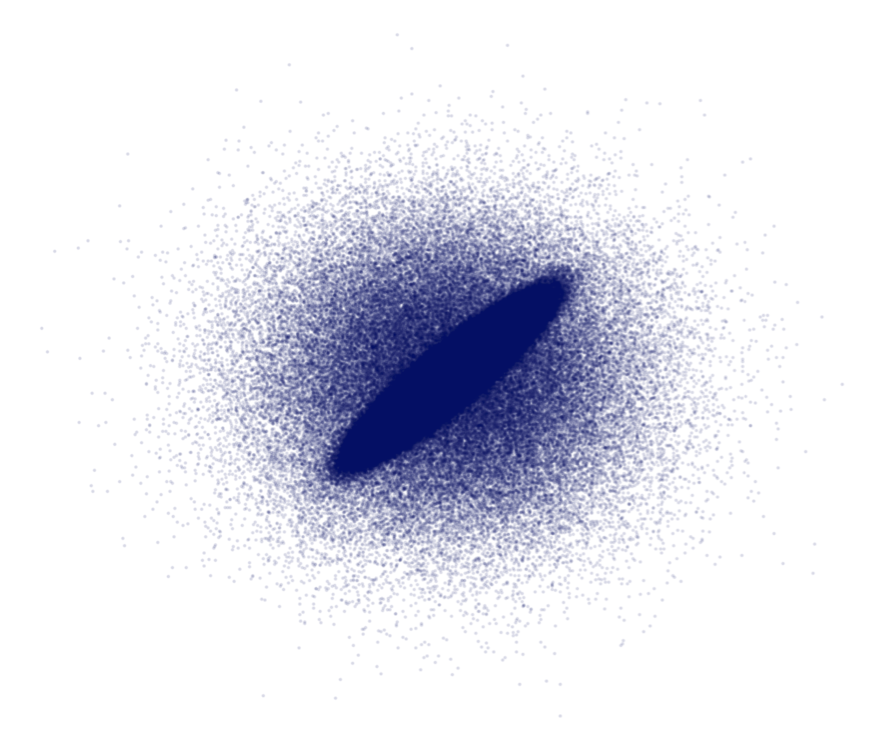}
      \vspace{-1mm}
     \caption{Hidden Correlation}
     \label{fig:hiddenCorr}
   \end{subfigure}
    \vspace{-3mm}
   \caption{Visualizations of datasets used in our evaluation.}
   \label{fig:datasetsVis}
\end{figure}

\noindent
\textbf{Overview of experimental design.} Figure~\ref{fig:summary} provides an overview of the experimental design choices we described in Section~\ref{sec:experiments}. Figure~\ref{fig:datasetsVis} shows the visualizations of the datasets used in our experiments.
We show the aggregate saliency maps derived using our approach and describe our methodology for computing perceptual similarity. Figure~\ref{fig:perceptionMaps} shows the perception weights for the datasets in our evaluation, assigning a higher perception value to highly salient or highly dense areas.

\begin{figure*}[h]
    \centering
    \includegraphics[width=0.7\textwidth]{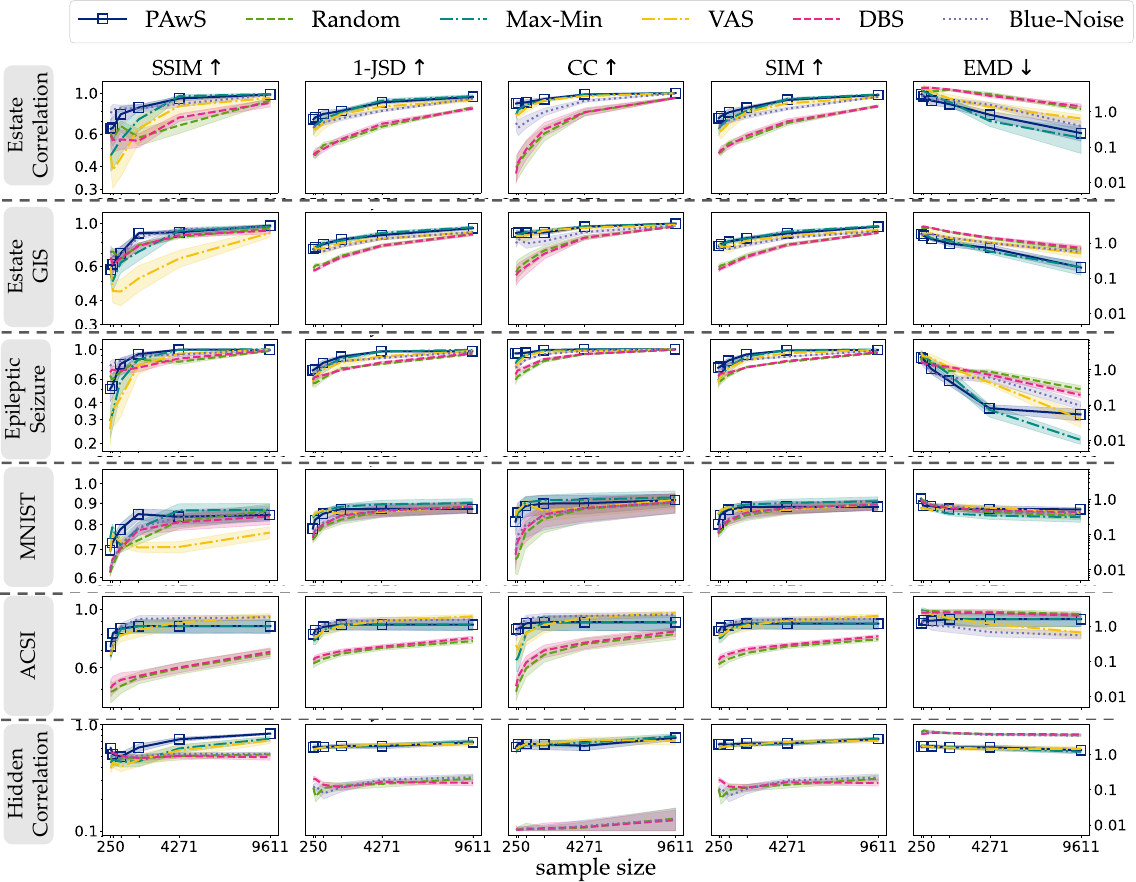} %
    \vspace{-4mm}
    \caption{Experimental results across all datasets and metrics.}
    \label{fig:datasets_across_metrics}
    \vspace{-1mm}
\end{figure*}

\smallskip
\noindent
\textbf{Perceptual similarity metrics.} Figure~\ref{fig:datasets_across_metrics} shows extensive results about the behavior of different sampling methods across six datasets and five evaluation metrics introduced in Section~5. The observations we make are consistent with the discussion in Section~6, and different metrics show similar results. \nocolorpa has strong behavior across all datasets and metrics, while probabilistic-based approaches consistently underperform. \nocolorMM and \nocolorvas typically perform better as the sample size increases. The behavior of \nocolorbn varies across datasets, while in hidden correlation performs as poorly as probabilistic-based methods with extremely low scores in SIM, CC, and 1-JSD metrics. We note EMD is a difficult metric to optimize for, and in some datasets (i.e., in epileptic seizure) all sampling methods reach lower scores.

\begin{figure*}[t]
    \begin{subfigure}[t]{0.5\textwidth}
        \centering
        \includegraphics[width=\textwidth]{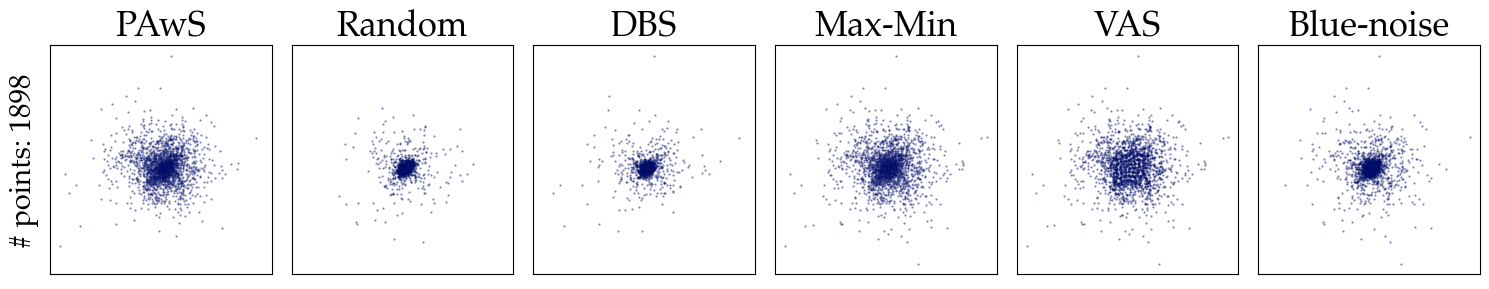}
        \vspace{-5mm}
        \caption{Epileptic Seizure}
    \end{subfigure}%
    \begin{subfigure}[t]{0.5\textwidth}
        \centering
        \includegraphics[width=\textwidth]{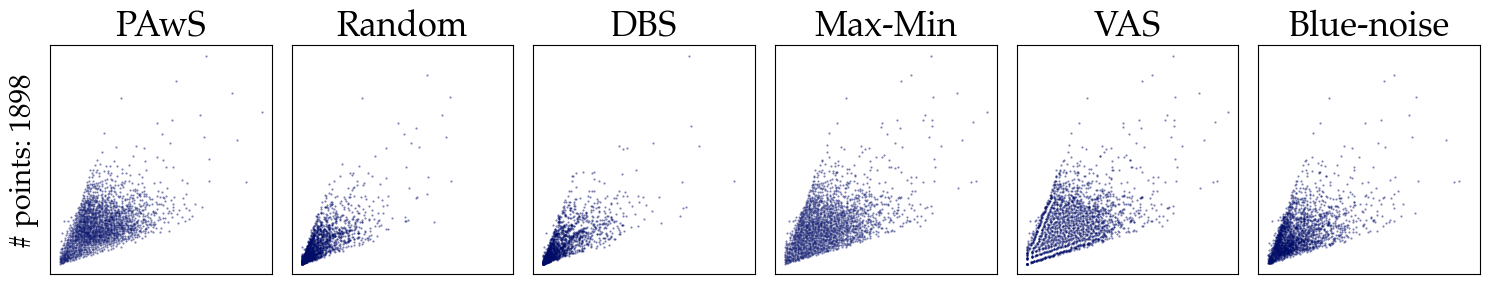}
        \vspace{-5mm}
        \caption{Estate Corr}
    \end{subfigure}
    \begin{subfigure}[t]{0.5\textwidth}
        \centering
        \includegraphics[width=\textwidth]{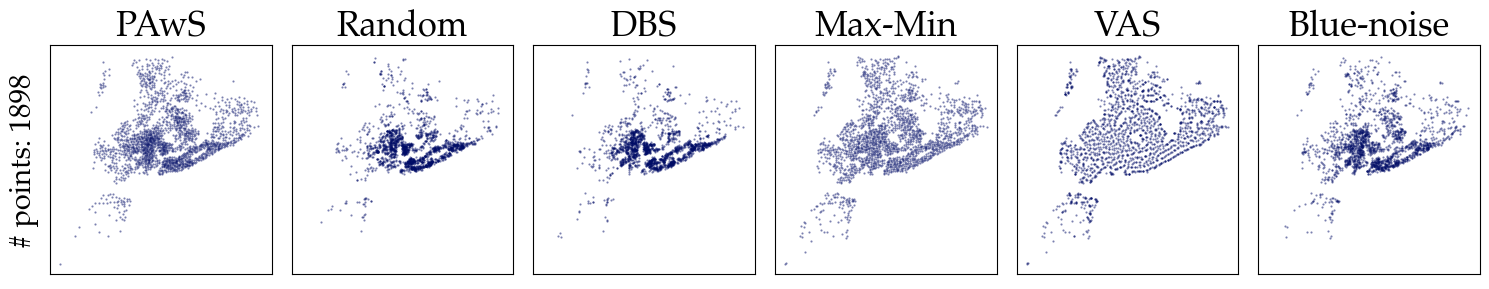}
        \vspace{-5mm}
        \caption{Estate Anomalies}
    \end{subfigure}%
    \begin{subfigure}[t]{0.5\textwidth}
        \centering
        \includegraphics[width=\textwidth]{images/mnist_user_study_samples.png}
        \vspace{-5mm}
        \caption{MNIST}
    \end{subfigure}
    \begin{subfigure}[t]{0.5\textwidth}
        \centering
        \includegraphics[width=\textwidth]{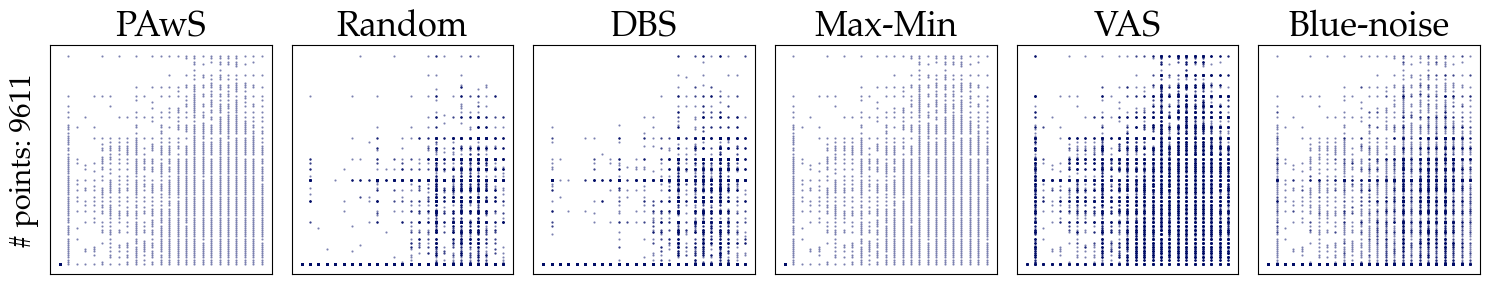}
        \vspace{-5mm}
        \caption{ACSI}
    \end{subfigure}%
    \begin{subfigure}[t]{0.5\textwidth}
        \centering
        \includegraphics[width=\textwidth]{images/hidden_corr_user_study_samples.png}
        \vspace{-5mm}
        \caption{Hidden Correlation}
    \end{subfigure}
    \vspace{-4mm}
    \caption{Visual stimuli used in the user study experiments for all datasets and sampling methods.}
    \label{fig:qualUserStudy}
\end{figure*}

\smallskip
\noindent
\textbf{Qualitative samples and user study results per dataset.} Figure~\ref{fig:qualUserStudy} shows the visualizations we used in our user study across four criteria. The study included 6 datasets, and for each dataset, participants compared a \nocolorpa sample to an alternative baseline sample at a time (five in total). The sample sizes were chosen according to the original dataset sizes: $1,898$ ($\sim$ 16\%) for Epileptic Seizure, $1,898$ ($\sim 11$\%) for Estate Correlation and Estate GIS; $4,571$ ($\sim 6$\%) for MNIST; and 9,611 ($\sim 4$\% and $\sim 0.3$\%) for the larger ACSI and Hidden Correlation datasets. In total there were: $6 \ \textup{datasets} \ \times \ 5 \ \textup{sample pairs} \ \times \ 4 \ \textup{tasks}$, resulting in 120 tasks. To distribute the workload, each participant completed tasks for 3 of the 6 datasets, and the datasets were evenly presented to participants.

\begin{figure*}[t]
    \begin{subfigure}[t]{0.19\textwidth}
        \centering
        \includegraphics[width=1.02\textwidth]{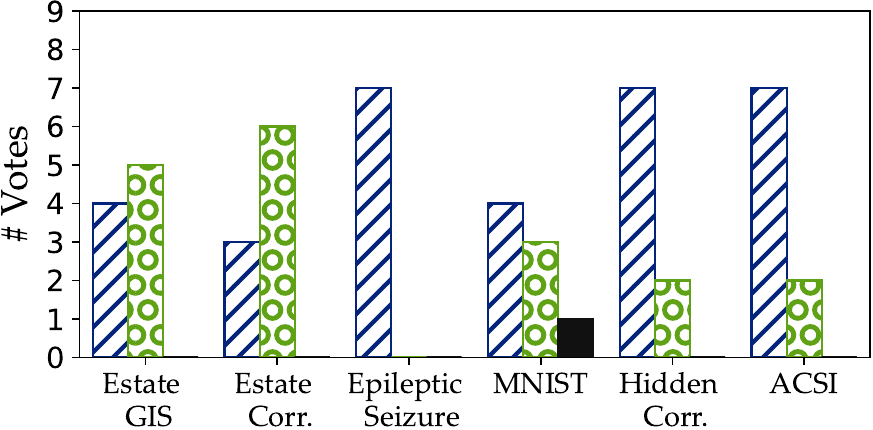}
        \vspace{-5mm}
        \caption{\nocolorpa vs \nocolorran}
    \end{subfigure}\hfill
    \begin{subfigure}[t]{0.19\textwidth}
        \centering
        \includegraphics[width=1.02\textwidth]{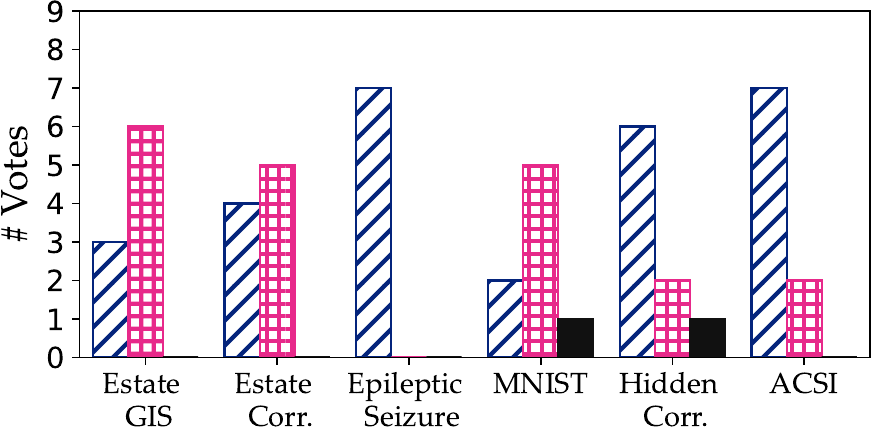}
        \vspace{-5mm}
        \caption{\nocolorpa vs \nocolordbs}
    \end{subfigure}\hfill
    \begin{subfigure}[t]{0.19\textwidth}
        \centering
        \includegraphics[width=1.02\textwidth]{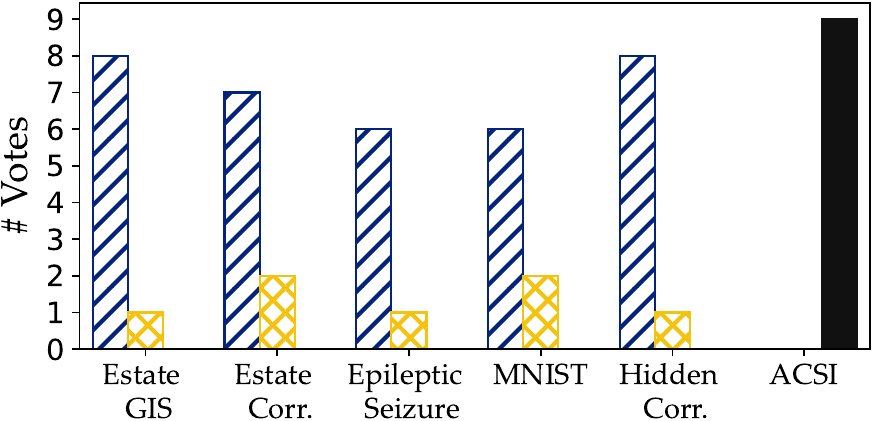}
        \vspace{-5mm}
        \caption{\nocolorpa vs \nocolorvas}
    \end{subfigure}\hfill
    \begin{subfigure}[t]{0.19\textwidth}
        \centering
        \includegraphics[width=1.02\textwidth]{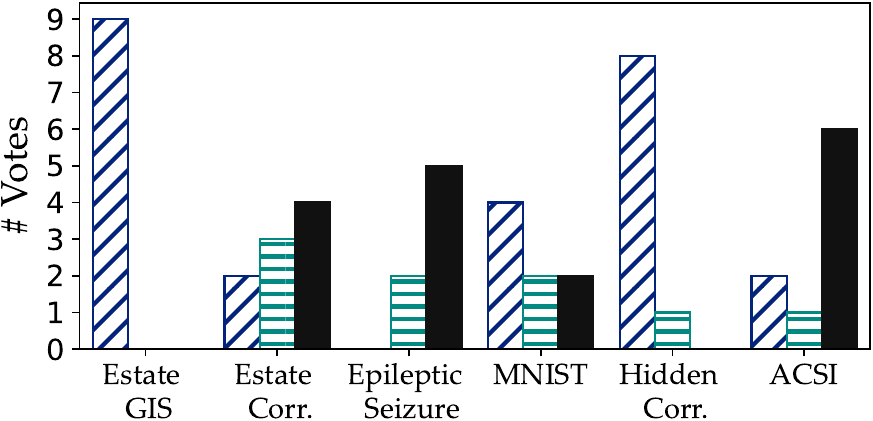}
        \vspace{-5mm}
        \caption{\nocolorpa vs \nocolorMM}
    \end{subfigure}\hfill
    \begin{subfigure}[t]{0.19\textwidth}
        \centering
        \includegraphics[width=1.02\textwidth]{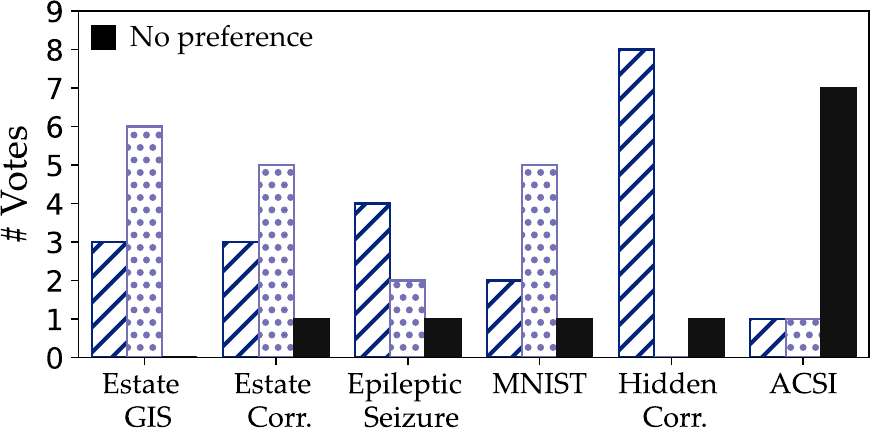}
        \vspace{-5mm}
        \caption{\nocolorpa vs \nocolorbn}
    \end{subfigure}
    \vspace{-4mm}
    \caption{A per-dataset analysis of users' preferences for the sample that better represents correlation trends, clusters, and other patterns.}
    \label{fig:studyRes1}
\end{figure*}

\begin{figure*}[t]
    \begin{subfigure}[t]{0.19\textwidth}
        \centering
        \includegraphics[width=1.02\textwidth]{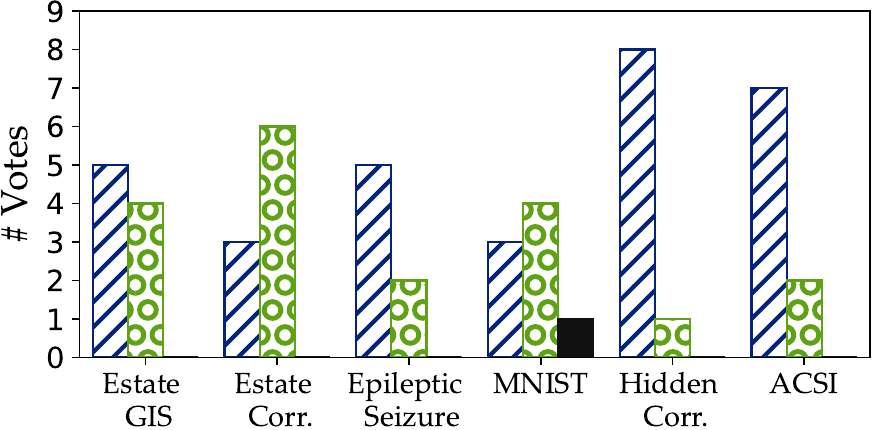}
        \vspace{-5mm}
        \caption{\nocolorpa vs \nocolorran}
    \end{subfigure}\hfill
    \begin{subfigure}[t]{0.19\textwidth}
        \centering
        \includegraphics[width=1.02\textwidth]{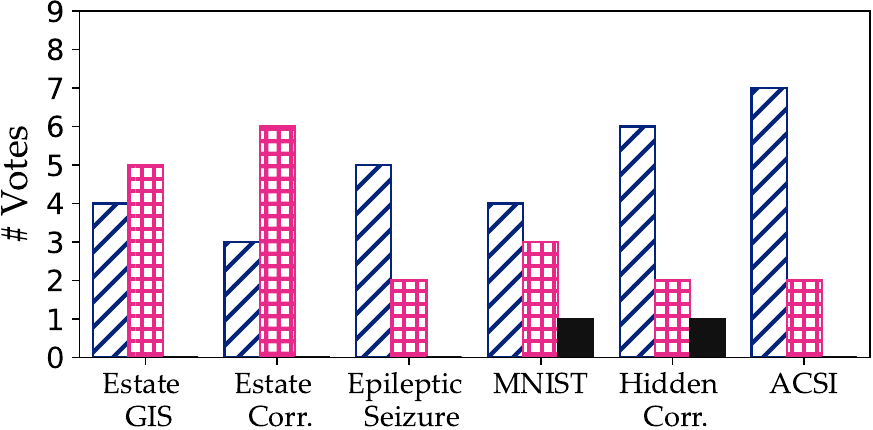}
        \vspace{-5mm}
        \caption{\nocolorpa vs \nocolordbs}
    \end{subfigure}\hfill
    \begin{subfigure}[t]{0.19\textwidth}
        \centering
        \includegraphics[width=1.02\textwidth]{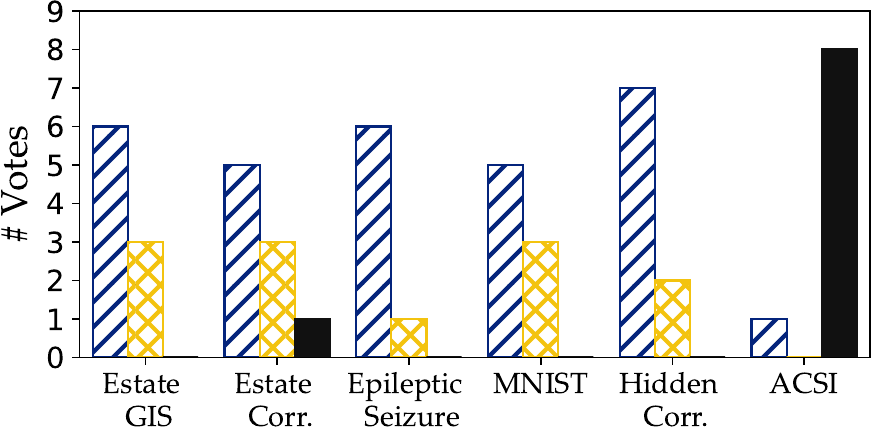}
        \vspace{-5mm}
        \caption{\nocolorpa vs \nocolorvas}
    \end{subfigure}\hfill
    \begin{subfigure}[t]{0.19\textwidth}
        \centering
        \includegraphics[width=1.02\textwidth]{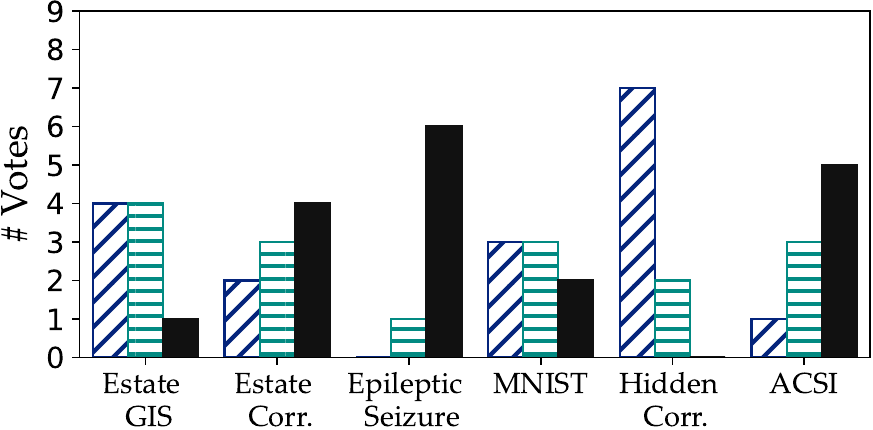}
        \vspace{-5mm}
        \caption{\nocolorpa vs \nocolorMM}
    \end{subfigure}\hfill
    \begin{subfigure}[t]{0.19\textwidth}
        \centering
        \includegraphics[width=1.02\textwidth]{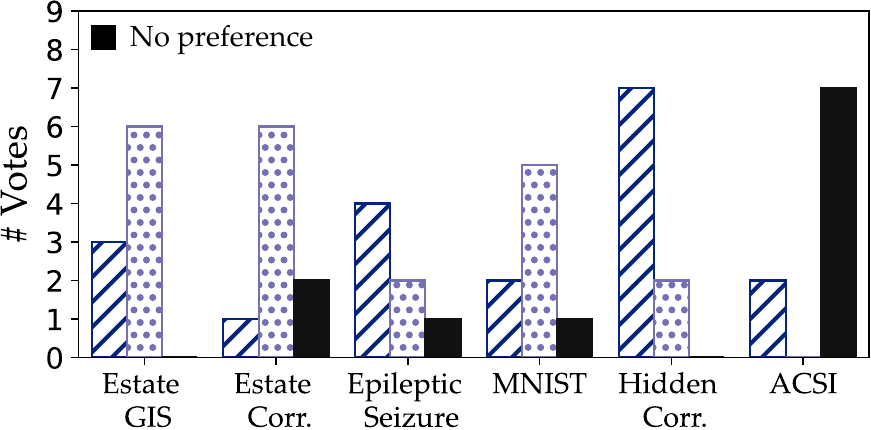}
        \vspace{-5mm}
        \caption{\nocolorpa vs \nocolorbn}
    \end{subfigure}
    \vspace{-4mm}
    \caption{A per-dataset analysis of users' preferences for the sample that better represents outliers and overall data shape. }
    \label{fig:studyRes2}
\end{figure*}

\begin{figure*}[t]
    \begin{subfigure}[t]{0.19\textwidth}
        \centering
        \includegraphics[width=1.02\textwidth]{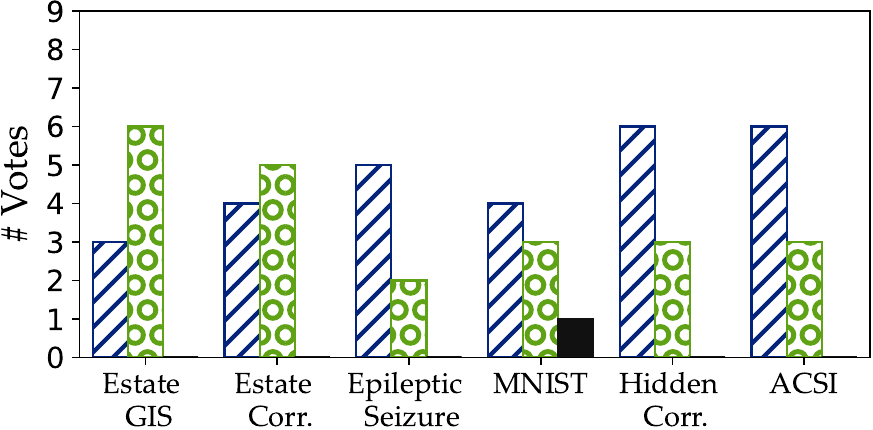}
        \vspace{-5mm}
        \caption{\nocolorpa vs \nocolorran}
    \end{subfigure}\hfill
    \begin{subfigure}[t]{0.19\textwidth}
        \centering
        \includegraphics[width=1.02\textwidth]{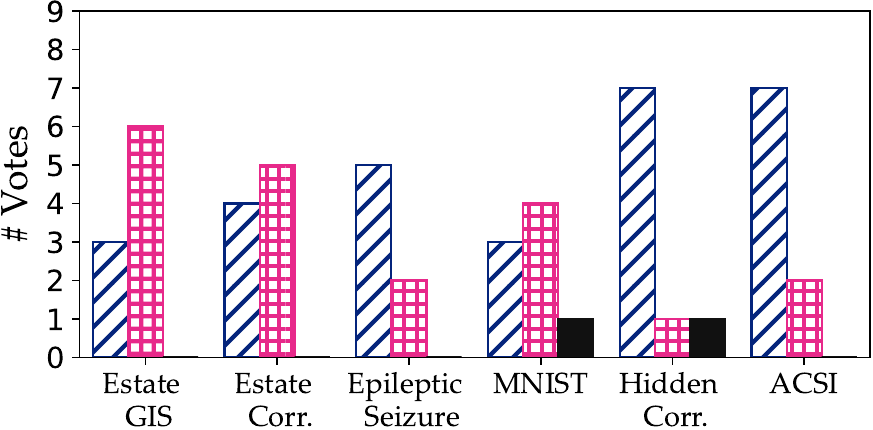}
        \vspace{-5mm}
        \caption{\nocolorpa vs \nocolordbs}
    \end{subfigure}\hfill
    \begin{subfigure}[t]{0.19\textwidth}
        \centering
        \includegraphics[width=1.02\textwidth]{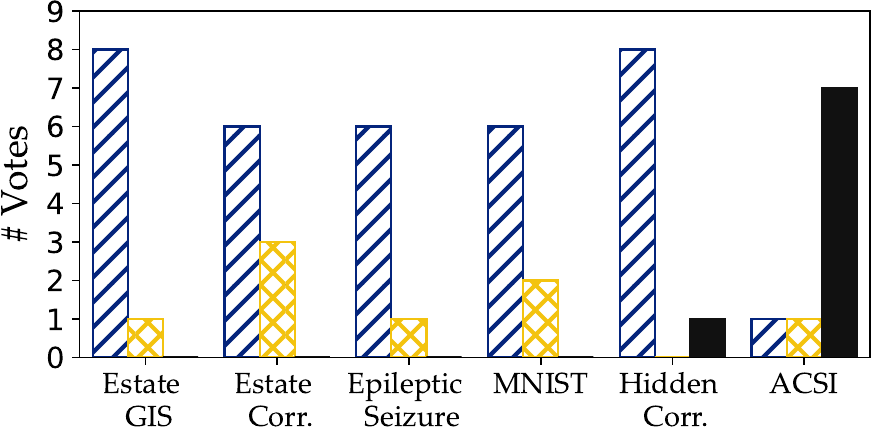}
        \vspace{-5mm}
        \caption{\nocolorpa vs \nocolorvas}
    \end{subfigure}\hfill
    \begin{subfigure}[t]{0.19\textwidth}
        \centering
        \includegraphics[width=1.02\textwidth]{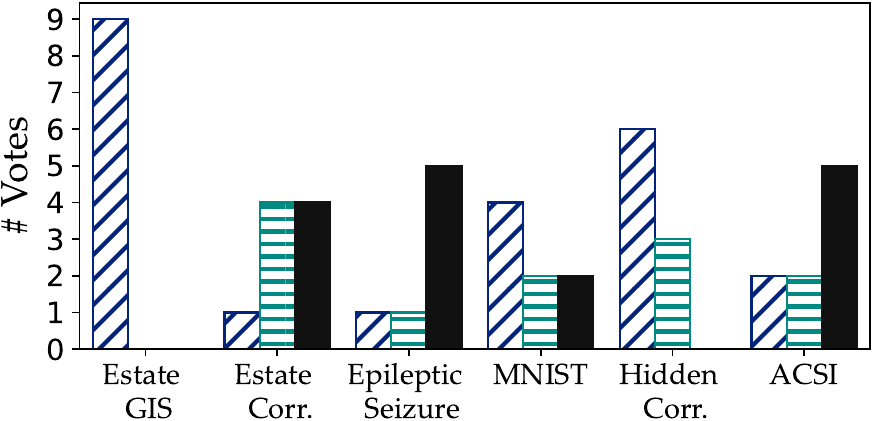}
        \vspace{-5mm}
        \caption{\nocolorpa vs \nocolorMM}
    \end{subfigure}\hfill
    \begin{subfigure}[t]{0.19\textwidth}
        \centering
        \includegraphics[width=1.02\textwidth]{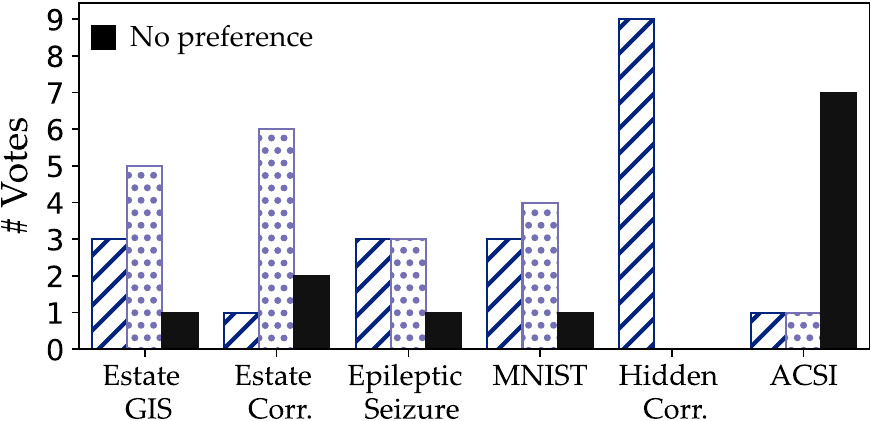}
        \vspace{-5mm}
        \caption{\nocolorpa vs \nocolorbn}
    \end{subfigure}
    \vspace{-4mm}
    \caption{A per-dataset analysis of users' preferences for the sample that better represents density variations.}
    \label{fig:studyRes3}
\end{figure*}

\begin{figure*}[t]
    \begin{subfigure}[t]{0.19\textwidth}
        \centering
        \includegraphics[width=1.02\textwidth]{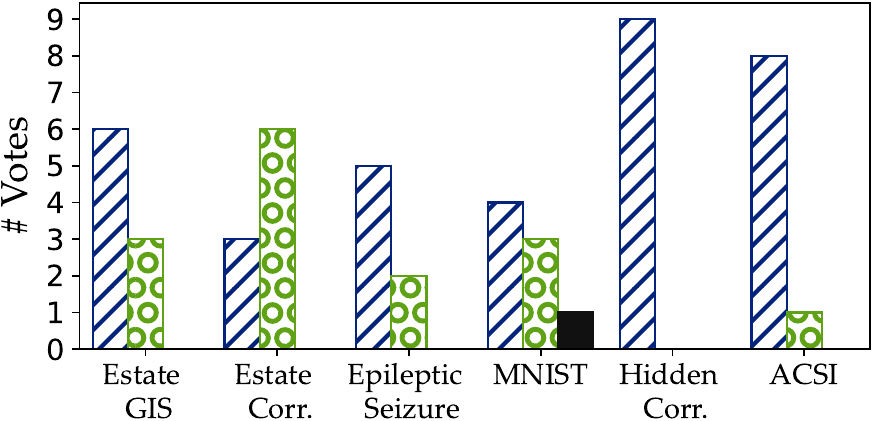}
        \vspace{-5mm}
        \caption{\nocolorpa vs \nocolorran}
    \end{subfigure}\hfill
    \begin{subfigure}[t]{0.19\textwidth}
        \centering
        \includegraphics[width=1.02\textwidth]{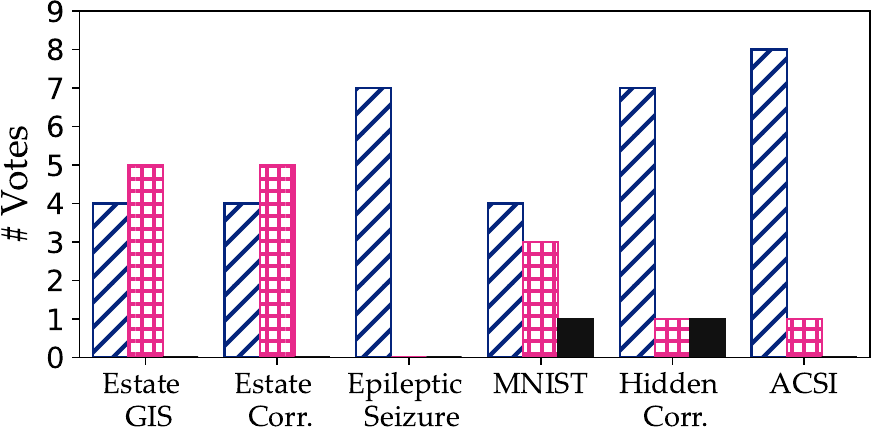}
        \vspace{-5mm}
        \caption{\nocolorpa vs \nocolordbs}
    \end{subfigure}\hfill
    \begin{subfigure}[t]{0.19\textwidth}
        \centering
        \includegraphics[width=1.02\textwidth]{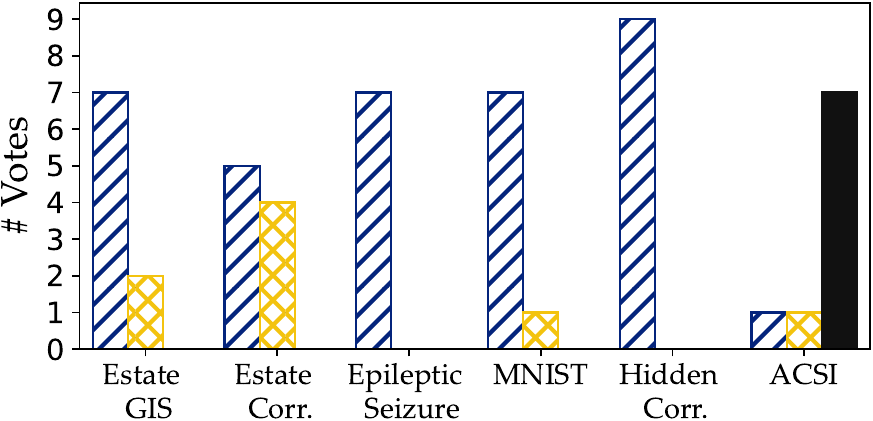}
        \vspace{-5mm}
        \caption{\nocolorpa vs \nocolorvas}
    \end{subfigure}\hfill
    \begin{subfigure}[t]{0.19\textwidth}
        \centering
        \includegraphics[width=1.02\textwidth]{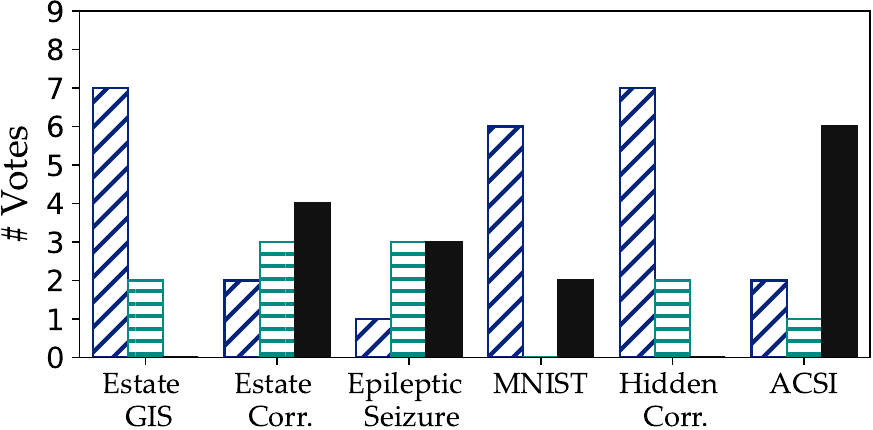}
        \vspace{-5mm}
        \caption{\nocolorpa vs \nocolorMM}
    \end{subfigure}\hfill
    \begin{subfigure}[t]{0.19\textwidth}
        \centering
        \includegraphics[width=1.02\textwidth]{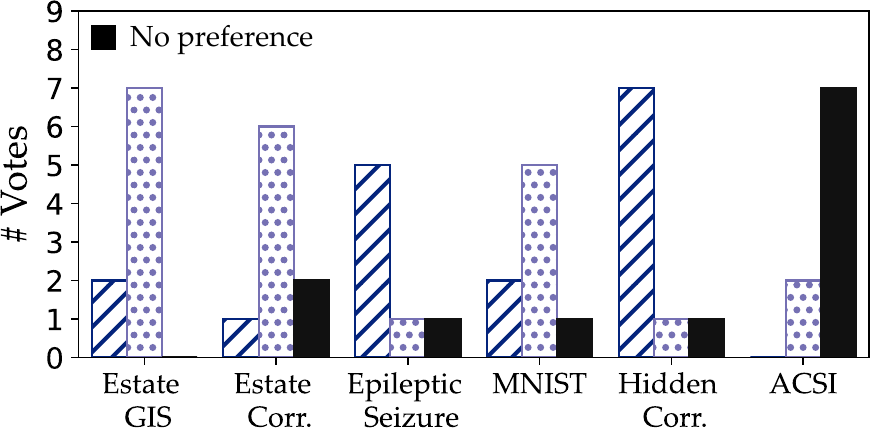}
        \vspace{-5mm}
        \caption{\nocolorpa vs \nocolorbn}
    \end{subfigure}
    \vspace{-4mm}
    \caption{A per-dataset analysis of users' preferences for the best overall sample.}
    \label{fig:studyRes4}
\end{figure*}

Figure~\ref{fig:studyRes1} reports how often participants preferred \nocolorpa over an alternative baseline concerning their efficacy in maintaining correlation trends, clusters, and other patterns; Figure~\ref{fig:studyRes2} reports how often participants preferred \nocolorpa over an alternative baseline concerning their efficacy in retaining outliers and overall data shape; and Figure~\ref{fig:studyRes3} report users' preference concerning density variations. Finally, Figure~\ref{fig:studyRes4} shows the result for the best overall sample per dataset.
We observe \nocolorpa is frequently preferred by humans compared to state-of-the-arts methods, while the preference becomes more evident in datasets where other sampling methods fail to represent some trend in the data; for instance, in hidden correlation, \nocolorpa is overly favored compared to all other methods, which fail to represent both the correlation trend and the overall data distribution. Similarly, in ACSI, humans overly prefer \nocolorpa compared to probabilistic-based approaches that create an evident underrepresented region in the sample.

\smallskip
\noindent
\textbf{Compression schemes and \nocolorapproxpa samples.} We report additional results for \nocolorapproxpa (Algorithm~2), which is our method for approximate visualizations.  \nocolorapproxpa uses perception-aware compressed data representations to derive samples without having access to the original data. Our compression approach (Section~4) uses two thresholds, $\lambda$ and $\sigma$, to derive these data representations and splits the canvas into non-overlapping bounding boxes that sufficiently approximate the data in them (as guided by $\lambda$), and cover areas of similar perception weights (as guided by $\sigma$). Then \nocolorapproxpa creates a sample by selecting points uniformly at random from these boxes. Lower values for the thresholds result in representations with a higher number of boxes that closely represent the underlying data (lower compression), and thus, the level of distortion observed in an \nocolorapproxpa sample is lower.

\begin{figure*}[h]
  \centering
  \begin{subfigure}[t]{0.33\textwidth}
    \centering
    \includegraphics[width=0.75\textwidth]{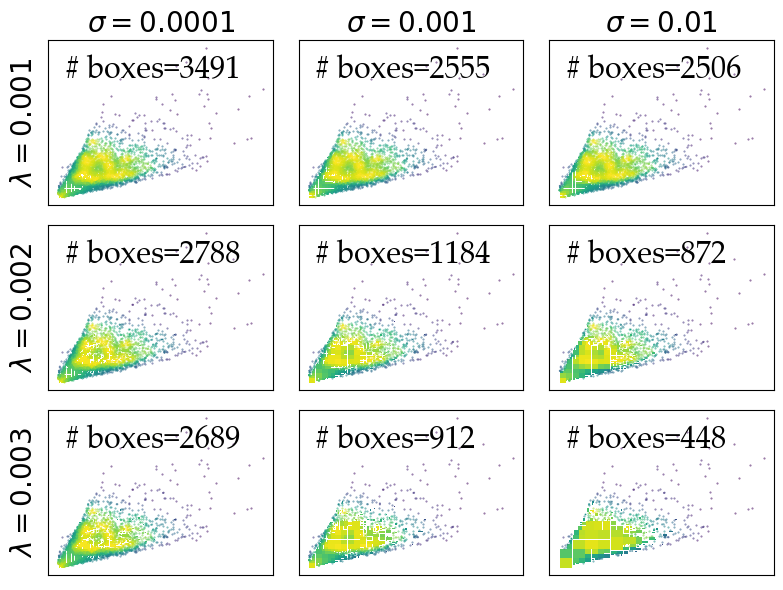}
    \vspace{-1mm}
    \caption{Estate Correlation}
    \label{fig:schemes_estate_corr}
  \end{subfigure}%
 \begin{subfigure}[t]{0.33\textwidth}
      \centering
      \includegraphics[width=0.75\textwidth]{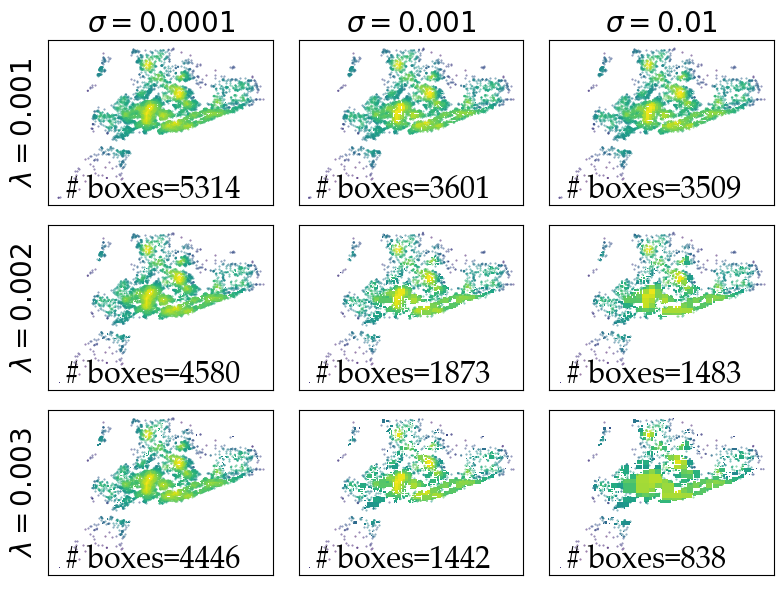}
      \vspace{-1mm}
      \caption{Estate GIS}
     \label{fig:schemes_estate_gis}
  \end{subfigure}%
  \begin{subfigure}[t]{0.33\textwidth}
      \centering
      \includegraphics[width=0.75\textwidth]{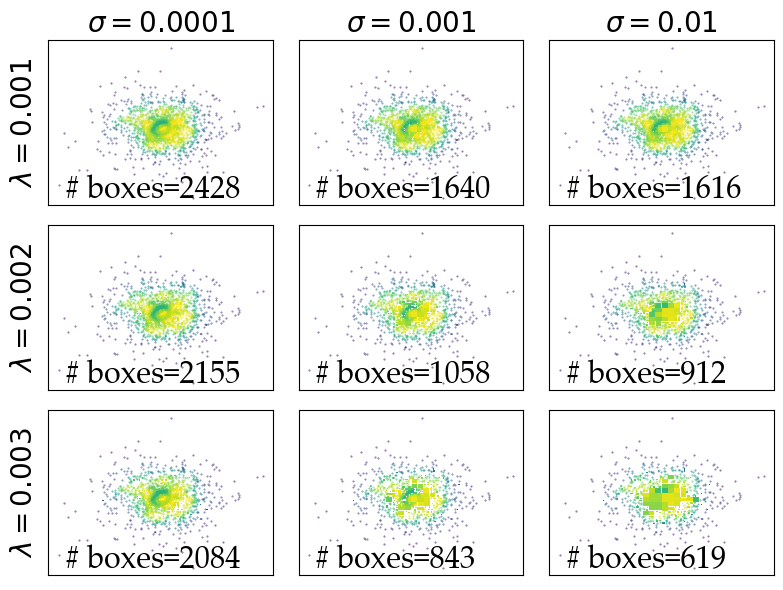}
      \vspace{-1mm}
      \caption{Epileptic Seizure}
      \label{fig:schemes_epileptic_corr}
  \end{subfigure}
  \begin{subfigure}[t]{0.33\textwidth}
    \centering
    \includegraphics[width=0.75\textwidth]{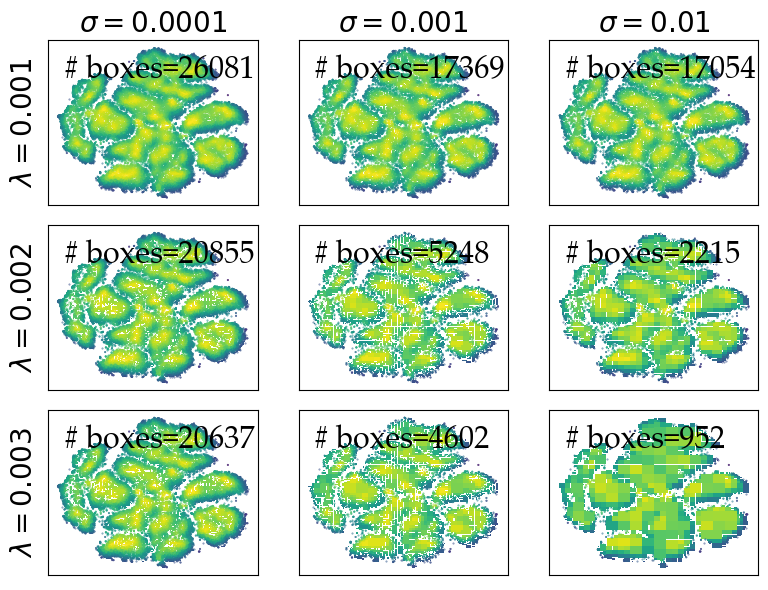}
    \vspace{-2mm}
    \caption{MNIST}
    \label{fig:schemes_mnist}
  \end{subfigure}%
 \begin{subfigure}[t]{0.33\textwidth}
      \centering
      \includegraphics[width=0.75\textwidth]{images/hidden_corr_approx_schemes_with_data.png}
      \vspace{-2mm}
      \caption{Hidden Correlation}
     \label{fig:schemes_acsi}
  \end{subfigure}%
  \begin{subfigure}[t]{0.33\textwidth}
      \centering
      \includegraphics[width=0.75\textwidth]{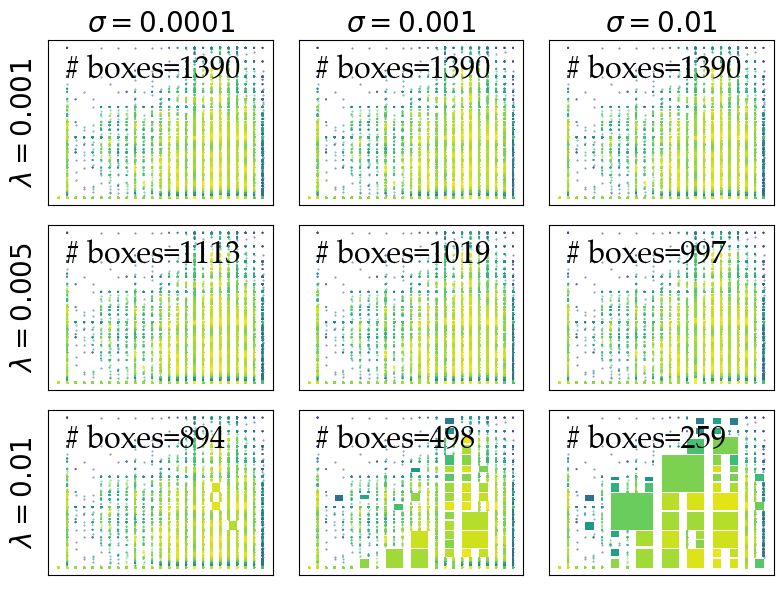}
      \vspace{-2mm}
      \caption{ACSI}
      \label{fig:schemes_hidden_corr}
  \end{subfigure}
  \vspace{-4mm}
  \caption{Perception-aware compression schemes for the datasets in our evaluation. We show how the parameters of the perception-aware compression approach affect the number of boxes of the scheme, and ultimately compression.}
  \label{fig:approxpa_schemes}
\end{figure*}

\begin{figure*}[h]
    \centering
    \begin{subfigure}[t]{0.33\textwidth}
        \centering
        \includegraphics[width=0.8\textwidth]{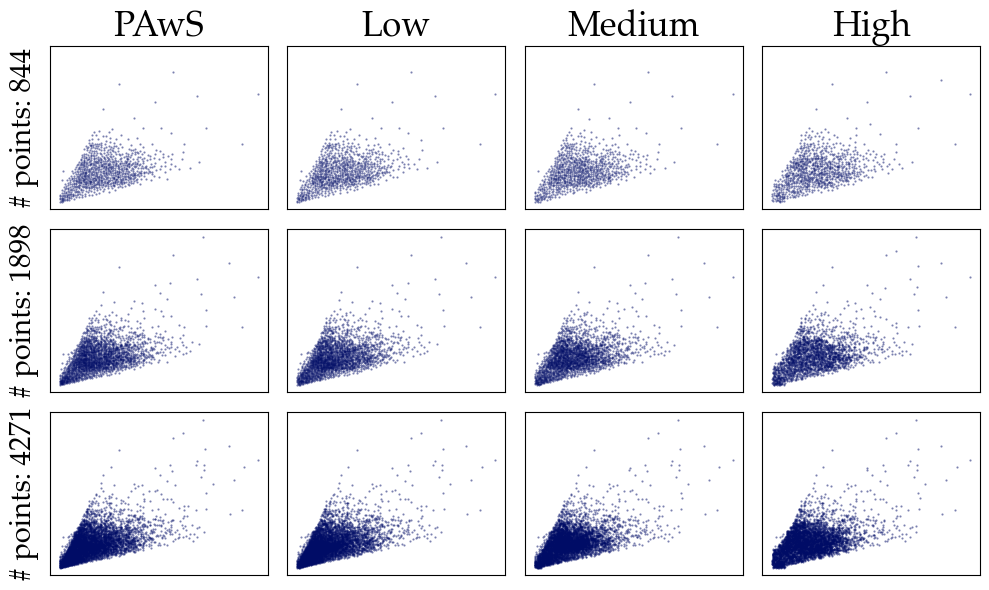}
        \vspace{-2mm}
        \caption{Estate Correlation}
        \label{fig:estate_corr_approx_qual}
    \end{subfigure}%
    \begin{subfigure}[t]{0.33\textwidth}
        \centering
        \includegraphics[width=0.8\textwidth]{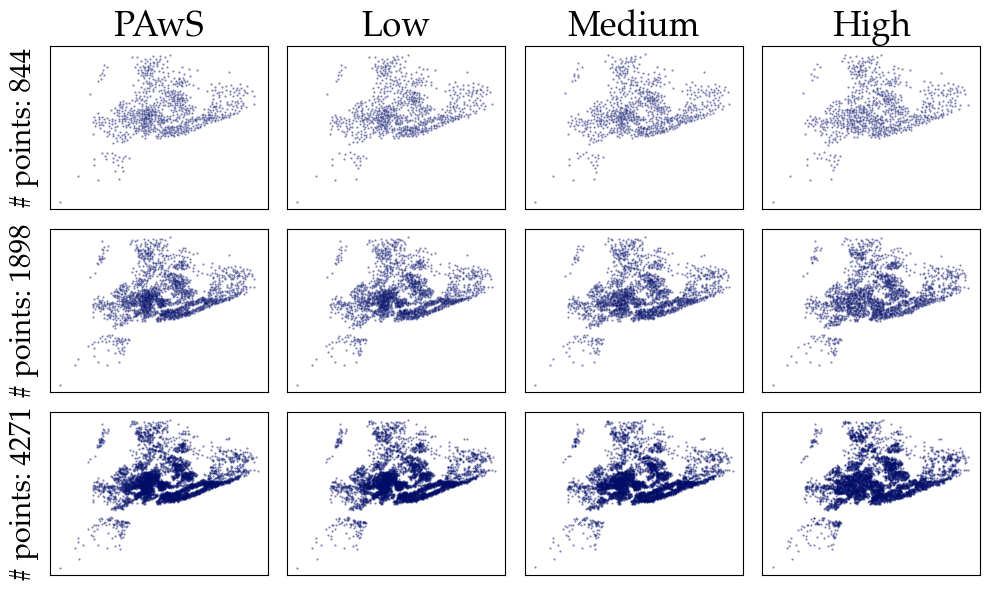}
        \vspace{-2mm}
        \caption{Estate GIS}
        \label{fig:estate_gis_approx_qual}
    \end{subfigure}%
    \begin{subfigure}[t]{0.33\textwidth}
        \centering
        \includegraphics[width=0.8\textwidth]{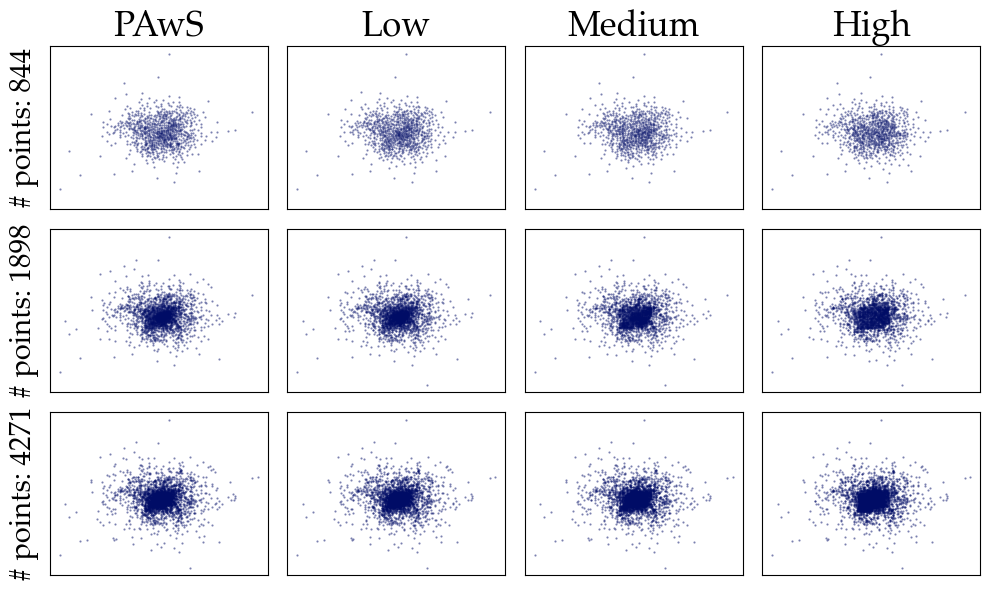}
        \vspace{-2mm}
        \caption{Epileptic Seizure}
        \label{fig:epileptic_corr_approx_qual}
    \end{subfigure}
    \\
    \begin{subfigure}[t]{0.33\textwidth}
        \centering
        \includegraphics[width=0.8\textwidth]{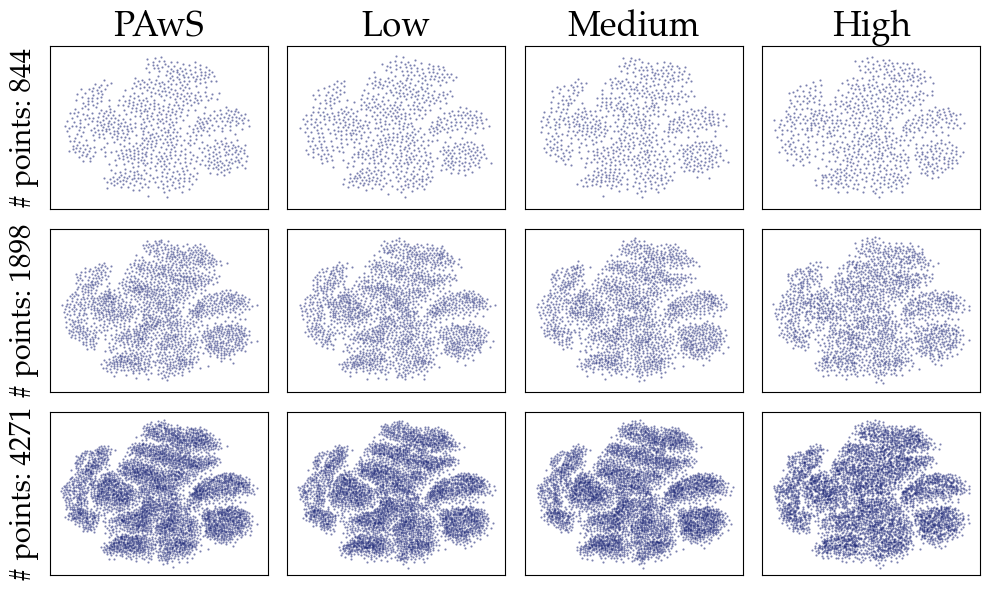}
        \vspace{-2mm}
        \caption{MNIST}
        \label{fig:mnist_approx_qual}
    \end{subfigure}%
    \begin{subfigure}[t]{0.33\textwidth}
         \centering
          \includegraphics[width=0.8\textwidth]{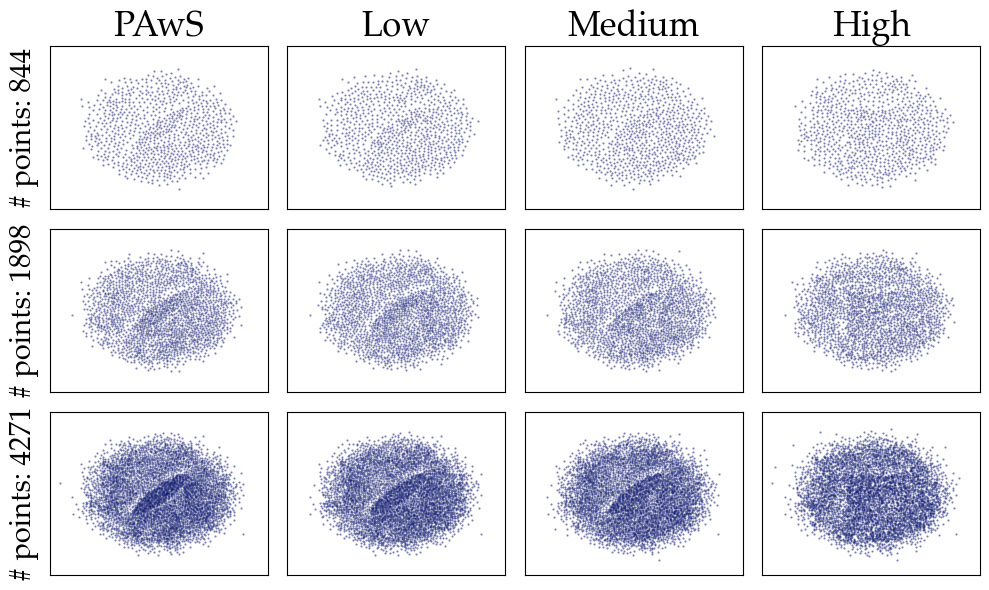}
          \vspace{-2mm}
          \caption{Hidden Correlation}
         \label{fig:acsi_approx_qual}
    \end{subfigure}%
    \begin{subfigure}[t]{0.33\textwidth}
          \centering
          \includegraphics[width=0.8\textwidth]{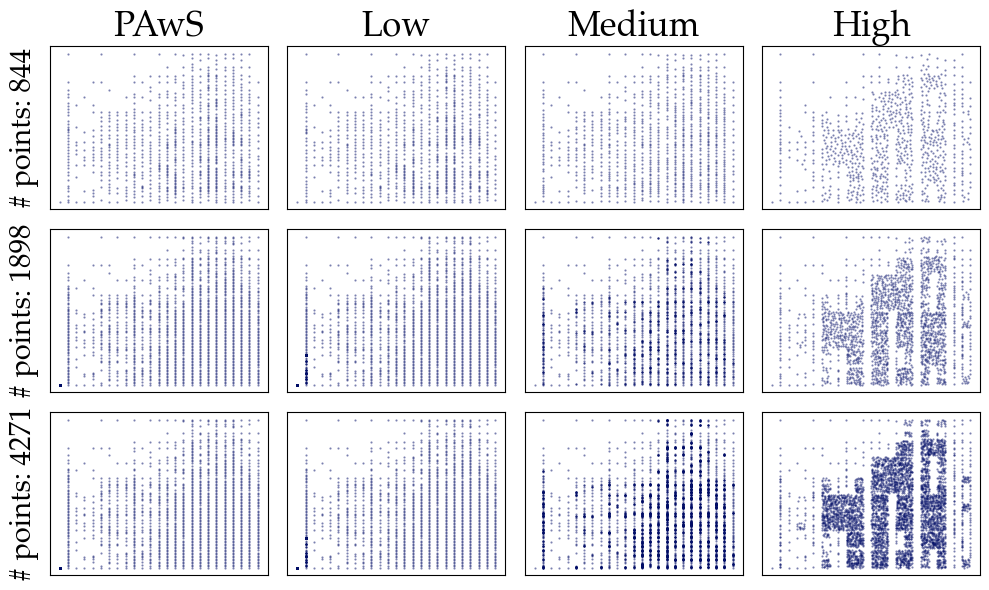}
          \vspace{-2mm}
          \caption{ACSI}
          \label{fig:hidden_corr_approx_qual}
    \end{subfigure}

  \vspace{-4mm}
  \caption{Approximate visualizations of various sample sizes, as derived by \nocolorapproxpa using three perception-aware compressed data representations. We show results for three levels of compression for the various datasets.}
  \label{fig:approxpa_qual1}
\end{figure*}

We present results at three compression levels: (1)~low: $\langle \lambda, \sigma \rangle= \langle 0.001,  0.001 \rangle$, (2)~medium: $\langle \lambda, \sigma \rangle= \langle 0.002, 0.001 \rangle$, and (3)~high: $\langle \lambda, \sigma \rangle= \langle 0.003,  0.01 \rangle$. For consistency and ease of exposition, we use the same threshold values for all datasets but ACSI. For ACSI, to observe notable differences across compression schemes, we vary the thresholds as: (1)~low: $\langle \lambda, \sigma \rangle= \langle 0.001,  0.001 \rangle$, (2)~medium: $\langle \lambda, \sigma \rangle= \langle 0.005, 0.001 \rangle$, and (3)~high: $\langle \lambda, \sigma \rangle= \langle 0.01,  0.01 \rangle$. Figure~\ref{fig:approxpa_schemes} shows the derived compression schemes, and Figure~\ref{fig:approxpa_qual1} shows samples, across three sampling sizes, derived by \nocolorapproxpa while using the corresponding data representations. We observe that qualitative degradation becomes more visible in high compression schemes, but \nocolorapproxpa still captures the main trends and shape of the datasets.

